\documentclass[12pt]{article}
\usepackage[english]{babel}
\usepackage{hyperref}
\usepackage{amsmath}
\usepackage{amsfonts}
\usepackage{amssymb}
\usepackage[dvips]{graphicx}
\usepackage{bm}

\begin{document}
\date{}

\begin{center}
{\Large\textbf{{}Covariant    Cubic Interacting Vertices for \\Massless
and Massive Integer Higher Spin Fields}} \vspace{18mm}

{\large I.L. Buchbinder$^{(a,b,c,d)}\footnote{E-mail:
buchbinder@theor.jinr.ru; joseph@tspu.edu.ru}$ ,\; A.A. Reshetnyak$^{(a,c,e)}\footnote{E-mail:
reshet@tspu.edu.ru}$}

\vspace{8mm}

\noindent ${{}^{(a)}}${\em Bogoliubov Laboratory of Theoretical
Physics,\\
Joint Institute for Nuclear Research,\\ 141980 Dubna, Moscow Region, Russia}

\noindent  ${{}^{(b)}} ${\em
Center of Theoretical Physics, \\
Tomsk State Pedagogical University,\\
634061 Tomsk, Russia}

\noindent  ${{}^{(c)}} ${\em
National Research Tomsk State  University,\\
634050 Tomsk, Russia}

\noindent ${{}^{(d)}}${\em Tomsk State University of Control
Systems\\
and Radioelectronics, 634050, Tomsk, Russia}

\noindent  ${{}^{(e)}} ${\em
National Research Tomsk Polytechnic   University,\\
634050 Tomsk, Russia} \vspace{20mm}

\begin{abstract}
We develop the BRST approach to construct the general off-shell
local Lorentz covariant cubic interaction vertices for irreducible
massless and massive higher spin fields on $d$-dimensional Minkowski space. We
consider two different cases for interacting higher spin fields:
with one massive and two massless; with two massive both with
coinciding and with different masses  and one massless  fields   of
spins $s_1, s_2, s_3$. Unlike the previous results on cubic vertices
we extend our earlier result  in \cite{BRcub} for massless fields
and employ  the complete BRST operator, including the trace constraints,
which is used to formulate an irreducible representation with definite
integer spin. We generalize the cubic vertices proposed for
reducible higher spin fields in \cite{BRST-BV3} in the form of
multiplicative and non-multiplicative BRST-closed constituents and
calculate the new contributions to the vertex, which contains
the additional terms with a smaller number of space-time derivatives.
We prove that without  traceless conditions for the cubic vertices in \cite{BRST-BV3}
it is impossible to provide the noncontradictory Lagrangian dynamics and find explicit
traceless solution for these vertices.  As the
examples, we explicitly construct the interacting Lagrangians for the
massive of spin  $s$ field and massless scalars both with and
without  auxiliary fields. The interacting models with different
combinations of triples higher spin  fields: massive of spin $s$
with massless scalar and vector fields and with two vector fields;
massless of helicity $\lambda$ with massless scalar and massive
vector fields; two massive fields of spins $s, 0$ and massless
scalar  are also considered.
\end{abstract}

\end{center}
\thispagestyle{empty}

\section{Introduction}

{The construction of interacting higher spin field theory attracts a significant attention both from a general theoretical
point of view and in connection with the possibilities of discovering new approaches to describe a gravity at the quantum level}
(see for a review, e.g. \cite{revvas},  \cite{revBCIV},
\cite{reviews3}, \cite{rev_Bekaert}, \cite{DidSk}, \cite{reviewsV},
\cite{Ponomarev} and the references therein). The extension of General Relativity on a base of local supersymmetry principle up to the supergravity models \cite{Nieuwenhuizen} with improved quantum properties and  a connection with  (Super)string  Field Theory permits one to include  massless fields of  spins $s>2$ in  Higher  Spin Gravity (see \cite{Snowmass} and references therein) with  respecting the string field theory properties, asymptotic safety and some others.  The AdS/CFT correspondence gives strong indications
that higher spin excitations  can be significant  to elaborate the quantum gravity challenges \cite{Giombi}.
Interacting massive and massless higher spin fields in
constant-curvature spaces  provide another   possible
insight into the origin of Dark Matter and Dark Energy \cite{LHC},  \cite{LHC2} beyond the
 models with  vector
massive fields \cite{vectbosDM} and sterile neutrinos \cite{sterilen} to be by reasonable  candidates for Dark
Matter, see  for reviews \cite{DM1}, \cite{DM2}, \cite{Odintsov1}.

{The simplest of higher spin interactions, the cubic vertex for various fields with higher spins,
has been studied by many authors} with use of different approaches
(see, e.g., the recent papers \cite{Manvelyan}, \cite{Manvelyan1},
\cite{Joung}, \cite{Joung1},  \cite{frame-like1}, \cite{Metsaev0712}, \cite{Tsulaiai2009}
\cite{BRST-BV3},  \cite{FranciaMM}, \cite{JoungMP}, \cite{frame-like2}, \cite{zinoviev2}, \cite{BKTW},
\cite{Metsaev-mass} and the references therein)\footnote{A complete
list of papers on cubic vertex on constant curvature spaces contains dozens of papers. Here we
cite only the recent papers containing a full list of references.}.
Note the results on the structure of cubic vertices
obtained  in terms of physical degrees of freedom in a concise form
in the light-cone {approach} in \cite{Metsaev0512},
\cite{Metsaev-mass}. In the covariant metric-like form the list of
cubic vertices for reducible representations of Poincare group with
discrete spins (being consistent with \cite{Metsaev0512}) are
contained in \cite{BRST-BV3}, where the cubic vertices were derived
  using the constrained  BRST approach, but without imposing on the
vertex  the algebraic constraints. The latter peculiarity
 leads to the violation of the irreducibility of the
representation for interacting higher spin fields and, hence, to a
{possible} change of the number of physical degrees of freedom\footnote{{Without finding the solution for the vertex respecting the algebraic constraints, the
number of physical degrees of freedom}, which is determined by one of independent initial data for the equations of motion   for the interacting model
is different (less) than
as one  from that for the undeformed model with vanishing algebraic constraints evaluated on respecting equations of motion,{ but with the deformed gauge symmetry not respecting these constraints}.}.
Also, we point out the constructions of cubic vertices
within the BRST approach without use of constraints responsible for
trace conditions in the BRST charge (see e.g. \cite{BKTW} and the
references therein). It means, in fact, that the vertex is obtained
in terms of reducible higher spin fields\footnote{{To avoid various misunderstandings, we emphasize that we use the term “unconstrained formulation” in the
sense that all possible constraints are consequences of the Lagrangian equations of motion. No additional restrictions, separate from the equations of motion, are imposed.}}.

In this paper, we derive the cubic vertices for irreducible {massless} and massive
higher spin fields focusing on the manifest Lorentz covariance.
Analysis is carried out within BRST approach with complete BRST
operator {that extends our earlier approaches \cite{BRcub},
\cite{Rcubmasless},\cite{BKStwis} and involves} a converted set of operator constraints
forming a first-class gauge algebra. The set of constraints includes
on equal-footing the on-shell condition $l_0$ and constraints $l_1,
\, l_{11}$, responsible for divergences and traces. {Unlike our consideration, in the constrained BRST approach,  the
operator $l_{11}$ is imposed as a constraint on the set of fields and gauge parameters outside of  the
Lagrangian formulation} for simplicity of calculations. Such an approach inherits the way of obtaining
the Lagrangian formulation for higher spin fields  from the
tensionless limit \cite{BRST-BFV1} for (super)string theory with
resulting BRST charge without presence of the algebraic (e.g. trace)
constraints. We have already noted \cite{BRcub} that {this way of consideration} is
correct but the actual Lagrangian description of irreducible fields
is achieved only after {additional} imposing the subsidiary
conditions which are not derived from the Lagrangian. Of course, the
Lagrangian formulations for the same {free} irreducible higher spin field in Minkowski space obtained in {constrained
and unconstrained BRST approaches are equivalent}
 \cite{Reshetnyak_con}\footnote{For irreducible massless and
massive field representations with half-integer spin the Lagrangians
with reducible gauge symmetries and compatible holonomic
constraints, were firstly obtained therein.}. However, the {corresponding} equivalence
has not yet been {proved} for interacting irreducible
higher-spin fields as it was recently demonstrated for massless case
\cite{BRcub}, \cite{Rcubmasless} for cubic vertices.
Aspects of the BRST approach with complete BRST operator for a
Lagrangian description of various free and interacting massive
higher spin field models in Minkowski and AdS spaces were
developed in many works (e.g., see the papers \cite{PT}, \cite{BPT},
\cite{BKP}, \cite{BKr}, \cite{BFPT}, \cite{BGK}, \cite{BG},
\cite{BR}, and the review \cite{reviews3}).

As a result we face the problem  when constructing the cubic vertex
for irreducible mass\-less and massive higher integer spin fields
on $d$-dimensional flat space-time within met\-ric-like formalism on
the base of complete BRST operator. It is exactly the problem that
we intend to {consider} in the paper. We  expect that the final cubic
vertices will contain new terms (as compared with \cite{BRST-BV3})
with the {traces} of the fields. {Such new terms may evidently have significance
when gauging away auxiliary gauge symmetry and fields to get a component Lagrangian formulation}.


The aim of the paper is
to present a complete solution of {above} problem for the cubic vertices for
unconstrained irreducible  massless and massive higher spin fields within
BRST approach and to obtain from general oscillator-like vertices
explicit tensor representations for Lagrangian formulations with
reducible gauge symmetry  for some  triples of interacting higher
spin fields.

The paper has the following organization.  Section~\ref{cBRSTBFV}
presents the basics of a BRST Lagrangian construction for free
massive higher spin field, with all constraints $l_0,\, l_1,\,
l_{11}$ taken into account. In Section~\ref{BRSTinter}, we deduce a
system of equations for a cubic (linear) deformation in fields of
the free action (free gauge transformations). A solution for the
deformed cubic vertices and gauge transformation is given in a
Section~\ref{BRSTsolgen} for one massive and two massless fields;
for two  massive with different and coinciding masses and one
massless fields. The number of examples for the fields with  special
set of spins are presented in the Section~\ref{examples}.  The main
result of the work is that the cubic vertices and deformed reducible
gauge transformations include both types of constraints: with
derivative $l_{1}$ and one with trace  $l_{11}$. In conclusion  a
final summary with comments are given. A derivation of Singh-Hagen
Lagrangian from free BRST Lagrangian formulation for massive field
of spin $s$ presented in appendix~\ref{Singhcvomp}.
Appendices~\ref{Singhcvomp1},~\ref{Singhcvomp12} contain   results of calculations for
component interacting Lagrangian and gauge transformations for
massive field of spin $s$ with massless scalars and with massless vector and scalar. In appendix~\ref{appconBRST}
we formulate conditions for the incomplete BRST operator, traceless constraints and cubic vertices
to get non-contradictory Lagrangian dynamics for a model with interacting fields with given spins.
{We find the form of projectors $\prod_{j=1}^3P^{(j)}_{0|11}$  for .respective cubic vertices
$\big|  V^{(3)}_c\rangle^{(m)_3}_{(s)_{3}}$  from \cite{BRST-BV3} to have the cubic vertices
$\big|  \overline{V}{}^{(3)}_c\rangle^{(m)_3}_{(s)_{3}} $, firstly determined by (\ref{projectors})
and (\ref{L11V+})  for irreducible interacting fields.
}
We use the usual definitions and notations  from the work
\cite{BRcub} for a metric tensor $ \eta_{\mu\nu} = diag (+,
-,...,-)$ with Lorentz indices $\mu, \nu = 0,1,...,d-1$ and the
respective notation $\epsilon(F)$, $gh(F)$, $[F,\,G\}$, $[x]$, $\theta_{k,l}$ $= 1(0)$, when $k>l (k\leq l)$,
$(s)_{3}$  for the values of Grassmann parity and ghost number of a
homogeneous quantity $F$, as well as the supercommutator,  the
integer part of a real-valued $x$, Heaviside $\theta$-symbol  and for  the  triple  $
(s_1,s_2,s_3)$.

\section{Lagrangian formulation for free massive higher spin  fields}

\label{cBRSTBFV}

Here, we shortly present the basics  of the BRST approach to free
massive higher integer spin field theory for its following use to
construct a general cubic interacting vertex.

The unitary massive irreducible representations of
Poincare $ISO(1,d-1)$ group with integer spins $s$ can be
realized using the real-valued totally symmetric tensor fields
$\phi_{\mu_1...\mu_s}(x)\equiv \phi_{\mu(s)}$ subject to the
conditions
\begin{eqnarray}\label{irrepint}
    &&  \big(\partial^\nu\partial_\nu + m^2,\, \partial^{\mu_1},\, \eta^{\mu_1\mu_2}\big)\phi_{\mu(s)}  = (0,0,0)  \  \ \ \   \Longleftrightarrow  \  \\
     &&       \big(l_0,\, l_1,\, l_{11}, g_0 -d/2\big)|\phi\rangle  = (0,0,0,s)|\phi\rangle. \nonumber
\end{eqnarray}
The basic vectors $|\phi\rangle$ and the operators $l_0,\,
l_1,\, l_{11}, g_0 -d/2$ above  are defined in the Fock space $\mathcal{H}$
with the Grassmann-even oscillators $a_\mu, a^+_\nu$, ($[a_\mu, a^+_\nu]= - \eta_{\mu\nu}$)  as follows
\begin{eqnarray}\label{FVoper}
&&   |\phi\rangle  =  \sum_{s\geq 0}\frac{\imath^s}{s!}\phi^{\mu(s)}\prod_{i=1}^s a^+_{\mu_i}|0\rangle, \\
&&   \big(l_0,\, l_1,\, l_{11}, g_0\big) = \big(\partial^\nu\partial_\nu+ m^2 ,\, - \imath a^\nu  \partial_\nu ,\, \frac{1}{2}a^\mu a_\mu ,  -\frac{1}{2}\big\{a^+_{\mu},\, a^{\mu}\big\}\big).\nonumber
\end{eqnarray}
The free dynamics of the field with
definite spin $s$ in the framework of  BRST approach  is described by the first-stage reducible gauge theory with the gauge invariant action given on the configuration space $M^{(s)}_{cl}$ whose dimension grows with the growth of $"s"$, thus, including the basic field $\phi_{\mu(s)}$ with many auxiliary
fields $\phi_{1\mu(s-1)},...$ of lesser than $s$ ranks. All these
fields are incorporated into the vector $|\chi\rangle_s$ and the dynamics is
encoded  by the action
\begin{eqnarray}
\label{PhysStatetot} \mathcal{S}^m_{0|s}[\phi,\phi_1,...]=
\mathcal{S}^m_{0|s}[|\chi\rangle_s] = \int d\eta_0 {}_s\langle\chi|
KQ|\chi\rangle_s,
\end{eqnarray}
where $\eta_0$ and $K$ be respectively a zero-mode ghost field and an operator defining the
inner product. The action (\ref{PhysStatetot}) is invariant under
the reducible gauge transformations
\begin{eqnarray}
\label{gauge trasnform}
\delta|\chi\rangle_s =  Q|\Lambda\rangle_s , \ \ \delta |\Lambda\rangle_s = Q|\Lambda^1\rangle_s
,  \ \ \delta |\Lambda^1\rangle_s =0,
\end{eqnarray}
with  $|\Lambda\rangle_s$, $|\Lambda^1\rangle_s$ to be the
vectors of zero-level and first-level gauge  parameters of the
abelian gauge transformations (\ref{gauge trasnform}). The
quantity $Q$ in (\ref{PhysStatetot}) is the  BRST operator having the same structure as one for massless case \cite{BRcub}
constructed on the base of the  constraints  $l_0,\, {l}_1,\, {l}{}^{+}_1,\,
{l}{}_{11},\, {l}{}^{+}_{11} = \frac{1}{2}a^{+\nu}a^{+}_{\nu}$ with the
Grassmann-odd  ghost operators $\eta_0,\, \eta_1^+,\, \eta_1,\,
\eta_{11}^+,\, \eta_{11}$,  $ {\cal{}P}_0$,  $ \mathcal{P}_{1}$,  $
\mathcal{P}^+_{1},$ $\mathcal{P}_{11},\, \mathcal{P}^+_{11},$
\begin{eqnarray}
&& {Q} =
\eta_0l_0+\eta_1^+\check{l}_1+\check{l}_1^{+}\eta_1+
\eta_{11}^+\widehat{L}_{11}+\widehat{L}_{11}^{+}\eta_{11} +
{\imath}\eta_1^+\eta_1{\cal{}P}_0,
\label{Qctotsym}
\end{eqnarray}
where
\begin{eqnarray}
\hspace{-0.5ex}&\hspace{-0.5ex}&\hspace{-0.5ex} \big( \check{l}_1,\, \check{l}_1^{+}  \big) =  \big( {l}_1 + m d,\, {l}_1^{+}+ m d^+  \big), \ \
\big(\widehat{L}_{11} ,\,\widehat{L}{}^+_{11}\big) =  \big(
\check{L}_{11}+\eta_{1} \mathcal{P}_{1} , \,
\check{L}{}^+_{11}+\mathcal{P}^+_{1}\eta^+_{1} \big).
\label{extconstsp2}
\end{eqnarray}
Here
\begin{eqnarray} \label{extconstsp21}
\check{L}_{11}={l}_{11}- (1/2)(d)^2+(b^+b+h)b,\,\, \ \  \check{L}{}^{+}_{11}={l}^+_{11}- (1/2)(d^+)^2 +b^+
\end{eqnarray}
and $(\epsilon, gh) Q = (1, 1).$ The algebra of the operators $l_0$ ,$l_1$, $l^{+}_1, \check{L}_{11}, \check{L}_{11}^+,
G_0$ looks like
\begin{equation}\label{subalgebr}
[l_0, l^{(+)}_1] = 0, \ [l_1,l_1^+]=l_0 -m^2 \quad \mathrm{and} \quad  [\check{L}_{11},  \check{L}_{11}^+] = G_0,\
[G_0, \check{L}_{11}^{+}] = 2\check{L}_{11}^+
\end{equation}
and their independent non-vanishing  cross-commutators are $[l_1,\check{L}_{11}^+]=-l_1^+$,  $[l_1,G_{0}]=l_1$.

The ghost operators satisfy the non-zero  anticommuting relations
\begin{equation}\label{ghanticomm}
  \{\eta_0, \mathcal{P}_0\}= \imath,\ \  \ \{\eta_1, \mathcal{P}_1^+\}=\{\eta^+_1, \mathcal{P}_1\}= \{\eta_{11},   \mathcal{P}_{11}^+\}=\{\eta_{11}^+,   \mathcal{P}_{11}\}=1.
\end{equation}

The theory is characterized by the spin operator
${\sigma}$, which is defined according to
\begin{eqnarray}
\hspace{-0.5ex}&\hspace{-0.5ex}&\hspace{-0.5ex}  {\sigma}  =   G_0+ \eta_1^+\mathcal{P}_{1}
-\eta_1\mathcal{P}_{1}^+  + 2(\eta_{11}^+\mathcal{P}_{11} -\eta_{11}\mathcal{P}_{11}^+)
\label{extconstsp3}\\
\hspace{-0.5ex}&\hspace{-0.5ex}&\hspace{-0.5ex} G_0=g_0 + d^+d +2b^+b+ \frac{1}{2}+ h. \nonumber
\end{eqnarray}
Here $d,\, d^+$,  $b,\, b^+$ ($[d,\, d^+]=1$, $[b,\, b^+]=1$) be two pairs of auxiliary Grassmann-even oscillators.
The operator ${\sigma}$ selects the vectors
with definite spin value $s$
\begin{eqnarray}
 \hspace{-0.5ex}&\hspace{-0.5ex}&\hspace{-0.5ex} {\sigma} (|\chi\rangle_s,\,
 |\Lambda\rangle_s,\, |\Lambda^1\rangle_s)  = (0,0,0),
\label{extconstsp}
\end{eqnarray}
where the standard distribution for  Grassmann parities and the ghost numbers of the these vectors are $(0,0)$, $(1,-1),$ $(0,-2)$
respectively.

All the operators  above act  in a total Hilbert space with the scalar product of the vectors depending
on all oscillators  $(A;A^+)$ = $(a^{\mu},b,d; a^{\mu+},b^+,d^+)$ and ghosts
\begin{eqnarray}
&& \langle\chi |\psi\rangle = \int d^d x \langle0|  \chi^*\big(A;\eta_0, \eta_1, \mathcal{P}_1,\eta_{11}, \mathcal{P}_{11}\big)\psi\big(A^+;\eta_0,\eta^+_1, \mathcal{P}^+_1,\eta^+_{11}, \mathcal{P}^+_{11}\big)|0\rangle.
\label{scalarprod}
\end{eqnarray}

The operators $Q, {\sigma}$ are supercommuting and  Hermitian with
respect to the scalar product (\ref{scalarprod}) including the
operator $K$ (see e.g., \cite{Reshetnyak_con}, \cite{BPT}, \cite{BR})  being equal to $1$ on Hilbert subspace not depending on auxiliary $b, b^+$ operators
  \begin{align}\label{geneq}
   & Q^2 =  \eta_{11}^+\eta_{11} \sigma ,\ &&  [Q,\, \sigma\} =0; \\
   & Q^+K =  KQ ,  \ && \sigma^+K =  K\sigma ,  \label{geneq1}\\
   &    K=1\otimes \sum_{n=0}^{\infty}\frac{1}{n!}(b^+)n|0\rangle\langle 0|b^n C(n,h(s)),
&&  C(n,h(s))\equiv \prod_{i=0}^{n-1}(i+h(s))
   \label{geneq2}
  \end{align}
The BRST operator $Q$ is nilpotent on the subspace with zero
eigenvectors for the spin operator $\sigma$ (\ref{extconstsp}).

The field $ |\chi\rangle_s$, the zero $|\Lambda\rangle_s$ and  the
first $|\Lambda^1\rangle_s$ level gauge parameters
 labeled by the symbol $"s"$ as eigenvectors
of the spin condition in  (\ref{extconstsp}) has the same decomposition as ones  in \cite{BRcub} but with
ghost-independent vectors $|\Phi_{...}\rangle_{s-...}$, $|\Xi_{...}\rangle_{s-...}$ instead of  $|\phi_{...}\rangle_{s-...}$, $|\Xi_{...}\rangle_{s-...}$
\begin{eqnarray}
\hspace{-1em}&\hspace{-1em}&\hspace{-1em} |\chi\rangle_s  =
|\Phi\rangle_s+\eta_1^+\Big(\mathcal{P}_1^+|\Phi_2\rangle_{s-2}+\mathcal{P}_{11}^+|\Phi_{21}\rangle_{s-3} +\eta_{11}^+\mathcal{P}_1^+\mathcal{P}_{11}^+|\Phi_{22}\rangle_{s-6}\Big) \label{spinctotsym}
\end{eqnarray}
\begin{eqnarray}
 \hspace{-1em}&\hspace{-1em}&\hspace{-1em} \phantom{ |\chi^0_c\rangle_s} +\eta_{11}^+\Big(\mathcal{P}_1^+|\Phi_{31}\rangle_{s-3}+\mathcal{P}_{11}^+|\Phi_{32}\rangle_{s-4}\Big)+  \eta_0\Big(\mathcal{P}_1^+|\Phi_1\rangle_{s-1}+\mathcal{P}_{11}^+|\Phi_{11}\rangle_{s-2}  \nonumber \\ \hspace{-1em}&\hspace{-1em}&\hspace{-1em} \phantom{ |\chi^0_c\rangle_s}  +  \mathcal{P}_1^+\mathcal{P}_{11}^+\Big[ \eta^+_{1} |\Phi_{12}\rangle_{s-4}+\eta^+_{11} |\Phi_{13}\rangle_{s-5}\Big]\Big),\nonumber \\
\hspace{-1em}&\hspace{-1em}&\hspace{-1em} |\Lambda\rangle_s  =  \mathcal{P}_1^+  |\Xi\rangle_{s-1}+\mathcal{P}_{11}^+|\Xi_{1}\rangle_{s-2} +\mathcal{P}_1^+\mathcal{P}_{11}^+\Big(\eta_1^+|\Xi_{11}\rangle_{s-4} \label{parctotsym}
\\
\hspace{-1em}&\hspace{-1em}&\hspace{-1em} \phantom{|\chi^1\rangle_s}
 + \eta_{11}^+|\Xi_{12}\rangle_{s-5}\Big) +  \eta_0\mathcal{P}_1^+\mathcal{P}_{11}^+|\Xi_{01}\rangle_{s-3} ,
  \nonumber\\
\hspace{-1em}&\hspace{-1em}&\hspace{-1em} {|\Lambda^1\rangle_s}  =
\mathcal{P}_1^+\mathcal{P}_{11}^+|\Xi^{1}\rangle_{s-3}.
\end{eqnarray}
Here
\begin{eqnarray}\label{Phiphi}
  |\Phi_{...}\rangle_{s-...} &=&  \sum_{l=0}^{[s-../2]-l}\frac{(b^+)^l}{l!}\sum_{k=0}^{s-2l...}\frac{(d^+)^k}{k!}|\phi_{...|l,k}(a^+)\rangle_{s-k-2l...} , \ \mathrm{for} \ |\phi_{|0,0}(a^+)\rangle_{s}\equiv |\phi\rangle_s,\\
   \label{Xiphi} |\Xi_{...}\rangle_{s-...} &=& \sum_{l=0}^{[s-.../2]-l}\frac{(b^+)^l}{l!}\sum_{k=0}^{s-2l...}\frac{(d^+)^k}{k!}|\Xi_{...|l,k}(a^+)\rangle_{s-k-2l...}.
\end{eqnarray}
We prove in the appendix~\ref{Singhcvomp}  that after imposing the
appropriate gauge conditions and eliminating the auxiliary fields
with help of the  equations of motion, the theory under
consideration is reduced to  ungauged form equivalent to Singh-Hagen action
\cite{SinghHagen} in terms of totally symmetric  traceful
tensor field $\phi^{\mu(s)}$ and auxiliary traceful
$\phi_1^{\mu(s-3)}$.

Now we turn to the interacting theory.

\section{System of equations for cubic vertex}
\label{BRSTinter}

Here, we follow  the general scheme developed for massless case in \cite{BRcub}  to find the cubic
interaction vertices for the models with one massive and two massless higher spin fields,  two massive
 and one massless higher spin fields with different mass values distribution  and  derive
the equations for these vertices.

 To include the cubic interaction  we introduce three vectors
$|\chi^{(i)}\rangle_{s_i}$, gauge parameters
$|\Lambda^{(i)}\rangle_{s_i}$, $|\Lambda^{(i)1}\rangle_{s_i}$ with
corresponding vacuum vectors $|0\rangle^i$ and oscillators, where
$i=1,2,3$.  It permits to  define the deformed action and the deformed
gauge transformations as follows
\begin{eqnarray}\label{S[n]}
  && S^{(m)_3}_{[1]|(s)_3}[\chi^{(1)},\chi^{(2)}, \chi^{(3)}] \ = \  \sum_{i=1}^{3} \mathcal{S}^{m_i}_{0|s_i}   +
  g  \int \prod_{e=1}^{3} d\eta^{(e)}_0  \Big( {}_{s_{e}}\langle \chi^{(e)} K^{(e)}
  \big|  V^{(3)}\rangle^{(m)_3}_{(s)_{3}}+h.c. \Big)  , \\
   && \delta_{[1]} \big| \chi^{(i)} \rangle_{s_i}  =  Q^{(i)} \big| \Lambda^{(i)} \rangle_{s_i} -
g \int \prod_{e=1}^{2} d\eta^{(i+e)}_0  \Big( {}_{s_{i+1}}\langle
\Lambda^{({i+1})}K^{(i+1)}\big|{}_{s_{i+2}}
   \langle \chi^{({i+2})}K^{(i+1)}\big|  \label{cubgtr}\\
   &&
\ \   \phantom{\delta_{[1]} \big| \chi^{(i)} \rangle_{s_i}} +(i+1 \leftrightarrow i+2)\Big)
\big|\widetilde{V}{}^{(3)}\rangle^{(m)_3}_{(s)_{3}}, \nonumber\\
&& \delta_{[1]} \big| \Lambda^{(i)} \rangle_{s_i}  =  Q^{(i)} \big| \Lambda^{(i)1} \rangle_{s_i}
 -g  \int \prod_{e=1}^{2} d\eta^{(i+e)}_0  \Big( {}_{s_{i+1}}\langle \Lambda^{(i+1)1}K^{(i+1)}\big|{}_{s_{i+2}}
\langle \chi^{({i+2})}K^{(i+1)}\big|  \label{cubggtr}\\
   &&
\ \   \phantom{\delta_{[1]} \big| \chi^{(i)} \rangle_{s_i}} +(i+1
\leftrightarrow i+2)\Big) \big|\widehat{V}{}^{(3)}\rangle^{(m)_3}_{(s)_{3}}
\nonumber
\end{eqnarray}
with some  unknown  three-vectors $\big| V^{(3)}\rangle^{(m)_3}_{(s)_{3}}, \,
\big|\widetilde{V}{}^{(3)}\rangle^{(m)_3}_{(s)_{3}}, \,
\big|\widehat{V}{}^{(3)}\rangle^{(m)_3}_{(s)_{3}}.$  Here
$\mathcal{S}^{m_i}_{0|s_i}$ is the free action (\ref{PhysStatetot}) for
the field $\big| \chi^{(i)} \rangle_{s_i}$,\, $Q^{(i)}$ is the BRST
charge corresponding to spin $s_{i},\, i=1,2,3$, $K^{(i)}$ is the
operator $K$ (\ref{geneq2}) corresponding to spin $s_{i},\, i=1,2,3$ for massive and with change $h(s) \to  h(s) +1/2$ for massless field
and $g$ is a deformation parameter (called usually as a coupling constant).  Also  we use the notation $(m)_{3}
\equiv (m_1,m_2,m_3)$ and convention $[i+3 \simeq i]$.

A concrete construction of the cubic interaction means finding the
concrete vectors $\big| V^{(3)}\rangle^{(m)_{3}}_{(s)_{3}}$, $\big|
\widetilde{V}{}^{(3)}\rangle^{(m)_{3}}_{(s)_{3}}$,
$\big|\widehat{V}{}^{(3)}\rangle^{(m)_{3}}_{(s)_{3}}$. For this purpose we can involve
the set of fields, the constraints, ghost operators related with
spins $s_1,s_2,s_3$ and the respective conditions of gauge invariance of the
deformed action under the deformed gauge transformations as well as the conservation of the form of the gauge transformations for the
fields $\big| \chi^{(i)} \rangle_{s_i} $ under the gauge
transformations $\delta_{[1]}\big| \Lambda^{(i)} \rangle_{s_i} $ at the first power in $g$\footnote{In this connection, note also the results of recent
works \cite{L1}, \cite{L2}, \cite{L3}, \cite{L4}
obtained on the base of the
deformation of general gauge theory \cite{BL1}, \cite{BL2}, \cite{L5}, \cite{BL3}.}.
\begin{eqnarray}
&&  g \int \prod_{e=1}^{3} d\eta^{(e)}_0   {}_{s_{j}}\langle
\Lambda^{(j)}K^{(j)} \big| {}_{s_{j+1}}\langle
\chi^{(j+1)}K^{(j+1)}\big|{}_{s_{j+2}}\langle
\chi^{(j+2)}K^{(j+2)}\big| \mathcal{Q}(V^3,\widetilde{V}^3) = 0,
\label{g1L}\\
&& \hspace{-1em}   g \int \prod_{e=1}^{2} d\eta^{(e)}_0   {}_{s_{j+1}}\langle \Lambda^{(j+1)1}K^{(j+1)}
\big| {}_{s_{j+2}}\langle \chi^{(j+2)}K^{(j+2)}\big|  \Big(\mathcal{Q}(\widetilde{V}^3,\widehat{V}^3)
- Q^{(j+2)} |\widehat{V}{}^{(3)}\rangle\Big) = 0,
\label{g1L1}
\end{eqnarray}
where
\begin{eqnarray}
\phantom{g^1:} && \mathcal{Q}(V^3,\widetilde{V}^3) = \sum_{k=1}^3
Q^{(k)}\big |\widetilde{V}{}^{(3)}\rangle^{(m)_3}_{(s)_3}
   +  Q^{(j)}\Big( \big|{V}{}^{(3)}\rangle^{(m)_3}_{(s)_{3}} -\big|\widetilde{V}{}^{(3)}\rangle^{(m)_3}_{(s)_3}\Big)
    , \ j=1,2,3.\label{g1operV3}%
\end{eqnarray}

Following to our  results \cite{BRcub}, \cite{Rcubmasless} we choose coincidence for the vertices:
$\big|{\widetilde{V}}{}^{(3)}\rangle$ =
$\big|{V}{}^{(3)}\rangle$ =
$\big|\widehat{V}{}^{(3)}\rangle$, which provides the validity of the operator equations at the  first order in $g$ (the highest orders are necessary  for finding the quartic and higher vertices)
\begin{equation}
\label{g1Lmod}
  Q^{tot}
\big|{V}{}^{(3)}\rangle^{(m)_3}_{(s)_{3}} =0, \qquad \sigma^{(i)}\big|{V}{}^{(3)}\rangle^{(m)_3}_{(s)_{3}}\ =\ 0,
\end{equation}
jointly with the spin conditions as the
consequence of the spin equation (\ref{extconstsp}) for each sample
(with $|\chi^{(i)}\rangle_{s_i},\,
 |\Lambda^{(i)}\rangle_{s_i},\, |\Lambda^{(i)1}\rangle_{s_i}$)
providing the nilpotency of total BRST operator $Q^{tot} \equiv
\sum_i Q^{(i)}$ when evaluated on the vertex due to the equations
(\ref{geneq}) and $\{Q^{(i)}, Q^{(j)}\}=0$ for $i\ne j$.

  A local dependence on
space-time coordinates in the vertices $\big |V^{(3)}\rangle$, $\big |\widetilde{V}{}^{(3)}\rangle$,
$\big |\widehat{V}{}^{(3)}\rangle$ means
\begin{equation}\label{xdep}
  \big |V^{(3)}\rangle^{(m)_3}_{(s)_3} = \prod_{i=2}^3 \delta^{(d)}\big(x_{1} -  x_{i}\big) V^{(3)|}{}_{(s)_3}^{(m)_3}
  \prod_{j=1}^3 \eta^{(j)}_0 |0\rangle , \ \  \  |0\rangle\equiv \otimes_{e=1}^3 |0\rangle^{e}
\end{equation}
(for $(\epsilon, gh)V^{(3)|}{}_{(s)_3}^{(m)_3}= (0,0)$). We have the conservation law: $\sum_{i=1}^3p^{(i)}_\mu  = 0$,  for the momenta associated with all vertices.
Again as for the massless case \cite{BRcub}, the deformed  gauge transformations  still form the closed
algebra, that means after the simple calculations
\begin{eqnarray}
&&  \big[\delta^{\Lambda_1}_{[1]},\delta^{\Lambda_2}_{[1]}\big]  |\chi^{(i)} \rangle \  =
\  - g \delta^{\Lambda_3}_{[1]}  |\chi^{(i)} \rangle
\label{closuregtr},\\
&&
 \big|\Lambda_3
\rangle \sim  \int \prod_{e=1}^{2} d\eta^{(i+e)}_0  \Big( \langle
\Lambda^{(i+1)}_2K\big|\langle \Lambda^{(i+2)}_1\big|K  +\big(i+1
\leftrightarrow i+2\big)\Big)   - \big({\Lambda_1} \leftrightarrow
{\Lambda_2}\big) \bigg\} \big|{V}{}^{(3)}\rangle
\nonumber
\end{eqnarray}
with  the Grassmann-odd  gauge parameter  $\Lambda_3$  being a
function of the parameters $\Lambda_1,\Lambda_2,$ $\Lambda_3 =
\Lambda_3(\Lambda_1,\Lambda_2).$.

The equations (\ref{g1Lmod})  (for coinciding vertices $V{}^{(3)}=\widetilde{V}{}^{(3)}=\widehat{V}{}^{(3)}$)
together  with the form of the commutator of the gauge
transformations (\ref{closuregtr}) determine the cubic  interacting
vertices for irreducible  massive and massless totally symmetric higher spin
fields.

\section{Solution for Cubic Vertices} \label{BRSTsolgen}

In this section we will construct the general solution for the cubic
vertices in following cases for interacting higher spin fields: with one massive and two massless;
 with two massive and one massless   of spins $s_1,
s_2, s_3$ according to \cite{BRST-BV3}, \cite{Metsaev-mass}:
we introduce fourth order polynomial $D\equiv D(m_1,m_2,m_3)$ and quantity $P_{\epsilon{}m} \equiv P(m_1,m_2,m_3)$:
\begin{eqnarray}\label{Drep}
&&  D = (m_1+m_2+m_3) (m_1-m_2+m_3) (m_1+m_2-m_3) (m_1-m_2-m_3), \\
&& P_{\epsilon{}m} = \epsilon_1{}m_1+\epsilon_2{}m_2+\epsilon_3{}m_3,\ \ \epsilon_i^2=1,\  i=1,2,3.\label{Prep}
\end{eqnarray}
With help of these quantities we have the classification
\begin{align}
 &  m_1=m_2=0,  m_3 \ne 0 & &\Rightarrow && D(0,0,m_3) >0, \label{m0}\\
 &  m_1=0,  m_2 = m_3 = m \ne 0 &&  \Rightarrow && D(0,m,m) =0, \  P_{\epsilon{}m} = 0 \label{m1}\\
   &m_1 =  0,  m_2\ne 0, m_3\ne 0,  m_3\ne  m_2   &&  \Rightarrow && D(0,m_2,m_3) >0, \label{m2}\\
    &m_1 \ne 0, \ m_2 \ne 0, \   m_3\ne 0 &&  \Rightarrow &&  D(m_1,m_2,m_3) >0,  \label{m3}\\
  & m_1 \ne 0, \ m_2 \ne 0, \   m_3\ne 0,    &&  \Rightarrow &&  D(m_1,m_2,m_3)<0, \label{m4}\\
& m_1 \ne 0, \ m_2 \ne 0, \   m_3\ne 0, &&  \Rightarrow && D(m_1,m_2,m_3)=0,   \  P_{\epsilon{}m} = 0     \label{m5}.
  \end{align}
Cases (\ref{m1})--(\ref{m5}) correspond to critical masses described in \cite{Metsaev-mass}, \cite{Metsaev3d}  respectively for $d=4$ and $d=3$,  on a base of use the
 conservation law for the momenta associated with vertices
\begin{equation}
\label{consermom}
 \sum_{i=1}^3p^{(i)}_\mu  = 0.
\end{equation}
and the process of decay of the
massive particle ($i = 1$) into the two massive particles ($i = 2,3$) in the rest frame of the first particle
\begin{equation}\label{consermom1}
 p^{(1)}_\mu =(m_1, {\mathbf{0}}). \ \ p^{(i)}_\mu  = (E_i, (-1)^i\mathbf{p})\ \ \mathrm{with} \ \ E_i=\sqrt{m_i^2+\mathbf{p}^2},
\end{equation}
from which it follows the well-known restrictions on  masses and $(d-1)$-space momentum $\mathbf{p}$
\begin{align}\label{consermom2}
 m_1> (=)m_2+m_3, \ \mathrm{for }\ \ {\mathbf{p}}\ne(=)0, && \mathbf{p}^2=D/ (4m_1^2).
\end{align}
Note, the case of equal masses $m_1=m_2=m_3$  corresponds to  (\ref{m4}) with $D(m,m,m)=-3m_1^4$, whereas the case of $m_i=m_{i+1} \ne  m_3$ may satisfy to any from the relations  (\ref{m3}),  (\ref{m4}),  (\ref{m5}) . The latter cases related to real ($D>0$), virtual ($D<0$) processes, and real process ($D=0$) with vanishing transfer of momentum\footnote{Note, for  the case (\ref{m5}) when $\epsilon_i=1$: $P_{\epsilon{}m} = m_1+m_2+m_3=0$, a consistent Lagrangian theory with reducible massive higher spin fields  in  $d=3$ flat space-time was derived in \cite{SkvortsovTungMiriam} in the light-cone approach.}.

\subsection{Cubic vertices for two massless fields and one massive field} \label{BRSTsolgen1}

For the case (\ref{m0}) with $ D >0$  we look for a general solution  of the equations (\ref{g1Lmod}) in the
form of products of specific operators, homogenous in oscillators.
From suggested in  \cite{BRST-BV3} two ways of vertex derivation known from light-cone approach \cite{Metsaev0512} as
{\emph{Minimal derivative scheme}} and { \emph{Massive field strength scheme}} (however due to uniqueness of the interaction vertex with given order $k$ of derivatives, we expect  the vertex obtained by one scheme should be differed from the vertex obtained by another scheme on  BRST-exact terms) we will consider the first one.

With use of the notations
\begin{equation}\label{notp}
  \widehat{p}{}^{(i)}_{\mu}\ =\ {p}^{(i+1)}_{\mu}-p^{(i+2)}_{\mu}, \  \widehat{\mathcal{P}}{}^{(i)}_0\ =\  \mathcal{P}^{(i+1)}_0- \mathcal{P}^{(i+2)}_0, \ \ \mathrm{a}{}^{(3)+}_{\mu}= a^{(3)+}_{\mu}-\frac{p^{(3)}_{\mu}}{m_3}d^{(3)+}
\end{equation}
the  massive field strength scheme corresponds to the set of monomials given on the constrained Fock space $\mathcal{H}_{tot|c}$
\begin{eqnarray}
  \label{LrZf}
  &&  L^{(i)} \ = \   \widehat{p}{}^{(i)}_{\mu}a^{(i)\mu+} - \imath  \widehat{\mathcal{P}}{}^{(i)}_0 \eta_1^{(i)+} , i=1,2; \qquad  L^{(3)} \ = \    \widehat{p}{}^{(3)}_{\mu}\mathrm{a}{}^{(3)\mu+}, \label{Lr3mfs}\\
 &&
L^{(12)+}_{11} \ = \  a^{(1)\mu+}a^{(2)+}_{\mu} +\frac{1}{2m_3^2} L^{(1)}L^{(2)}-
\frac{1}{2}\mathcal{P}^{(1)+}_1\eta_1^{(2)+} -
\frac{1}{2}\mathcal{P}^{(2)+}_1\eta_1^{(1)+}\label{Lrr+1m} \\
&& L^{(i3)+}_{11} \ = \  a^{(i)\mu+}\mathrm{a}^{(3)+}_{\mu} +(-1)^i\frac{1}{m_3^2} L^{(i)}{p}{}^{(i)}_{\mu}\mathrm{a}{}^{(3)\mu+}, \ \ i=1,2.
\label{Lrr+1m2}.
\end{eqnarray}
In turn, the minimal derivative scheme contains the monomials
\begin{eqnarray}
  \label{LrZ}
  && \hspace{-0.5em}  L^{(i)}\   = \    \widehat{p}{}^{(i)}_{\mu}a^{(i)\mu+} - \imath \widehat{\mathcal{P}}{}^{(i)}_0\eta_1^{(i)+} , \ \ \  i=1,2,3; \\
 &&\hspace{-0.5em}
L^{(1{}2)+}_{11}  = a^{(1)\mu+}a^{(2)+}_{\mu} +\frac{1}{2m_3^2} L^{(1)}L^{(2)} -
\frac{1}{2}\mathcal{P}^{(1)+}_1\eta_1^{(2)+} -
\frac{1}{2}\mathcal{P}^{(2)+}_1\eta_1^{(1)+}; \label{Lrr+1} \\
 &&\hspace{-0.5em}
L^{(23)+}_{11} =  a^{(2)\mu+}a^{(3)+}_{\mu} -\frac{1}{2m_3^2} L^{(2)}L^{(3)}+\frac{1}{2m_3} L^{(2)}d^{(3)+}
-
\frac{1}{2}\mathcal{P}^{(2)+}_1\eta_1^{(3)+} -
\frac{1}{2}\mathcal{P}^{(3)+}_1\eta_1^{(2)+};\label{Lrr+2}
\\
 &&\hspace{-0.5em}
L^{(31)+}_{11}  =  a^{(3)\mu+}a^{(1)+}_{\mu}  - \frac{1}{2m_3^2}L^{(3)} L^{(1)}-\frac{1}{2m_3}d^{(3)+} L^{(1)}
-
\frac{1}{2}\mathcal{P}^{(1)+}_1\eta_1^{(3)+} -
\frac{1}{2}\mathcal{P}^{(3)+}_1\eta_1^{(1)+};\label{Lrr+3}
 \\
  &&\hspace{-0.5em} Z \  = \ L^{(12)+}_{11}L^{(3)} + L^{(23)+}_{11}L^{(1)} + L^{(31)+}_{11}L^{(2)}. \label{LrZLLf}
\end{eqnarray}
        The operators above { do not introduce the divergences into the vertices},  are Grassmann-even with vanishing ghost number  and have the distributions in powers of creation oscillators $A^{(i)+}$ and momenta
         \begin{equation*}
\begin{tabular}{|c|c|c|c|c|c|c|}
\hline
& $L^{(i)}$ & $L^{(12)+}_{11}$ & $L^{(23)+}_{11}$ & $L^{(31)+}_{11}$& $\widehat{L}{}^{(i)+}_{11}$& $Z$ \\ \hline
$\mathrm{deg}_{A^{(j)+}}$ & $\delta_{ij}$ & $ \left( 1,1,0\right) $
& $ \left( 0,1,1\right)  $ & $ \left( 1,0,1\right)  $  & $2\delta_{ij}$& $(1,1,1)$\\ \hline
$\mathrm{deg}_{p}$ & $1 $ & $\leq 2$ & $\leq 2 $ & $\leq 2$%
& $0$& $\leq 3$ \\ \hline
\end{tabular}%
\end{equation*}
Note, first, that for massless case the latter row ($\mathrm{deg}_{p}$) for  massless analogs of  operators is filled as: $\big(1,0,0,0,0,1 \big)$.
Second, the operators (\ref{LrZf})
   $L^{(i)}$ for $i=1,2$ are not BRST-closed with respect to the constrained BRST operator $Q^{tot}_c \equiv  Q^{tot}\vert_{\eta^{(+)i}_{11}=0}$
as compared to      $L^{(3)}$, $Q^{tot}_c L^{(3)}|0\rangle =0$. Namely we have,  :
\begin{eqnarray}
  && Q^{tot}_c L^{(i)}|0\rangle = (-1)^i m_3^2 \eta_1^{(i)+}|0\rangle \ne 0, \quad i=1,2. \label{notBRSTL12}
\end{eqnarray}
and therefore the  operator $Z$ (\ref{LrZLLf}) is not $Q^{tot}_c$-  BRST-closed
\begin{eqnarray}\label{BRSTZnot}
  && Q^{tot}_c Z |0\rangle \ne 0   \Rightarrow  Q^{tot}Z |0\rangle  \ne 0
  \end{eqnarray}
  In turn, the operators $L^{(ii+1)+}_{11}$, $\widehat{L}{}^{(i)+}_{11}$ are $Q^{tot}_c$-  BRST-closed, but not $Q^{tot}$-  BRST-closed due to
\begin{eqnarray}&&
\widehat{L}{}^{(i)}_{11}L^{(ij)+}_{11}|0\rangle = 0, \quad i,j=1,2,3\\
&&
\widehat{L}{}^{(i)}_{11}(L^{(ij)+}_{11})^2|0\rangle  \ne 0
\label{LrL13}
 \end{eqnarray}
Then, following to \cite{BRcub} we have the respective trace operators (massless for $i=1,2$ and massive $i=3$)
\begin{equation}\label{trform}
U^{(s_i)}_{j_i}\big(\eta_{11}^{(i)+},\mathcal{P}_{11}^{(i)+} \big) \
: = \
(\widehat{L}{}^{(i)+}_{11})^{(j_i-2)}\big\{(\widehat{L}{}^{+(i)}_{11})^2
- j_i(j_i-1)\eta_{11}^{(i)+}\mathcal{P}_{11}^{(i)+}\big\}, \ \
i=1,2,3.
\end{equation}
Indeed,
the $Q^{tot}$- BRST closeness for the operator $L^{(3)}$ is reduced
to the fulfillment of the equations at the terms linear in
$\eta_{11}^{(3)+}$
 \begin{eqnarray}\label{Lr11}
&&       \widehat{L}{}^{(i)}_{11}(L^{(i+j)})^k |0\rangle  \equiv 0, j=1,2,  \  \  \forall k \in \mathbb{N}\,,  \\
&&     \widehat{L}{}^{(i)}_{11}L^{(i)} |0\rangle =  \Big(-\widehat{p}{}^{(i)}_{\mu}a^{(i)\mu}+ \imath \widehat{\mathcal{P}}^{(i)}_0\eta_1^{(i)} + L^{(i)}\widehat{L}{}^{(i)}_{11}\Big)|0\rangle =0 , \label{Lr11Lr1} \\
&&    \widehat{L}{}^{(i)}_{11}(L^{(i)})^2 |0\rangle =  \eta^{\mu\nu}\widehat{p}{}^{(i)}_{\mu} \widehat{p}^{(i)}_{\nu}|0\rangle = (\widehat{p}{}^{(i)})^2|0\rangle \ne 0, i=1,2,3.\label{Lr11Lr2}
 \end{eqnarray}
The last relations and ones for $(L^{(3)})^k $ do not vanish under the sign of inner
products and justify the introduction of the BRST-closed  forms, first for $k \leq 5$
\begin{eqnarray}
\label{LrZ1}
  &&  \mathcal{L}^{(3)}_{1} \ = \  {L}^{(3)}- [\widehat{L}{}^{(3)}_{11},{L}^{(3)}\}\frac{b^{(3)+}}{h^{(3)}}       , \qquad   \mathcal{L}^{(3)}_{2} \ = \  ({L}^{(3)})^2 - (\hat{p}^{(3)})^2\frac{b^{(3)+}}{h^{(3)}}     ,   \\
  \label{LrZ2}
  &&  \mathcal{L}^{(3)}_{3} \ = \  {L}^{(i)}\Big(({L}^{(3)})^2 - 3(\hat{p}^{(3)})^2\frac{b^{(3)+}}{h^{(3)}}   \Big)  ,   \\
 &&  \mathcal{L}^{3)}_{4} \ = \  ({L}^{(3)})^2\Big(({L}^{(3)})^2 - 6(\hat{p}^{(3)})^2\frac{b^{(3)+}}{h^{(3)}}   \Big)+
 3(\hat{p}^{(3)})^4\frac{(b^{(3)+})^2}{h^{(3)}(h^{(3)}+1)} ,  \label{LrL4} \\
 &&  \mathcal{L}^{(3)}_{5} \ = \  ({L}^{(3)})^3\Big(({L}^{(3)})^2 - 10(\hat{p}^{(3)})^2\frac{b^{(3)+}}{h^{(3)}}   \Big)+
 \frac{3\cdot 10}{2}{L}^{(3)}(\hat{p}^{(3)})^4\frac{(b^{(3)+})^2}{h^{(3)}(h^{(3)}+1)} ,  \label{LrL5}
\end{eqnarray}
then,  by   induction for arbitrary $k\in \mathbb{N}$
 \begin{eqnarray}
 && \label{LrL2e} \mathcal{L}^{(3)}_{k} \ = \ \sum_{j=0}^{[k/2]} (-1)^{j}({L}^{(3)})^{k-2j}(\hat{p}^{(3)})^{2j} \frac{k!}{j! 2^j(k-2j))!} \frac{(b^{(3)+})^j}{C(j,h^{(3)}(s))}   .
\end{eqnarray}
(the equivalent polynomial representation for BRST-closed operator  $ \mathcal{L}^{(3)}_{k}$ is also found, see Subsection~\ref{BRSTsolgen1mm}).
 By the same reason, the  any power  of the
forms $L^{(ii+1)+}_{11}$, (hence $(L^{(ii+1)+}_{11})^k$, $k>1$) are not BRST-closed as well due to equations (\ref{Ltr11Lr1}),
 \begin{eqnarray}\label{Ltr11}
&&   \hspace{-0.5em}    \eta_{11}^{(i)+}\widehat{L}{}^{(i)}_{11}(L^{(i+1i+2)+}_{11})^k |0\rangle  \equiv 0,   \  \  \forall k \in \mathbb{N}\,,  \\
&&   \hspace{-0.5em}  \sum_{i}\eta_{11}^{(i)+}\widehat{L}{}^{(i)}_{11} L^{(12)+}_{11} |0\rangle =  \sum_{i}\eta_{11}^{(i)+}\Big(-a^{(2)\mu+}a^{(i)}_{\mu} +\frac{1}{2m_3^2} \big( -\hat{p}^{(i)}a^{(i)}_{\mu} - \imath \widehat{\mathcal{P}}{}^{(i)}_0 \eta^{(i)}_1\big)L^{(2)}\label{Ltr11Lr10}
 \end{eqnarray} \begin{eqnarray}
&& \hspace{-0.5em}
-\frac{1}{2}\mathcal{P}^{(2)+}_1\eta_1^{(i)} -
\frac{1}{2}\eta_1^{(2)+}\mathcal{P}^{(i)}_1\Big)|0\rangle =0 , \nonumber\\
&&    \hspace{-0.5em}  \sum_{i}\eta_{11}^{(i)+}\widehat{L}{}^{(i)}_{11} (L^{(12)+}_{11})^2 |0\rangle = \sum_{i}\eta_{11}^{(i)+}[[\widehat{L}{}^{(i)}_{11}, L^{(12)+}_{11}\}, L^{(12)+}_{11}\} |0\rangle \ne 0,\label{Ltr11Lr1}\\
&&\hspace{-0.5em}  \sum_{i=1}^2\eta_{11}^{(i)+} [[\widehat{L}{}^{(i)}_{11}, L^{(12)+}_{11}\}, L^{(12)+}_{11}\} = \Big[\eta_{11}^{(1)+}\Big(a^{(2)\mu+}a^{(2)+}_{\mu} +\frac{1}{2m_3^2} \big( \hat{p}^{(1)}a^{(2)+}_{\mu} + \frac{\imath}{2} \widehat{\mathcal{P}}{}^{(1)}_0 \eta^{(2)+}_1\big)L^{(2)}\nonumber \\
&&\hspace{-0.5em}  +
\frac{1}{2}\mathcal{P}^{(2)+}_1\eta_1^{(2)+} \Big) + \eta_{11}^{(2)+}\Big(a^{(1)\mu+}a^{(1)+}_{\mu} +\frac{1}{2m_3^2} \big(\hat{p}^{(2)}a^{(1)+}_{\mu} + \frac{\imath}{2} \widehat{\mathcal{P}}{}^{(2)}_0 \eta^{(1)+}_1\big)L^{(1)} +
\frac{1}{2}\mathcal{P}^{(1)+}_1\eta_1^{(1)+} \Big)\Big] \nonumber,
 \end{eqnarray}
To compensate this term in  $L^{(12)+}_{11}$  we add the modified summand  for it and find BRST-closed completion in the form
\begin{eqnarray}\label{Lr12312}
  &&  \mathcal{L}^{(12)+}_{11|1} \ = \    {L}^{(12)+}_{11} -\sum_{i_0} {W}^{(i_0)}_{(12)|0}\frac{b^{(i_0)+}}{h^{(i_0)}} +\frac{1}{2}\Big(\sum_{i_0\ne j_0}[\widehat{L}{}^{(j_0)}_{11}, {W}^{(i_0)}_{(12)|0}\}\frac{b^{(i_0)+}}{h^{(i_0)}} \frac{b^{(j_0)+}}{h^{(j_0)}} \\
 && \quad +\sum_{i_0}[\widehat{L}{}^{(i_0)}_{11}, {W}^{(i_0)}_{(12)|0}\}\frac{(b^{(i_0)+})^2}{h^{(i_0)}(h^{(i_0)}+1)} \Big)  ,   \nonumber
 \end{eqnarray}
 for $({L}^{(12)+}_{11})^2$
 \begin{eqnarray}
\label{LrL122}
  &&   \mathcal{L}^{(12)+}_{11|2} \ = \  {L}^{(12)+}_{11}\mathcal{L}^{(12)+}_{11|1}  -\sum_{i_1} {W}^{(i_1)}_{(12)|1}\frac{b^{(i_1)+}}{h^{(i_1)}} +\frac{1}{2}\Big(\sum_{i_1\ne j_1}[\widehat{L}{}^{(j_1)}_{11}, {W}^{(i_1)}_{(12)|1}\}\frac{b^{(i_1)+}}{h^{(i_1)}} \frac{b^{(j_1)+}}{h^{(j_1)}}   \\
 && \quad +\sum_{i_1}[\widehat{L}{}^{(i_1)}_{11}, {W}^{(i_1)}_{(12)|1}\}\frac{(b^{(i_1)+})^2}{h^{(i_1)}(h^{(i_1)}+1)} \Big) -\frac{1}{6}\Big(3\sum_{i_1\ne j_0}[\widehat{L}{}^{(j_1)}_{11},[\widehat{L}{}^{(i_1)}_{11}, W^{(i_1)}_{(12)|1}\}\}\times \nonumber \\
 && \quad \times \frac{b^{(j_1)+}}{h^{(j_1)}} \frac{(b^{(i_1)+)^2}}{h^{(i_1)}(h^{(i_1)}+1)} +\sum_{i_1}[\widehat{L}{}^{(i_1)}_{11},[\widehat{L}{}^{(i_1)}_{11}, W^{(i_1)}_{(12)|2}\}\}\frac{(b^{(i_1)+})^3}{C(3,h^{(i_1)}(s))} \Big),   \nonumber ,
  \end{eqnarray}
then  for $({L}^{(12)+}_{11})^3$
 \begin{eqnarray}
\label{LrL123}
 &&   \mathcal{L}^{(12)+}_{11|3} \ = \  {L}^{(12)+}_{11}\mathcal{L}^{(12)+}_{11|2} -  \sum_{i_2} W^{(i_2)}_{(12)|2}\frac{b^{(i_2)+}}{h^{(i_2)}}+\frac{1}{2}\Big(\sum_{i_2\ne j_1}[\widehat{L}{}^{(j_2)}_{11}, W^{(i_2)}_{(12)|2}\}\frac{b^{(i_2)+}}{h^{(i_2)}} \frac{b^{(j_2)+}}{h^{(j_2)}} \\
 && \quad +\sum_{i_2}[\widehat{L}{}^{(i_2)}_{11}, W^{(i_2)}_{(12)|2}\}\frac{(b^{(i_2)+})^2}{h^{(i_2)}(h^{(i_2)}+1)}   \Big)-\frac{1}{6}\Big(3\sum_{i_2\ne j_2}[\widehat{L}{}^{(j_2)}_{11},[\widehat{L}{}^{(i_2)}_{11}, W^{(i_2)}_{(12)|2}\}\} \nonumber \\
 && \quad \times \frac{b^{(j_2)+}}{h^{(j_2)}} \frac{(b^{(i_2)+)^2}}{h^{(i_2)}(h^{(i_2)}+1)}   +\sum_{i_2}[\widehat{L}{}^{(i_2)}_{11},[\widehat{L}{}^{(i_2)}_{11}, W^{(i_2)}_{(12)|2}\}\}\frac{(b^{(i_2)+})^3}{\prod_{p=0}^{2}(h^{(i_2)}+p)} \Big)
  \nonumber \\
 && \quad +\frac{1}{4!}\Big(4\sum_{i_2\ne j_2}\mathrm{ad}_{\widehat{L}{}^{(j_{2})}_{11}}\mathrm{ad}^{2}_{\widehat{L}{}^{(i_{2})}_{11}} W^{(i_2)}_{(12)|2}
 \frac{b^{(j_2)+}}{h^{(j_2)}} \frac{(b^{(i_2)+})^3}{C(3,h^{(i_2)})}+3\sum_{i_2\ne j_2}\mathrm{ad}^2_{\widehat{L}{}^{(j_{2})}_{11}}\mathrm{ad}_{\widehat{L}{}^{(i_{2})}_{11}} W^{(i_2)}_{(12)|2}
 \nonumber \\
  && \quad \times \frac{(b^{(j_2)+})^2}{C(2,h^{j_2)})} \frac{(b^{(i_2)+})^2}{C(2,h^{(i_2)})}  +\sum_{i_2}\mathrm{ad}^{3}_{\widehat{L}{}^{(i_{2})}_{11}} W^{(i_2)}_{(12)|2}\frac{(b^{(i_2)+})^4}{C(4,h^{(i_2)})} \Big) ,     \nonumber
 \end{eqnarray}
 and by  induction for $({L}^{(12)+}_{11})^{k+1}$ at $k=1 ,\ldots , \min(s_1,s_2)-1$:
 \begin{eqnarray} \label{LrL12k1}
  &&   \mathcal{L}^{(12)+}_{11|k+1} \ = \ {L}^{(12)+}_{11}\mathcal{L}^{(12)+}_{11|k} -  \sum_{i_{k}} W^{(i_{k})}_{(12)|k}\frac{b^{(i_{k})+}}{h^{(i_{k})}} +\frac{1}{2}\Big(\sum_{i_{k}\ne j_{k}}[\widehat{L}{}^{(j_{k})}_{11},W^{(i_{k})}_{(12)|k}\}\frac{b^{(i_{k})+}}{h^{(j_{k})}} \frac{b^{(j_{k})+}}{h^{(j_{k})}} \\
 && \qquad
   +\sum_{i_{k}}[\widehat{L}{}^{(i_{k})}_{11},W^{(i_{k})}_{(12)|k}\}\frac{(b^{(i_{k})+})^2}{\prod_{p=0}^{1}(h^{(i_{k})}+p)}\Big)
     + \sum_{e=3}^{k+2}\frac{(-1)^e}{e!}\sum_{ l = 0,\,  j\ne i}^{[e/2]}\frac{e!}{[e/(e-l)](e-l)!l!} \times \nonumber \\
 && \qquad  \times\mathrm{ad}^{l}_{\widehat{L}{}^{(j_{k})}_{11}}\mathrm{ad}^{e-l-1}_{\widehat{L}{}^{(i_{k})}_{11}} W^{(i_{k})}_{(12)|k}\frac{(b^{(j_{k})+})^l}{C(l,h^{(j_k)})}\frac{(b^{(i_{k})+})^{e-l}}{C(e-l, h^{(i_{k})})}
   .\nonumber
\end{eqnarray}
In (\ref{Lr12312})--(\ref{LrL12k1})  the indices $i_0,...,i_k$ are running two values:  $1,2$; and we have used the notations
\begin{eqnarray}
&& [\widehat{L}{}^{(i_0)}_{11}, L^{(12)+}_{11}\} \ \equiv \ W^{(i_0)}_{(12)|0}\,, \  \ [\widehat{L}{}^{(i_k)}_{11}, L^{(12)+}_{11}\mathcal{L}^{(12)+}_{11|k}\} \ \equiv \ W^{(i_k)}_{(12)|k}\,,  k=1,2,...,\  \label{defW}\\
  && \mathrm{ad}^{l+1}_{\widehat{L}{}^{(j_{k})}_{11}} B \ \equiv \  \big[\widehat{L}{}^{(j_{k})}_{11},\stackrel{l \ \mathrm{times}}{\overbrace{\big[\widehat{L}{}^{(j_{k})}_{11}, \ldots ,\big[\widehat{L}{}^{(j_{k})}_{11},\,B\big\}\big\}}}\big\}.\nonumber
\end{eqnarray}
By the construction the calligraphic operators are traceless
\begin{eqnarray}
&&    [\widehat{L}{}^{(i)}_{11  },\,\mathcal{L}^{(12)+}_{11|k+1} \}  = 0, \ \ k=0,1,\ldots , \min(s_1,s_2)-1 \label{Ltr11Lr2},
 \end{eqnarray}
   because of the last terms  in (\ref{Lr12312})--(\ref{LrL12k1}) in front of the maximal power in $b^{(i)+}$, i.e.  $[\widehat{L}{}^{(j_0)}_{11} \hspace{-0.13em}, \hspace{-0.12em} {W}^{(i_0)}_{(12)|0}\}$, ... , $\mathrm{ad}^{l}_{\widehat{L}{}^{(j_{k})}_{11}}\mathrm{ad}^{k-l+1}_{\widehat{L}{}^{(i_{k})}_{11}} W^{(i_{k})}_{(12)|k}$
depend on only $a^{(i)}_{\mu}, d^{(i)}, \eta/^{(i)}_1,  \mathcal{P}_1^{(i)}$  (annihilation) oscillators and therefore the compensation procedure is finalized.

 In deriving (\ref{Lr12312})--(\ref{LrL12k1}) we have used the permutation properties
 \begin{eqnarray}
 \label{permpr-1}
&&     [\widehat{L}{}^{(i_3)}_{11},W^{(j_3)}_{(12)|k}\}= [\widehat{L}{}^{(j_3)}_{11},W^{(i_3)}_{(12)|k}\}\,,\\
&&     [\widehat{L}{}^{(i_3)}_{11},[\widehat{L}{}^{(j_3)}_{11}, W^{(i_3)}_{(12)|k}\}\}= [\widehat{L}{}^{(j_3)}_{11},[\widehat{L}{}^{(i_3)}_{11}, W^{(i_3)}_{(12)|k}\}\}= [\widehat{L}{}^{(i_3)}_{11},[\widehat{L}{}^{(i_3)}_{11}, W^{(j_3)}_{(12)|k}\}\}\,,  \label{permpr0} \\
&&
\mathrm{ad}^{p}_{\widehat{L}{}^{(i_{k-1})}_{11}}\mathrm{ad}^{l}_{\widehat{L}{}^{(j_{k-1})}_{11}}\mathrm{ad}^{e}_{\widehat{L}{}^{(i_{k-1})}_{11}} W^{(i_{k-1})}_{(12)|k}=\mathrm{ad}^{l}_{\widehat{L}{}^{(j_{k-1})}_{11}}\mathrm{ad}^{e+p}_{\widehat{L}{}^{(i_{k-1})}_{11}} W^{(i_{k-1})}_{(12)|k},\label{permpr1}
 \end{eqnarray}
 which follows from  the Jacobi identity, first, for  triple $\widehat{L}{}^{(i_3)}_{11}$, $\widehat{L}{}^{(j_3)}_{11}$, $W^{(i_3)}_{(12)|k}$, second, of  its repeated application for $\widehat{L}{}^{(i_3)}_{11}$, $\widehat{L}{}^{(j_3)}_{11}$, $\mathrm{ad}^{p}_{\widehat{L}{}^{(i_{k-1})}_{11}} W^{(i_{k-1})}_{(12)|k}$ with account for commuting of  two first trace operators,  e.g.
 \begin{equation}\label{permpr2}
   \mathrm{ad}_{\widehat{L}{}^{(i_{k-1})}_{11}}\mathrm{ad}_{\widehat{L}{}^{(j_{k-1})}_{11}}\mathrm{ad}^{e}_{\widehat{L}{}^{(i_{k-1})}_{11}} W^{(i_{k-1})}_{(12)|k}=\mathrm{ad}_{\widehat{L}{}^{(j_{k-1})}_{11}}\mathrm{ad}^{e+1}_{\widehat{L}{}^{(i_{k-1})}_{11}} W^{(i_{k-1})}_{(12)|k}.
 \end{equation}
 Thus, all calligraphic operators $\mathcal{L}^{(12)+}_{11|k+1}$ are  BRST-closed.

 Analogously, we have the same BRST-closed completions for
 $\mathcal{L}^{(23)+}_{11|1}$ and $\mathcal{L}^{(31)+}_{11|1}$,  and respective BRST-closed  forms $ \mathcal{L}^{(23)+}_{11|k+1}$, $ \mathcal{L}^{(31)+}_{11|k+1}$ to be   uniquely written as follows, for $i=2,3$:
\begin{eqnarray}\label{Lr1231}
  &&\hspace{-0.5em}  \mathcal{L}^{(i{}i+1)+}_{11|1} \ = \    {L}^{(i{}i+1)+}_{11} -\sum_{i_0=2}^3 {W}^{(i_0)}_{(i{}i+1)|0}\frac{b^{(i_0)+}}{h^{(i_0)}} +\frac{1}{2}\Big(\sum_{i_0\ne j_0}[\widehat{L}{}^{(j_0)}_{11}, {W}^{(i_0)}_{(i{}i+1)|0}\}\frac{b^{(i_0)+}}{h^{(i_0)}} \frac{b^{(j_0)+}}{h^{(j_0)}}  \\
 &&\hspace{-0.5em} \quad +\sum_{i_0}[\widehat{L}{}^{(i_0)}_{11}, {W}^{(i_0)}_{(i{}i+1)|0}\}\frac{(b^{(i_0)+})^2}{h^{(i_0)}(h^{(i_0)}+1)} \Big)    ,   \nonumber
 \end{eqnarray}
 \begin{eqnarray}
   && \hspace{-0.5em}  \mathcal{L}^{(i{}i+1)+}_{11|k+1} \ = \ {L}^{(i{}i+1)+}_{11}\mathcal{L}^{(i{}i+1)+}_{11|k} -  \hspace{-0.3em}\sum_{i_{k}} W^{(i_{k})}_{(i{}i+1)|k}\frac{b^{(i_{k})+}}{h^{(i_{k})}}  +\frac{1}{2}\Big(\hspace{-0.4em}\sum_{i_{k}\ne j_{k}}[\widehat{L}{}^{(j_{k})}_{11},W^{(i_{k})}_{(i{}i+1)|k}\} \label{LrL23k1} \\
 && \hspace{-0.5em} \quad \times \frac{b^{(i_{k})+}}{h^{(j_{k})}} \frac{b^{(j_{k})+}}{h^{(j_{k})}}
   +\sum_{i_{k}}[\widehat{L}{}^{(i_{k})}_{11},W^{(i_{k})}_{(i{}i+1)|k}\}\frac{(b^{(i_{k})+})^2}{\prod_{p=0}^{1}(h^{(i_{k})}+p)}\Big)
     + \sum_{e=3}^{k+2}\frac{(-1)^e}{e!} \times \nonumber \\
 && \hspace{-0.5em} \quad  \times \sum_{ l = 0,\,  j\ne i}^{[e/2]}\frac{e!}{[e/(e-l)](e-l)!l!}\mathrm{ad}^{l}_{\widehat{L}{}^{(j_{k})}_{11}}\mathrm{ad}^{e-l-1}_{\widehat{L}{}^{(i_{k})}_{11}} W^{(i_{k})}_{(i{}i+1)|k}\frac{(b^{(j_{k})+})^l}{C(l,h^{(j_k)})}\frac{(b^{(i_{k})+})^{e-l}}{C(e-l,h^{(i_k)})}
     \nonumber
\end{eqnarray}
(for $.k=1,...,\min(s_i,s_{i+1})-1$).
Note, in the expressions for BRST-closed  forms $ \mathcal{L}^{(23)+}_{11|k},  \mathcal{L}^{(31)+}_{11|k}$ indices $i_0, j_0,...,i_{k-1}, j_{k-1}$  are  ranging respectively from  $\{2, 3\}$ and $\{3,1\}$.

\begin{figure}[h]
{\footnotesize\begin{picture}(10,3)
\put(125.5, -30.5){\includegraphics[scale=0.35]{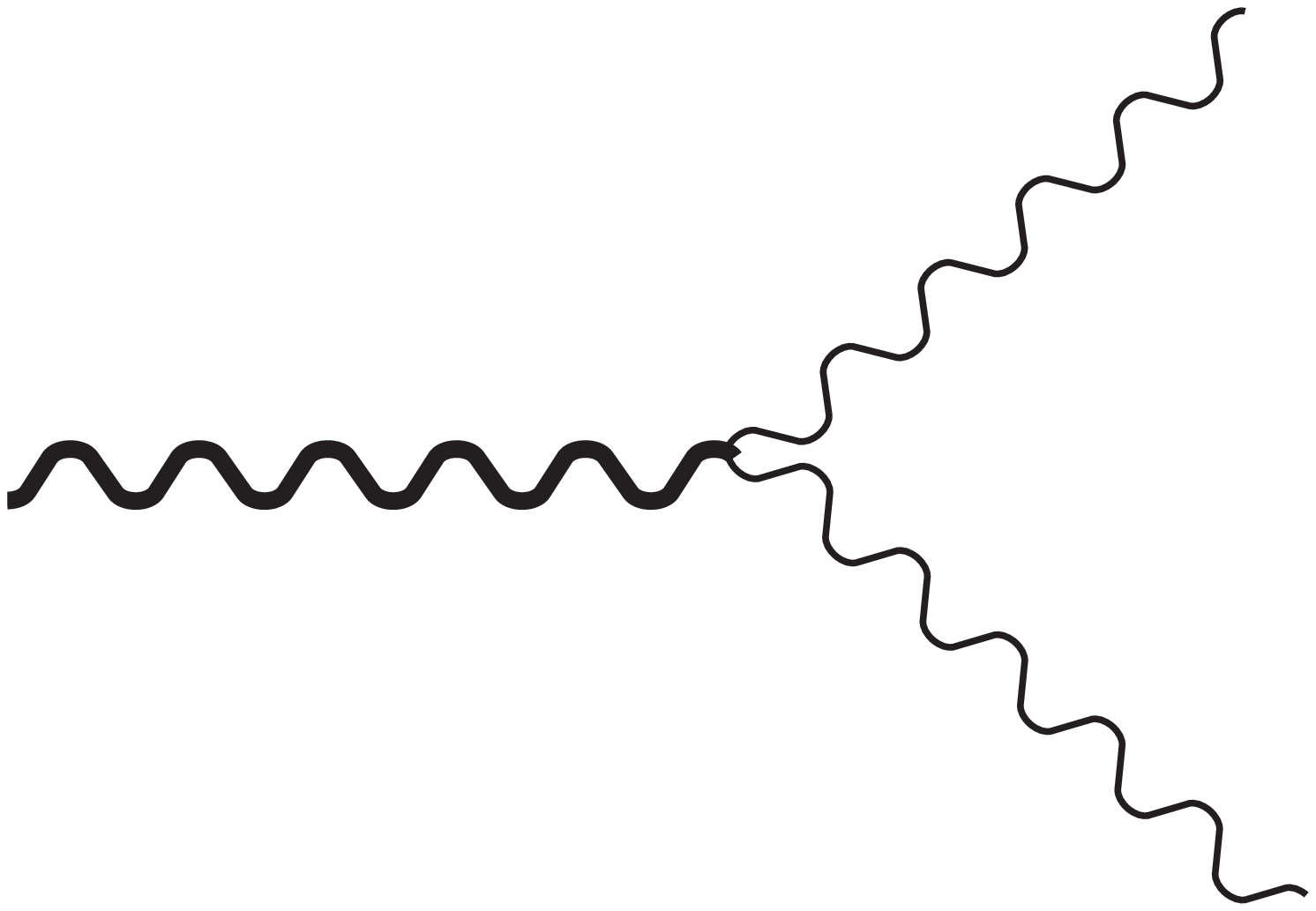}}
\put(40.5, 25.5){$ |{V}{}^{(3)}\rangle^{(0,0,m)}_{(\lambda_1 ,\lambda_2,  s_3)}\ \ \equiv $ }
\put(140.5, 43.5){$ \big(m, s_3\big)$ }
\put(285.5, 73.5){$\big(0,\lambda_1\big)$ }
\put(285.5, -13.5){$\big(0,\lambda_2\big)$}
\put(290.5, 30.5){$+ \ \ \ldots\ldots $ }
\end{picture}}
\vspace{3ex}
\caption{Interaction vertex  $ |{V}{}^{(3)}\rangle^{m}_{(s)_3}$   for the  massive  field
$\phi^{(3)}_{\mu(s_3)}$  of spin $s_3$ and two massless fields $\phi^{(i)}_{\mu(\lambda_i)}$  of  helicities $\lambda_i$  for  $i=1,2$. The terms in $"\ldots"$ correspond to the auxiliary fields  from $|\Phi^{(i)}\rangle_{s_i}$, $|\Phi^{(3)}\rangle_{s_3}$}\label{m0m0m1}
\end{figure}

As the result, the \emph{solution} for the  parity invariant vertex (given by Figure~\ref{m0m0m1})

\vspace{15ex}
~has the form
\begin{eqnarray}\label{genvertex}
   |{V}{}^{(3)}\rangle^{m}_{(s)_3} &=& |{V}{}^{M(3)}\rangle^m_{(s)_3}  + \sum_{(j_1,j_2,j_3) >0}^{([s_{1}/2],[s_{2}/2],[s_{3}/2])} U^{(s_1)}_{j_3}U^{(s_2)}_{j_3}U^{(s_3)}_{j_3}|{V}{}^{M(3)}\rangle^m_{(s)_3-2(j)_3},
   \end{eqnarray}
where the vertex $\big|{V}{}^{M(3)}\rangle^m_{(s)_3-2(j)_3}$ was
defined in Metsaev's paper \cite{BRST-BV3} with account for (\ref{xdep})  but with modified forms $\mathcal{L}^{(3)}_{k}$,  (\ref{LrL2e})  and $\mathcal{L}^{(ii+1)+}_{11|{\sigma^{(i+2)}}}$ instead of $\big({L}^{(ii+1)+}_{11}\big)^{\sigma^{(i+2)}}$ (\ref{Lrr+1})--(\ref{Lrr+3})
\begin{eqnarray}\label{Vmets}
{V}{}^{M(3)|m}_{{(s)_3-2(j)_3}} & = &   \sum_{k}\mathcal{L}^{(3)}_{k}\prod_{i=1}^3 \mathcal{L}^{(ii+1)+}_{11|{\sigma_{i+2}}} , \ \  (s,J) \ = \  \sum_{i}\big(s_i , \ j_i\big),
\end{eqnarray}
and is $(3+1)$-parameters family to be  enumerated by the natural parameters $(j)_3$, and $k$ subject to the
relations
 \begin{eqnarray}
 && \sigma_{i} =   \frac{1}{2}(s-2J-k)-s_i  ,\ i=1,2; \qquad   \sigma_{3} =   \frac{1}{2}(s+k)-s_3  , \\
 && {\max}\big( 0, (s_3-2j_{3})-\sum_{i=1}^2(s_i-2j_{i})\big) \leq k\leq s_{3}-2j_{3}- \big|s_{1}-2j_{1}- (s_{2}-2j_{2})\big| , \\
 &&  0\leq  2j_i \leq 2 [s_i/2] , \qquad  s-2J-k= 2p, \ p\in \mathbb{N}_0.
\end{eqnarray}
Note, that the vertex  in the unconstrained formulation depends on 3 additional parameters $(j)_3$ enumerated the number of  traces in the respective set
of fields and usual one $k$ respecting the minimal order of derivatives in ${V}{}^{M(3)|m}_{{(s)_3-2(j)_3}}$. For vanishing $(j)_3$ remaining parameterp corresponds to one in constrained BRST formulation \cite{BRST-BV3}.

\subsubsection{Trace-deformed vertex generalization } \label{BRSTsolgen1mm}

The standard trace restriction $\widehat{L}{}^{(i)}_{11}|\chi^{(i)}\rangle=\widehat{L}{}^{(i)}_{11}|\Lambda^{(i)}\rangle=0$ imposed  off-shell in constrained BRST approach may be deformed (in the scheme with complete BRST operator)  on the interacting  level when deriving from resolution of deformed equations of motion  and gauge transformations by following $Q^{tot}$-closed modification of the $\mathcal{L}^{(3)}_{k}$ as compared to (\ref{LrZ1})--(\ref{LrL2e})
\begin{eqnarray}
\label{LrZ10}
  &&  \mathcal{L}^{(3)}_{1} \ = \  {L}^{(3)} - [\widehat{L}{}^{(3)}_{11},{L}^{(3)}\}\frac{b^{(3)+}}{h^{(3)}}     ,   \\
\label{LrZ2m0}
  && \widetilde{ \mathcal{L}}{}^{(3)}_{2} \ = \  (\mathcal{L}^{(3)}_1)^2 -i\widehat{\mathcal{P}}{}_0^{(3)}\eta^{(3)+}_{11}-{\widehat{l}}{}_0^{(3)}  \frac{b^{(3)+}}{h^{(3)}}     ,   \\
  \label{LrZ2m}
  &&  \widetilde{\mathcal{L}}{}^{(3)}_{2k} \ = \ (\widetilde{\mathcal{L}}^{(3)}_{2})^k, \qquad \widetilde{ \mathcal{L}}{}^{(3)}_{2k-1} \ = \ (\widetilde{\mathcal{L}}^{(3)}_{2})^{k-1}\mathcal{L}^{(3)}_{1}  ,
\end{eqnarray}
Note, the representation  (\ref{LrZ10})--(\ref{LrZ2m}) contains the term linear in $\eta^{(3)+}_{11}$ without  $\mathcal{P}^{(3)+}_{11}$.
Thus, the another (more general) solution for the vertex $   |{V}{}^{(3)}\rangle^m_{(s)_3} $ (\ref{genvertex}) is obtained after substitution of new $ \widetilde{\mathcal{L}}{}^{(3)}_{k}$  instead of old ones $ \mathcal{L}^{(3)}_{k}$. It is used in Subsection~\ref{BRSTsolgen2mmm31} for the example with the vertex $   |{V}{}^{(3)}\rangle^m_{(1,0,s)} $ of interacting massive field of spin $s$ with massless scalar and vector fields.

\subsection{Cubic vertices for one massless field and two massive fields} \label{BRSTsolgen12}
In this section we consider the cases of coinciding   and different masses for massive fields

\subsubsection{One massless and two massive fields  with coinciding  masses} \label{BRSTsolgen1mmm}

\begin{figure}[h]
{\footnotesize\begin{picture}(10,3)
\put(125.5, -40.5){\includegraphics[scale=0.35]{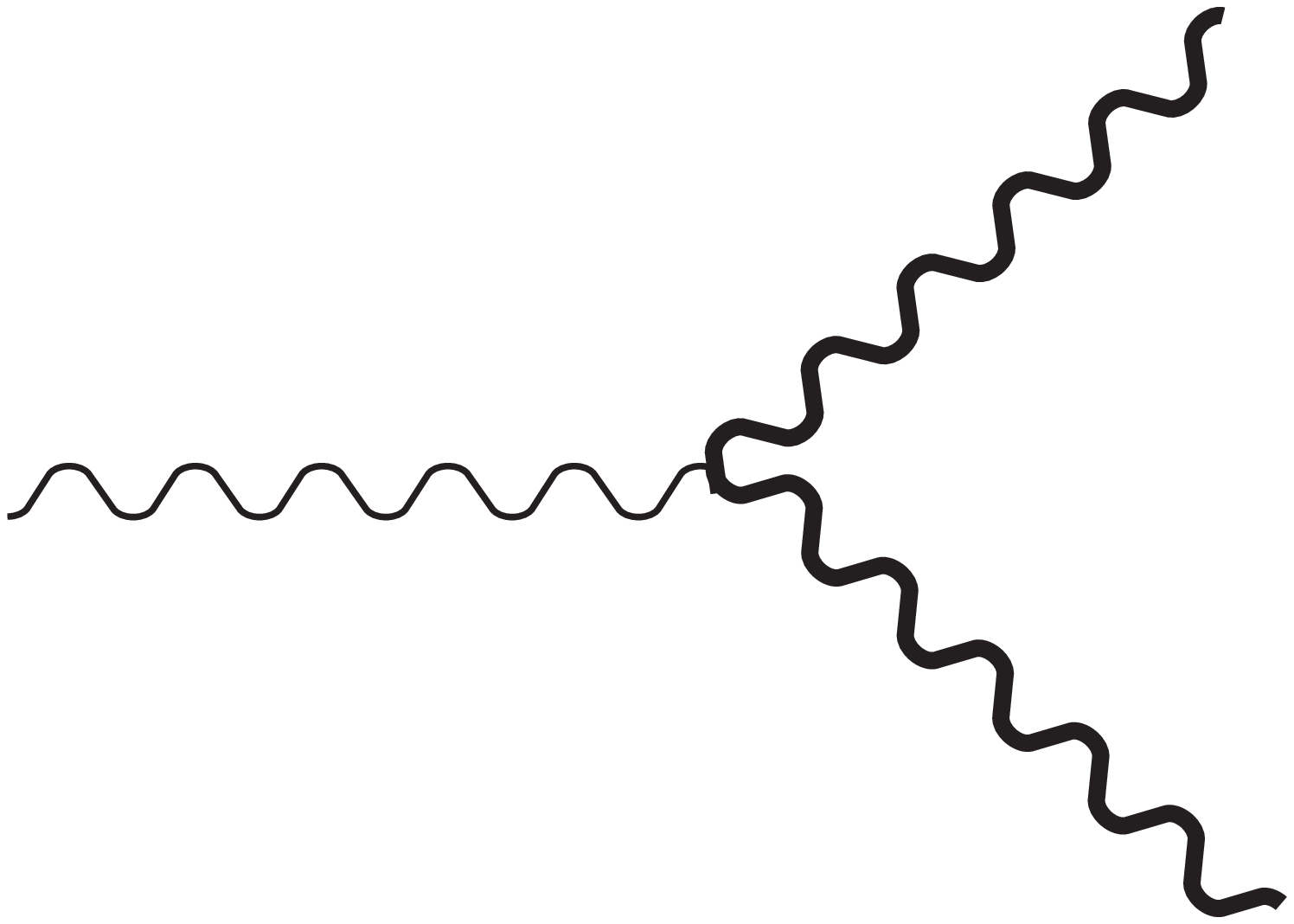}}
\put(45.5, 25.5){$ |{V}{}^{(3)}\rangle^{(0,m,m)}_{(\lambda_1 ,s_2,  s_3)}\ \ \equiv $ }
\put(150.5, 39.5){$ \big(0, \lambda_1\big)$ }
\put(285.5, 63.5){$\big(m, s_2\big)$ }
\put(285.5, -3.5){$\big(m, s_3\big)$}
\put(290.5, 30.5){$+ \ \ \ldots\ldots $ }
\end{picture}}
\vspace{3ex}
\caption{Interaction vertex  $ |{V}{}^{(3)}\rangle^{(\bar{m})_2}_{(s)_3}$   for two  massive  fields
$\phi^{(i)}_{\mu(s_i)}$ with coinciding masses  $m_2=m_3=m$ of spins $s_i$  for  $i=1,2$ and  massless field $\phi^{(1)}_{\mu(\lambda_1)}$  of  helicity $\lambda_1$. The terms in $"\ldots"$ correspond to the auxiliary fields  from $|\Phi^{(1)}\rangle_{s_1}$, $|\Phi^{(i)}\rangle_{s_i}$}\label{m0m1m1}
\end{figure}

In  the case (\ref{m1}) with $ D=0$, $P_{\epsilon m}=0$  of  massive fields of the same masses
$m_2 = m_3 = m \ne 0 $ (critical case)
there are following  $Q^{tot}_c$ BRST-closed  operators (in
\emph{minimal derivative scheme})
\begin{eqnarray}
  \label{LrZi1011}
  &&\hspace{-0.5em}  \check{L}{}^{(i)} \ = \   \widehat{p}{}^{(i)}_{\mu}a^{(i)\mu+} +(-1)^i m d^{(i)+}\theta_{i,1}- \imath \widehat{\mathcal{P}}^{(i)}_0\eta_1^{(i)+} ,\quad  i=1,2,3\\
 &&\hspace{-0.5em}
L^{(23)+}_{11} \  = \ a^{(2)\mu+}a^{(3)+}_{\mu} +\frac{1}{2m}( d^{(2)+}L^{(3)}-d^{(3)+}L^{(2)})+d^{(2)+}d^{(3)+}\label{Lrr+m11011} \\
&&  \phantom{L^{(23)+}_{11} \  =}-
\frac{1}{2}\mathcal{P}^{(2)+}_1\eta_1^{(3)+} - \frac{1}{2}\mathcal{P}^{(3)+}_1\eta_1^{(2)+}, \nonumber
\end{eqnarray}
\vspace{-2ex}
\begin{eqnarray}
  && \hspace{-0.5em} Z \ = \  \widetilde{L}^{(12)+}_{11}\check{L}^{(3)} + cycl.perm.(1,2,3) . \label{LrZLLfq011}\\
 &&\hspace{-0.5em}
\widetilde{L}^{(23)+}_{11} \ = \  L^{(23)+}_{11} - \frac{1}{2m}( d^{2)+}\check{L}^{(3)})-d^{3)+}\check{L}^{(2)})  \label{Lrr+12011}
\\
 &&\hspace{-0.5em}
\widetilde{L}^{(i{}i+1)+}_{11} \ = \ a^{(i)\mu+}a^{(i+1)+}_{\mu}
-
\frac{1}{2}\mathcal{P}^{(i)+}_1\eta_1^{(i+1)+} -
\frac{1}{2}\mathcal{P}^{(i+1)+}_1\eta_1^{(i)+}, \ i=1,3.\label{Lrr+3q011}
\end{eqnarray}
  The trace operators  $\widehat{L}{}^{(i)+}_{11}$, $\widehat{L}{}^{(i)}_{11}$  look for massless field
\begin{eqnarray}
&&\hspace{-0.5em}  \Big(\widehat{L}{}^{(1)+}_{11},\, \widehat{L}{}^{(1)}_{11}\Big)  =  \Big({l}^{(1)+}_{11} +b^{(1)+}+\mathcal{P}^{(1)+}_{1}\eta^{(1)+}_{1}, \, {l}^{(1)}_{11} +(b^{(1)+}b^{(1)}+h^{(1)})b^{(1)}+ \eta_{1}^{(1)} \mathcal{P}^{(1)}_{1}\Big)
\label{extconstsp25}
\end{eqnarray}
and  according to (\ref{extconstsp2}), (\ref{extconstsp21})  for massive  case  when $i=2,3$
\begin{eqnarray}
&& \widehat{L}{}^{(i)+}_{11} =  {l}^{(i)+}_{11}- (1/2)(d^{(i)+})^2 + b^{(i)+}+\mathcal{P}^{(i)+}_{1}\eta^{(i)+}_{1},  \label{extconstsp26} \\
&& \widehat{L}{}^{(i)}_{11} =  {l}^{(i)}_{11}- (1/2)(d^{(i)})^2 +(b^{(i)+}b^{(i)}+h^{(i)})b^{(i)}+
 \eta_{1}^{(i)} \mathcal{P}^{(i)}_{1} .
\label{extconstsp27}
\end{eqnarray}
The general solution  for the  parity invariant cubic vertex describing interaction for irreducible massless field with helicity $s_1$ and two massive  with spins $s_2$, $s_3$  with the same masses $(\bar{m})_2= (0, m,m)$
is shown  by  Figure~\ref{m0m1m1}

\vspace{15ex}
~\noindent and
  has the form
\begin{eqnarray}\label{genvertexm2mn}
   |{V}{}^{(3)}\rangle^{(\bar{m})_2}_{(s)_3} &=& |{V}{}^{M(3)}\rangle^{(\bar{m})_2}_{(s)_3}  + \sum_{(j_3,j_3,j_3) >0}^{([s_{1}/2],[s_{2}/2],[s_{3}/2])} U^{(s_1)}_{j_3}U^{(s_2)}_{j_3}U^{(s_3)}_{j_3}|{V}{}^{M(3)}\rangle^{(\bar{m})_2}_{(s)_3-2(j)_3},
   \end{eqnarray}
where the vertex $\big|{V}{}^{M(3)}\rangle^{(\bar{m})_2}_{(s)_3-2(j)_3}$ was
defined in Metsaev's paper \cite{BRST-BV3} for $(j)_3=0$ with account for (\ref{LrZi1011})--(\ref{Lrr+3q011})  but with modified forms $\mathcal{\check{L}}^{(i)}_{k_i}$,
\begin{eqnarray}
  && \label{LrL2ie} \mathcal{\check{L}}^{(i)}_{k_i} \ = \ \sum_{j=0}^{[k_i/2]} (-1)^{j}({\check{L}}^{(i)})^{k_i-2j}(\hat{p}^{(i)})^{2j} \frac{k_i!}{j! 2^j(k_i-2j))!} \frac{(b^{(i)+})^j}{C(j,h^{(i)})} , \\
   && \qquad \mathrm{ for} \ (\hat{p}^{(i)})^{2} =   \eta^{\mu\nu}(p^{(i+1)}_{\mu}-p^{(i+2)}_{\mu})  (p^{(i+1)}_{\nu}-p^{(i+2)}_{\nu}) -(-1)^i m^2 \theta_{i,1}   ,\  i=1,2,3 .
   \nonumber
\end{eqnarray}
and forms $\mathcal{L}^{(i{}i+1)+}_{11|{\sigma_{(i+2)}}}$ instead of polynomials  $\big({L}^{(i{}i+1)+}_{11}\big)^{\sigma_{(i+2)}}$ given in (\ref{Lrr+12011}), (\ref{Lrr+3q011}) for $i=1,2,3$:
\begin{eqnarray}\label{Lr1ii+11mm1}
  &&   \hspace{-0.5em} \mathcal{L}^{(i{}i+1)+}_{11|1} \ = \    {L}^{(i{}i+1)+}_{11} -\sum_{i_0=i}^{i+1} {W}^{(i_0)}_{(i{}i+1)|0}\frac{b^{(i_0)+}}{h^{(i_0)}} +\frac{1}{2}\Big(\sum_{i_0\ne j_0}[\widehat{L}{}^{(j_0)}_{11}, {W}^{(i_0)}_{(i{}i+1)|0}\}\frac{b^{(i_0)+}}{h^{(i_0)}} \frac{b^{(j_0)+}}{h^{(j_0)}}   \\
 && \hspace{-0.5em} \quad +\sum_{i_0}[\widehat{L}{}^{(i_0)}_{11}, {W}^{(i_0)}_{(i{}i+1)|0}\}\frac{(b^{(i_0)+})^2}{h^{(i_0)}(h^{(i_0)}+1)} \Big)    ,   \nonumber
  \\
   &&  \hspace{-0.5em}  \mathcal{L}^{(i{}i+1)+}_{11|k+1} \ = \ {L}^{(i{}i+1)+}_{11}\mathcal{L}^{(i{}i+1)+}_{11|k} -  \hspace{-0.3em}\sum_{i_{k}=i}^{i+1} W^{(i_{k})}_{(i{}i+1)|k}\frac{b^{(i_{k})+}}{h^{(i_{k})}}  +\frac{1}{2}\Big(\hspace{-0.3em}\sum_{i_{k}\ne j_{k}}[\widehat{L}{}^{(j_{k})}_{11},W^{(i_{k})}_{(i{}i+1)|k}\} \label{LrL23k1k1} \\
 &&  \hspace{-0.5em} \quad \times\frac{b^{(i_{k})+}}{h^{(j_{k})}} \frac{b^{(j_{k})+}}{h^{(j_{k})}}
   +\sum_{i_{k}}[\widehat{L}{}^{(i_{k})}_{11},W^{(i_{k})}_{(i{}i+1)|k}\}\frac{(b^{(i_{k})+})^2}{\prod_{p=0}^{1}(h^{(i_{k})}+p)}\Big)
     + \sum_{e=3}^{k+2}\frac{(-1)^e}{e!} \times \nonumber \\
 &&  \hspace{-0.5em} \quad  \times\sum_{ l = 0,\,  j\ne i}^{[e/2]}\frac{e!}{[e/(e-l)](e-l)!l!}\mathrm{ad}^{l}_{\widehat{L}{}^{(j_{k})}_{11}}\mathrm{ad}^{e-l-1}_{\widehat{L}{}^{(i_{k})}_{11}} W^{(i_{k})}_{(i{}i+1)|k}\frac{(b^{(j_{k})+})^l}{C(l,h^{(j_k)})}\frac{(b^{(i_{k})+})^{e-l}}{C(e-l,h^{(i_k)})}
   . \nonumber
\end{eqnarray}
(for $ k=1,...,\min(s_i,s_{i+1})-1$) and with $Q^{tot}$- BRST-closed polynomials  $\mathcal{Z}^{\lambda}$, for  $\lambda=1,2,..., \min(s_1,s_2,s_3)  $ constructed from BRST-closed forms $\widetilde{\mathcal{L}}^{(i{}i+1)+}_{11|{1}}$ according to (\ref{Lr1ii+11mm1}),  subject to change of ${\mathcal{L}}^{(i{}i+1)+}_{11|{1}}$ on one with tilde and $\mathcal{\tilde{L}}^{(i)}_{1}$:
\begin{eqnarray}
  && \mathcal{Z} \ = \  \widetilde{\mathcal{L}}^{(12)+}_{11|{1}}\mathcal{\check{L}}^{(3)}_{1} + cycl.perm.(1,2,3) . \  \mathrm{with} \ \ \mathcal{\tilde{L}}^{(i)}_{1}=  \check{L}{}^{(i)}- [\widehat{L}{}^{(i)}_{11},\check{L}^{(i)}\}\frac{b^{(3)+}}{h^{(3)}} .\label{LrZLLtot}
\end{eqnarray}
Explicitly, the vertex  ${V}{}^{M(3)|{(\bar{m})_2}}_{{(s)_3-2(j)_3}} $ is determined by
\begin{eqnarray}\label{Vmets0mm}
{V}{}^{M(3)|{(\bar{m})_2}}_{{(s)_3-2(j)_3}} & = &\sum_{\sigma, \lambda} \mathcal{ L}^{(23)+}_{11|\sigma}  \mathcal{Z}^{\lambda} \prod_{i=1}^3\mathcal{\check{L}}^{(i)}_{k_i} , \ \   \mathrm{for}\  \ 0\leq  k_1\leq s_{1}-2j_1
\end{eqnarray}
 and is $(3+2)$-parameter family to be enumerated by the natural parameters (corresponding for traces) $(j)_3$, and $k_{\min}$, $k_{\max}$   (corresponding for order of derivatives)  subject to the equations
 \begin{eqnarray}
 &&  k_{1} = k_{\min},\ \ \  k_{2} =   k_{\max}-k_{\min} - (s_{3}-2j_3),  \qquad   k_{3} =   k_{\max}-k_{\min} - (s_{3}-2j_3),\\
     &&  \sigma =s -2  (s_{1}-2j_1) - k_{\max}+2 k_{\min}, \qquad  \lambda = s_{1}-2j_1 - k_{\min}, \\
 &&\big[\max(s_2-2j_2,\,s_3-2j_3)\big] \leq  k_{\max}-k_{\min}  \leq \big[ s -2  (s_{1}-2j_1)  +  k_{\min}\big].
.\nonumber
\end{eqnarray}
The representation for the vertex (\ref{genvertexm2mn}), (\ref{Vmets0mm}) for irreducible massless and massive fields with the same masses  presents the basic results of this subsection.
For vanishing $(j)_3$ remaining parameters correspond to ones in constrained BRST formulation \cite{BRST-BV3}.

\subsubsection{One massless field and two massive fields with different masses} \label{BRSTsolgen1m1m2}

For the case (\ref{m2}) with $ D >0$  with different non-vanishing masses  $m_3\ne  m_2$ we start from    $Q^{tot}_c$ BRST-closed  operators, with except for $L^{(1)}$ (in
\emph{minimal derivative scheme})
\begin{eqnarray}
\label{LrZi231}
  &&  \hspace{-0.5em} L^{(i)}  =    \widehat{p}{}^{(i)}_{\mu}a^{(i)\mu+}+(-1)^i \frac{\delta_{i2} m_3^2 +\delta_{i3} m_2^2}{m_i} d^{(i)+}- \imath \widehat{\mathcal{P}}{}^{(i)}_0\eta_1^{(i)+} ,\ \  i=1,2,3;\\
 &&  \hspace{-0.5em}
L^{(23)+}_{11}   =  a^{(2)\mu+}a^{(3)+}_{\mu} +\frac{d^{(2)+}}{2m_2} L^{(3)}-\frac{d^{(3)+}}{2m_3}L^{(2)}+\frac{m_2^2+m_3^2}{2m_2m_3}d^{(2)+}d^{(3)+} \nonumber
\end{eqnarray}

\vspace{15ex}
\begin{figure}[h]
{\footnotesize\begin{picture}(10,3)
\put(125.5, -40.5){\includegraphics[scale=0.35]{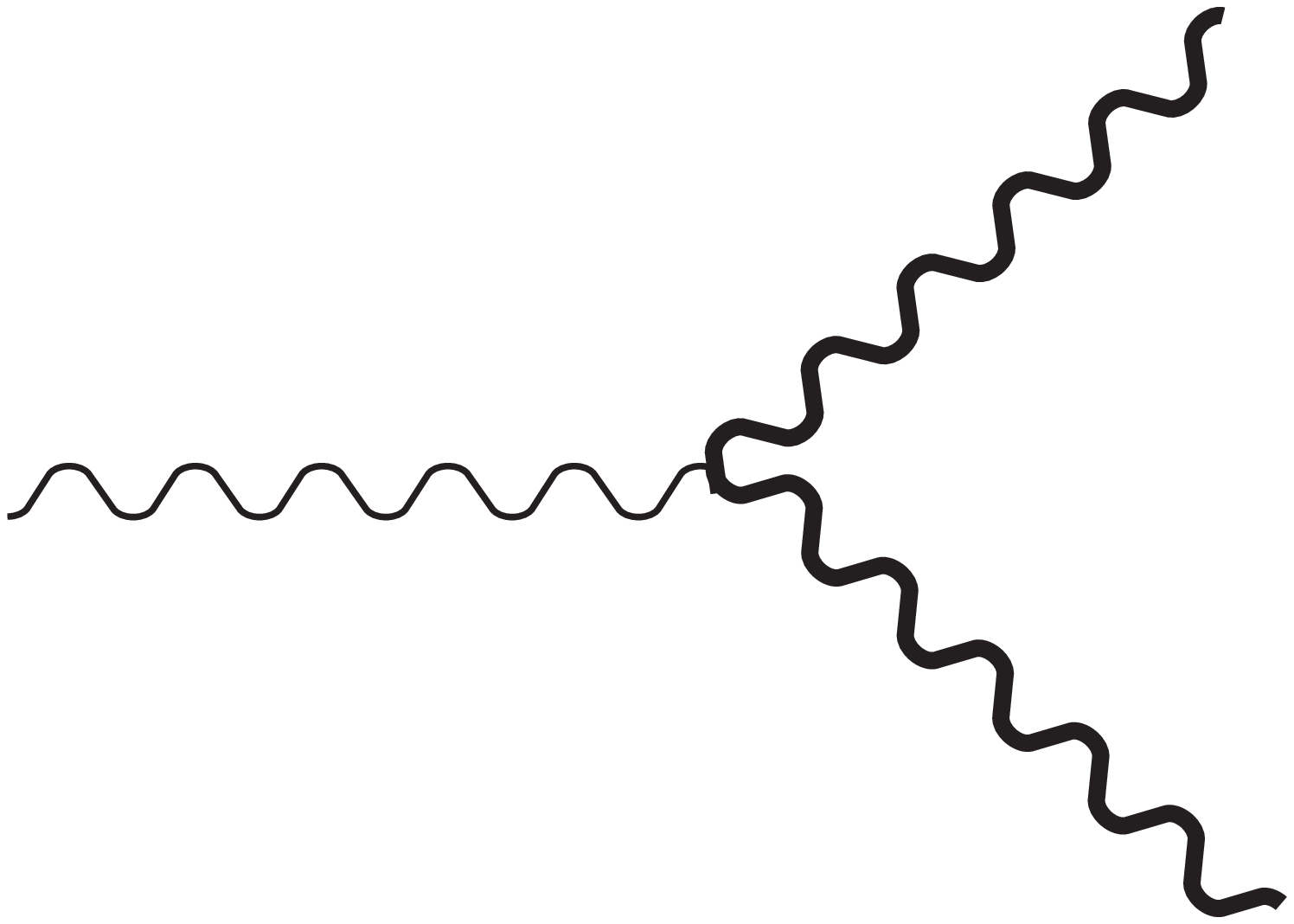}}
\put(45.5, 25.5){$ |{V}{}^{(3)}\rangle^{(0,m_2,m_3)}_{(\lambda_1 ,s_2,  s_3)}\ \ \equiv $ }
\put(150.5, 39.5){$ \big(0, \lambda_1\big)$ }
\put(285.5, 63.5){$\big(m_2, s_2\big)$ }
\put(285.5, -3.5){$\big(m_3, s_3\big)$}
\put(290.5, 30.5){$+ \ \ \ldots\ldots $ }
\end{picture}}
\vspace{3ex}
\caption{Interaction vertex  $ |{V}{}^{(3)}\rangle^{(\bar{m})_2}_{(s)_3}$   for two  massive  fields
$\phi^{(i)}_{\mu(s_i)}$ with different masses  $m_i$ of spins $s_i$  for  $i=1,2$ and  massless field $\phi^{(1)}_{\mu(\lambda_1)}$  of  helicity $\lambda_1$. The terms in $"\ldots"$ correspond to the auxiliary fields  from $|\Phi^{(1)}\rangle_{s_1}$, $|\Phi^{(i)}\rangle_{s_i}$}\label{m0m1m2}
\end{figure}~\noindent
  \begin{eqnarray}
&& \hspace{-0.5em} \phantom{L^{(23)+}_{11}}-
\frac{1}{2}\mathcal{P}^{(2)+}_1\eta_1^{(3)+} - \frac{1}{2}\mathcal{P}^{(3)+}_1\eta_1^{(2)+}, \label{Lrpr+m111} \\
  &&  \hspace{-0.5em}  L^{(12)+}_{11}   =  a^{(1)\mu+}a^{(2)+}_{\mu} \hspace{-0.15em} -  \hspace{-0.15em}\Big( \frac{d^{(2)+}}{2m_2} +\frac{L^{(2)}}{2(m_3^2-m_2^2)} \hspace{-0.2em}\Big)L^{(1)} \hspace{-0.15em}-
\frac{1}{2}\mathcal{P}^{(1)+}_1\eta_1^{(2)+} \hspace{-0.15em} - \frac{1}{2}\mathcal{P}^{(2)+}_1\eta_1^{(1)+}, \label{Lrpr+m112}\\
 &&  \hspace{-0.5em}  L^{(31)+}_{11}   =  a^{(3)\mu+}a^{(1)+}_{\mu} \hspace{-0.15em}+  \hspace{-0.15em}\Big( \frac{d^{(3)+}}{2m_3} +\frac{L^{(3)}}{2(m_3^2-m_2^2)} \hspace{-0.15em}\Big)L^{(1)} \hspace{-0.15em}-
\frac{1}{2}\mathcal{P}^{(3)+}_1\eta_1^{(1)+} \hspace{-0.15em} - \frac{1}{2}\mathcal{P}^{(1)+}_1\eta_1^{(3)+}, \label{Lrpr+m113}
\end{eqnarray}
The general solution  for the  parity invariant  cubic vertex describing interaction for irreducible massless field with helicity $s_1$ and two massive  with spins $s_2$, $s_3$  with different masses
is shown  by  Figure~\ref{m0m1m2}
and
 has the representation in terms of the product of BRST-closed (with respective complete BRST operator $Q^{tot}$ in question)  forms
\begin{eqnarray}\label{genvertexm2m3n}
   |{V}{}^{(3)}\rangle^{(m)_2}_{(s)_3} &=& |{V}{}^{M(3)}\rangle^{(m)_2}_{(s)_3}  + \sum_{(j_3,j_3,j_3) >0}^{([s_{1}/2],[s_{2}/2],[s_{3}/2])} U^{(s_1)}_{j_3}U^{(s_2)}_{j_3}U^{(s_3)}_{j_3}|{V}{}^{M(3)}\rangle^{(m)_2}_{(s)_3-2(j)_3},
   \end{eqnarray}
where the vertex $\big|{V}{}^{M(3)}\rangle^{(m)_2}_{(s)_3-2(j)_3}$ was
given in  \cite{BRST-BV3} with account for (\ref{LrZi231})--(\ref{Lrpr+m113})  but with modified forms $\mathcal{L}^{(i)}_{k_i}$,  (\ref{LrL2idiff})  and $\mathcal{L}^{(jj+1)+}_{11|{\sigma_{(j+2)}}}$ (\ref{Lr1ii+11mmT}), (\ref{LrL23k1}) instead of $\big({L}^{(jj+1)+}_{11}\big)^{\sigma_{(j+2)}}$
\begin{eqnarray}\label{Vmetsm2}
{V}{}^{M(3)|{(m)_2}}_{(s)_3-2(j)_3} & = &  \sum_{\tau_2, \tau_3}\prod_{i=2}^3\mathcal{L}^{(i)}_{\tau_i}\prod_{r=1}^3 \mathcal{L}^{(rr+1)+}_{11|{\sigma_{r+2}}} ,
\end{eqnarray}
were the ranges for $\tau_2, \tau_3, \sigma_1, \sigma_2, \sigma_3$ are specified below. Here,
\begin{eqnarray}
   && \label{LrL2idiff} \mathcal{L}^{(i)}_{k_i} \ = \ \sum_{j=0}^{[k_i/2]} (-1)^{j}({L}^{(i)})^{k_i-2j}(\hat{p}^{(i)})^{2j} \frac{k_i!}{j! 2^j(k_i-2j))!} \frac{(b^{(i)+})^j}{C(j,h^{(i)})} , \\
   &&  \mathrm{ for} \ (\hat{p}^{(i)})^{2} =   \eta^{\mu\nu}(p^{(i+1)}_{\mu}-p^{(i+2)}_{\mu})  (p^{(i+1)}_{\nu}-p^{(i+2)}_{\nu}) - \frac{(\delta_{i2} m_3^2 +\delta_{i3} m_2^2)^2}{m_i^2}    ,\  i=2,3 .
\end{eqnarray}
and for mixed trace operators, when $i=1,2,3$
\begin{eqnarray}\label{Lr1ii+11mmT}
  && \hspace{-0.5em} \mathcal{L}^{(i{}i+1)+}_{11|1}  =     {L}^{(i{}i+1)+}_{11} -\sum_{i_0=i}^{i+1} {W}^{(i_0)}_{(i{}i+1)|0}\frac{b^{(i_0)+}}{h^{(i_0)}} +\frac{1}{2}\Big(\sum_{i_0\ne j_0}[\widehat{L}{}^{(j_0)}_{11}, {W}^{(i_0)}_{(i{}i+1)|0}\}\frac{b^{(i_0)+}}{h^{(i_0)}} \frac{b^{(j_0)+}}{h^{(j_0)}}  \\
 && \hspace{-0.5em} \quad +\sum_{i_0}[\widehat{L}{}^{(i_0)}_{11}, {W}^{(i_0)}_{(i{}i+1)|0}\}\frac{(b^{(i_0)+})^2}{h^{(i_0)}(h^{(i_0)}+1)} \Big)    ,   \nonumber
  \\
   && \hspace{-0.5em}  \mathcal{L}^{(i{}i+1)+}_{11|k+1}  =  {L}^{(i{}i+1)+}_{11}\mathcal{L}^{(i{}i+1)+}_{11|k} - \hspace{-0.25em} \sum_{i_{k}=i}^{i+1}\hspace{-0.15em} W^{(i_{k})}_{(i{}i+1)|k}\frac{b^{(i_{k})+}}{h^{(i_{k})}}+\frac{1}{2}\Big(\hspace{-0.25em}\sum_{i_{k}\ne j_{k}}\hspace{-0.1em}[\widehat{L}{}^{(j_{k})}_{11},W^{(i_{k})}_{(i{}i+1)|k}\} \label{LrL23k1k2}\\
 && \hspace{-0.5em} \quad  \times \frac{b^{(i_{k})+}}{h^{(j_{k})}} \frac{b^{(j_{k})+}}{h^{(j_{k})}}
   +\sum_{i_{k}}[\widehat{L}{}^{(i_{k})}_{11},W^{(i_{k})}_{(i{}i+1)|k}\}\frac{(b^{(i_{k})+})^2}{\prod_{p=0}^{1}(h^{(i_{k})}+p)}\Big)
     + \sum_{e=3}^{k+2}\frac{(-1)^e}{e!} \times \nonumber \\
 &&\hspace{-0.5em} \quad  \times\sum_{ l = 0,\,  j\ne i}^{[e/2]}\frac{e!}{[e/(e-l)](e-l)!l!}\mathrm{ad}^{l}_{\widehat{L}{}^{(j_{k})}_{11}}\mathrm{ad}^{e-l-1}_{\widehat{L}{}^{(i_{k})}_{11}} W^{(i_{k})}_{(i{}i+1)|k}\frac{(b^{(j_{k})+})^l}{C(l,h^{(j_k)})}\frac{(b^{(i_{k})+})^{e-l}}{C(e-l,h^{(i_k)})} \nonumber
\end{eqnarray}
(for $k=1,...,\min(s_i,s_{i+1})-1$).
The vertex (\ref{Vmetsm2}) represents the  $(3+2)$-parameter family to be  enumerated by the natural parameters $(j)_3$, and  $\tau_2, \tau_3$ subject to the
equations
 \begin{eqnarray}
 && \sigma_{i} =   \frac{1}{2}[s-2J-2(s_i-2j_i)-(-1)^{\delta_{i1}}\tau_2 -(-1)^{\delta_{i2}}\tau_3] ,\ i=1,2,3;  \\
 && \big| \sum_{i=2}^3(-1)^{i}(s_i-2j_{i}- \tau_i)\big| \leq s_1-2j_1 \leq \sum_{i=2}^3(s_i-2j_{i}- \tau_i) , \\
 && s-2J - \tau_2- \tau_3= 2p, \ p\in \mathbb{N}_0.
\end{eqnarray}
The relations for the vertex (\ref{genvertexm2m3n}), (\ref{Vmetsm2}) for irreducible massless and massive fields with different masses  presents our basic results in the subsection.
Again, for vanishing $(j)_3$ remaining parameters correspond to ones in constrained BRST formulation \cite{BRST-BV3}.

\section{Examples for HS fields with special spin values} \label{examples}

Here, we consider in ghost-independent and component (tensor) forms the cubic vertices  for special cases of interacting higher spin fields.

\subsection{Vertices for Fields with  $(m,s)$, $(0,\lambda_1)$, $(0,\lambda_2)$} \label{BRSTsolgen2mmm}

In the subsection we derive the cubic vertices ${V}{}^{M(3)|m}_{(s)_3}$ for one massive HS field with $(m,s)$ and two massless HS fields with
$(0,\lambda_1)$, $(0,\lambda_2)$ with small values of two spin parameters,

\subsubsection{Case $(m,s)$, $(0,\lambda_i)$ for $\lambda_i=0$} \label{BRSTsolgen2mmm3}
First, for the interaction of 2 massless scalars with massive HS field we have according to (\ref{genvertex}), (\ref{Vmets}) the $j$-parameter family of vertices for $j=1,...,[s/2]$ with restoring  the dimensional coupling constants $t_j$ ($\dim t_j$=$s+d/2 -3-2j$, in metric units providing a dimensionless of the action)
\begin{eqnarray}\label{genvertex1}
   {V}{}^{(3)|m}_{(0,0,s)} &=&\sum_{j \geq0}^{[s/2]}t_j U^{(s)}_{j}  \mathcal{L}^{(3)}_{s-2j} = \sum_{j \geq0}^{[s/2]}t_j U^{(s)}_{j}\sum_{i=0}^{[(s-2j)/2]} (-1)^{i}({L}^{(3)})^{s-2j-2i}\times \\
   && \times (\hat{p}^{(3)})^{2i} \frac{(s-2j)!}{i! 2^i(s-2j-2i)!} \frac{(b^{(3)+})^i}{C(i,h^{(3)})} ,\nonumber
   \end{eqnarray}
   with following decomposition in powers of $\eta_1^{(3)+}$  for the operators
  \begin{eqnarray}\label{genvertexaux0}
  & \hspace{-0.9em}& \hspace{-0.9em} ({L}^{(3)})^{k}=\  ({L}^{(3)})^{k}_0 - \imath k \widehat{\mathcal{P}}^{(3)}_0\eta_1^{(3)+} ({L}^{(3)})^{k-1}_0  \equiv \big(\widehat{p}{}^{(3)}_{\mu}a^{(i)\mu+}\big)^{k-1} \hspace{-0.15em}\Big(\widehat{p}{}^{(3)}_{\mu}a^{(3)\mu+} - \imath k \widehat{\mathcal{P}}^{(3)}_0 \eta_1^{(3)+}\hspace{-0.15em}\Big),\\
\label{genvertexaux1}
  & \hspace{-0.9em}& \hspace{-0.9em}  \mathcal{L}^{(3)}_{k} \ = \ \mathcal{L}^{(3)0}_{k} - \imath  \widehat{\mathcal{P}}^{(3)}_0\eta_1^{(3)+}\big(\mathcal{L}^{(3)}_{k-1}\big)^{\prime}\ \equiv \ \mathcal{L}^{(3)}_{k}|_{\eta_1^{(3)+}=0}\\
    & \hspace{-0.9em}& \hspace{-0.9em}  \ \qquad  - \imath  \widehat{\mathcal{P}}^{(3)}_0\eta_1^{(3)+}\sum_{i=0}^{[k/2]} (-1)^{i}({L}^{(3)})_0^{k-1-2i} (\hat{p}^{(3)})^{2i} \frac{k!}{i! 2^i(k-1-2i)!} \frac{(b^{(3)+})^i}{C(i,h^{(3)})}.\nonumber
\end{eqnarray}
 The interacting part of the action $S^{(m)_3}_{[1]|(s)_3}$ (\ref{S[n]}) for $(m)_3=(0,0,m)$, ${(s)_3}=(0,0,s)$ depends only on basic $\mathbb{R}$-valued  fields $\phi^{(1)}$, $\phi^{(2)}$ $\phi^{(3)}_{\mu(s)}$ and on auxiliary ones $|\phi^{(3)}_{\mu(s-k-2l)|k,l}$. for $k=0,...,s$; $l=0,...,[s-k/2]$,   ($k+l>0$) according to (\ref{Phiphi}):
  \begin{equation}\label{S[n]1}
  S^{(m)_3}_{1|(s)_3}[\phi^{(1)}\hspace{-0.1em},\phi^{(2)}\hspace{-0.1em}, \chi^{(3)}]  =     g  \int  \prod_{i=1}^3 d\eta^{(i)}_0   \Big( {}_{s}\langle \chi^{(3)} K^{(3)}
  \big| {}_{0}\langle \phi^{(2)}
  \big| {}_{0}\langle \phi^{(1)}
  \big|{V}{}^{(3)}\rangle^{m}_{(s)_3} +h.c. \Big)  ,
  \end{equation}
  whereas the ghost-independent form for initial free action looks as
  \begin{eqnarray}\label{Sungh}
&&\sum_{i=1}^{3} \mathcal{S}^{m_i}_{0|s_i}  \ = \  \int  d^d x \Big[\sum_{i=1}^2  \phi^{(i)}\Box \phi^{(i)}+
 \mathcal{S}^{m}_{0|s} [\chi^{(3)}] \Big] ,
       \end{eqnarray}
      where  the functional $ \mathcal{S}^{m}_{0|s} [\chi^{(3)}]$  is explicitly determined in (\ref{Sungh0m}), (\ref{Scon0}) and  invariant with respect to the initial gauge transformations for the field $\big|\chi^{(3)} \rangle_{s}$ (\ref{0gtrind})--(\ref{0gtrind11})  and for the gauge parameter $\big| \Lambda^{(3)} \rangle_{s}$ (\ref{1gtrind}).
       The gauge symmetry is
                      untouched under deformation for interacting massive field $ \big| \Phi^{(3)}_{...} \rangle_{s-...}$, whereas in the sector of scalar fields they admit non-linear deformation
\begin{eqnarray}   && \delta_{[1]} \Big(\big| \chi^{(3)} \rangle_{s}, \big| \Lambda^{(3)} \rangle_{s}\Big)  =  Q^{(3)} \Big(\big| \Lambda^{( 3)} \rangle_{s}, \big| \Lambda^{(3)1} \rangle_{s}\Big) , \label{cubgtrex1}\\
         && \delta_{[1]} \big| \phi^{(1)} \rangle_{0}  =  -
g \int  d\eta^{(2)}_0 d\eta^{(3)}_0  {}_{s}\langle
\Lambda^{({3})}K^{(3)}\big|{}_{0}
   \langle \phi^{(2)}  \big|{V}{}^{(3)}\rangle^{m}_{(s)_3} , \label{cubgtrex3}\\
    && \delta_{[1]} \big| \phi^{(2)} \rangle_{0}  =  -
g \int  d\eta^{(3)}_0 d\eta^{(1)}_0 \Big( {}_{s}\langle
\Lambda^{({3})}K^{(3)}\big|{}_{0}
   \langle \phi^{({1})}  \big|{V}{}^{(3)}\rangle^{m}_{(s)_3}. \label{cubgtrex2}
\end{eqnarray}
The values of parameters $h^{(i)}(s_i)$ are equal to $h^{(1)}(0)=h^{(2)}(0)= -(d-6)/2$ and  $h^{(3)}(s)=-s-(d-5)/2$.

The interacting part of action $S^{(m)_3}_{1|(s)_3}$ (\ref{S[n]1})
 is written in the ghost-independent form
  \begin{eqnarray}\label{S[n]1ind9}
  && S^{(m)_3}_{1|(s)_3}[\phi^{(1)},\phi^{(2)}, \chi^{(3)}] \ = \    - g \prod_{i=2}^3 \delta^{(d)}\big(x_{1} -  x_{i}\big) \Big( \Big\{{}_{0}\langle \phi^{(2)}
  \big| {}_{0}\langle \phi^{(1)}\big|\Big({}_{s}\langle \Phi^{(3)} K^{(3)}
    \big|\\
    && \quad  \times\sum_{j \geq0}^{[s/2]}t_j (\check{L}{}^{(3)+}_{11})^j  \mathcal{L}^{(3)0}_{s-2j}
    + {}_{s-2}\langle \Phi^{(3)}_2 K^{(3)}
    \big|\sum_{j \geq0}^{[s/2]-1}t_j(j+1) (\check{L}{}^{(3)+}_{11})^j  \mathcal{L}^{(3)0}_{s-2(j+1)} \nonumber \\
    && \quad + {}_{s-4}\langle \Phi^{(3)}_{32} K^{(3)}
    \big|\sum_{j \geq0}^{[s/2]-2}t_j(j+1)(j+2) (\check{L}{}^{(3)+}_{11})^j  \mathcal{L}^{(3)0}_{s-2(j+2)}\nonumber \\
    && \quad - {}_{s-6}\langle \Phi^{(3)}_{22} K^{(3)}
    \big|\sum_{j \geq0}^{[s/2]-3}t_j(j+1)(j+2)(j+3) (\check{L}{}^{(3)+}_{11})^j  \mathcal{L}^{(3)0}_{s-2(j+3)}\Big\}|0\rangle +  h.c. \Big) \nonumber  ,
  \end{eqnarray}
  jointly with the gauge transformation
 \begin{eqnarray}
         &&\hspace{-0.5em} \delta_{[1]} \big| \phi^{(1)} \rangle_{0}  =  -
g \prod_{i=2}^3 \delta^{(d)}\big(x_{1} -  x_{i}\big){}_{0}\langle \phi^{(2)}\big| \Big\{{}_{s-1}\langle \Xi^{(3)}
   K^{(3)}
    \big|\sum_{j \geq0}^{[s-1/2]}t_j (\check{L}{}^{(3)+}_{11})^j \big(\mathcal{L}^{(3)}_{s-1-2j}\big)^\prime\label{cubgtrex3gi} \\
    &&\hspace{-0.5em} \quad  - {}_{s-5}\langle \Xi^{(3)}_{12}
   K^{(3)}
    \big|  \sum_{j \geq0}^{[s-5/2]}t_j(j+1)(j+2) (\check{L}{}^{(3)+}_{11})^j\big(\mathcal{L}^{(3)}_{s-5-2j}\big)^\prime\Big\}|0\rangle , \nonumber \\
    &&\hspace{-0.5em} \delta_{[1]} \big| \phi^{(2)} \rangle_{0}  = - \delta_{[1]} \big| \phi^{(1)} \rangle_{0}\vert_{\textstyle \big(\phi^{(1)}(x_1)\to \phi^{(2)}(x_2)\big)}\label{cubgtrex2gi} .
\end{eqnarray}
In deriving the action (\ref{S[n]1ind9}) we have taken of the vanishing  of  the  term  $\widehat{\mathcal{P}}{}^{(3)}_0$, whereas for the transformations (\ref{cubgtrex3gi}), (\ref{cubgtrex2gi})  its necessary presence  in ${L}^{(3)}$ by the rule (\ref{genvertexaux1}). In addition, we have used for $ U^{(s)}_{j}$
 the decomposition in powers of ghost oscillators, according to  (\ref{extconstsp2}), (\ref{trform})
 \begin{eqnarray}\label{Usj}
 && \hspace{-0.7em} U^{(s)}_{j}
 = \
\big(\check{L}{}^{(3)+}_{11}\big)^{(j-3)}\Big[\big(\check{L}{}^{+(3)}_{11}\big)^3+j \big(\check{L}{}^{+(3)}_{11}\big)^2\mathcal{P}_{1}^{(3)+}\eta_{1}^{(3)+}
+ j(j-1)\mathcal{P}_{11}^{(3)+}\eta_{11}^{(3)+}\\
&&  \phantom{U^{(s)}_{j}} \times\Big[\check{L}{}^{+(3)}_{11}+\big(j-2\big)\mathcal{P}_{1}^{(3)+}\eta_{1}^{(3)+}\Big\}\Big]. \nonumber
\end{eqnarray}
Presenting the above expressions in the oscillator forms, first, for cubic in fields action (\ref{S[n]1ind1}), second, for linear  in fields generators of the  gauge transformations (\ref{cubgtrex3gicomp1}), (\ref{cubgtrex2gicomp2})) and calculating the underlying scalar products by the rules  (\ref{A1})--(\ref{A3})
we get  finally, for the action (\ref{S[n]1ind1})  with accuracy up to overall factor $(-1)^ss!$ the representation (\ref{S[n]1ind1fa})
 \begin{eqnarray}\label{S[n]1ind1f}
  &&  \hspace{-0.5em} S^{(m)_3}_{1|(s)_3}=S^{(0)}_1\left[\phi^{(a)},  \Phi^{(3)}\right] +S^{(2)}_1\left[\phi^{(a)},  \Phi^{(3)}_2\right]+S^{(32)}_1\left[\phi^{(a)}, \Phi^{(3)}_{32}\right]+S^{(22)}_1\left[\phi^{(a)},  \Phi^{(3)}_{22}\right],
  \end{eqnarray}
(for $\phi^{(a)}=(\phi^{(1)},\phi^{(2)})$) also for the  gauge transformations  (\ref{cubgtrex3gicomp1}),
(\ref{cubgtrex2gicomp2}) given by (\ref{cubgtrex3gicomp11a}), (\ref{cubgtrex2gicomp21a})
 \begin{eqnarray}\label{cubgtrex3gicomp11}
         && \delta_{[1]} \phi^{(a)}(x_a) = (-1)^{a+1}\big(\delta_{1| \Xi^{(3)}} \phi^{(a)}(x_a) +\delta_{1| \Xi^{(3)}_{12}} \phi^{(a)}(x_a)\big),\ a=1,2.
 \end{eqnarray}
Let's discuss obtained solution.  The part of  vertex $\widehat{S}{}^{(m)_3}_{1|(s)_3}= S^{(m)_3}_{1|(s)_3}|_{j=0}$ (\ref{S[n]1ind1f}) without traces (for $j=0$ and therefore for $l=k=0$)  with only initial $\phi^{(3)\nu({s})}_{0,0}  $ and auxiliary $\phi^{(3)\nu({s-2i})}_{i,0}  $, $i=1,..., [s/2]$ fields
  reads
    \begin{eqnarray}\label{S[n]1ind1fmain}
  && \widehat{S}{}^{(m)_3}_{1|(s)_3} =  -2g t_0\int d^dx \bigg[\sum_{i=0}^{[s/2]}
     \frac{s!}{i!}  \sum_{u=0}^{s-2i} \frac{(-1)^u}{u!(s-2i-u)!}\sum_{q=0}^i \sum_{t=0}^{i-q}C_i^{q,t} \frac{ (-1)^t}{2^{i-t}} \\
   && \times \Big[\partial_{\nu_0}...\partial_{\nu_u}\Big(\partial_{\nu_{u+1}}...\partial_{\nu_{u+t}}\Box^{q}\phi^{(1)}\Big)  \Big]\Big[\partial_{\nu_{u+t+1}}...\partial_{\nu_{s-2i+t}}\partial^{\nu_{u+1}}...\partial^{\nu_{u+t}}\Box^{i-q-t}\phi^{(2)}\Big]
      \phi^{(3)\nu({s-2i})}_{i,0} \bigg\} .\nonumber
  \end{eqnarray}
  The respective gauge transformations $ \widehat{\delta}_{[1]} \phi^{(1)} ={\delta}_{[1]} \phi^{(1)} |_{j=0}$ from (\ref{cubgtrex3gicomp11}) depending only on the coupling constant $t_0$ take the form
   \begin{eqnarray} \label{cubgtrex3gicomp11red}
         &\hspace{-0.5em}&\hspace{-0.5em} \widehat{\delta}_{[1]} \phi^{(1)}(x_1) =
          -g t_0\int d^dx \bigg\{\sum_{i=0}^{[s-1/2]}  \sum_{u=0}^{s-1-2i} \frac{(s-1)!}{i!u!(s-1-2i-u)!}\sum_{q=0}^i \sum_{t=0}^{i-q}C_i^{q,t} \frac{ (-1)^t}{2^{i-t}} \\
   &\hspace{-0.5em} &\hspace{-0.5em} \times \Xi^{(3)\nu({s-1-2i})}_{i,0} (x)\Big[\partial_{\nu_{u+t+1}}...\partial_{\nu_{s-1-2i+t}}\Big(\partial^{\nu_{u+1}}...\partial^{\nu_{u+t}}\Box^{i-q-t}\phi^{(2)}\Big)\Big] \Big[\partial_{\nu_0}...\partial_{\nu_u}\Big(\Box^{q}\partial_{\nu_{u+1}}...\partial_{\nu_{u+t}}\Big)  \Big]
     \nonumber\\
  &\hspace{-0.5em}&\hspace{-0.5em} \quad  -
  \sum_{i=0}^{[s-5/2]}\sum_{u=0}^{s-5-2i} \frac{(s-5)!}{i!u!(s-5-2i-u)!}\sum_{q=0}^i \sum_{t=0}^{i-q}C_i^{q,t} \frac{ (-1)^t}{2^{i-t}}\Xi^{(3)\nu({s-5-2i})}_{12|{}i,0} (x) \nonumber\\
   &\hspace{-0.5em}&\hspace{-0.5em} \times \Big[\partial_{\nu_{u+t+1}}...\partial_{\nu_{s-5-2i+t}}\Big(\partial^{\nu_{u+1}}...\partial^{\nu_{u+t}}\Box^{i-q-t}\phi^{(2)}\Big)\Big] \Big[\partial_{\nu_0}...\partial_{\nu_u}\Big(\Box^{q}\partial_{\nu_{u+1}}...\partial_{\nu_{u+t}}\Big)  \Big]
    \bigg\}\delta^{(d)}\big(x -  x_{1}\big)
; \nonumber \\
    &\hspace{-0.5em}&\hspace{-0.5em}  \widehat{\delta}_{[1]} \phi^{(2)} (x_2) = -  \widehat{\delta}_{[1]} \phi^{(1)} (x_1)|_{[ \phi^{(1)} (x_1) \to  \phi^{(2)} (x_2)]}. \label{cubgtrex2gicomp21red}
\end{eqnarray}
We stress  the  action (\ref{S[n]1ind1fmain} and  gauge transformations (\ref{cubgtrex3gicomp11red}), (\ref{cubgtrex2gicomp21red}) coincide with ones for interacting massless fields with helicities $0,0$ and $\lambda\ \equiv  s$. The same is true for all interacting
terms in the action (\ref{S[n]1ind1f}) without traces, i.e for $j=0$, although it is not true  for the case  $j>0$ with traces  due to massive modes  presence when $k>0$.

Note, that for the \emph{Massive field strength scheme}, where    operator $\mathcal{L}^{(3)}$ (\ref{Lr3mfs}) (to be modified according to (\ref{LrZ10})) depends both on   $\widehat{p}{}^{(3){\mu}}$ and on  $d^{(3)+}$ the respective part of the action and deformed gauge transformations  without traces  contain  additional fields $\phi^{(3)\nu({s-2i-1})}_{i,1} $ related to massive modes.

Now, we may apply gauge-fixing procedure developed in the appendix~\ref{Singhcvomp} of  auxiliary fields elimination, due the procedure  independence from the scalars $\phi^{(1)},\phi^{(2)}$.
As the result, the interacting part of action (\ref{S[n]1ind1f}) will contain two terms with fields $\phi^{(3)\nu({s-2k})}_{0,2k}$, $\phi^{(3)\nu({s-2-2k})}_{2|0,2k}$ without $b^{(3)+}$-generated  fields,
 so that, the action
     \begin{eqnarray}\label{S[n]1ind1fred}
  && S^{(m)_3}_{1|(s)_3} =  -2g \int d^dx \bigg[\sum_{j \geq0}^{[s/2]}t_j\sum_{k \geq0}^{j}
     \frac{j!(s-2j)!}{2^{2(j-k)}k!(j-k)!}   \bigg\{\sum_{u=0}^{s-2j} \frac{(-1)^u}{u!(s-2j-u)!} \\
   && \times\Big[\partial_{\nu_0}...\partial_{\nu_u}\phi^{(1)}  \Big] \Big[\partial_{\nu_{u+1}}...\partial_{\nu_{s-2j}}\phi^{(2)}\Big]
    \bigg\}
      \phi^{(3)\nu({s-2k})}_{0,2k}\prod_{r=1}^{j-k}\eta_{\nu_{s-2j+2r-1}\nu_{s-2j+2r}}  \nonumber\\
   && +  \sum_{j \geq0}^{[s/2]-1}t_j\sum_{k \geq0}^{j}
     \frac{(j+1)!(s-2(j+1))!}{2^{2(j-k)}k!(j-k)!}     \bigg\{\sum_{u=0}^{s-2(j+1)} \frac{(-1)^u}{u!(s-2(j+1)-u)!} \Big[\partial_{\nu_0}...\partial_{\nu_u}\phi^{(1)}  \Big]  \nonumber\\
   && \times \Big[\partial_{\nu_{u+1}}...\partial_{\nu_{s-2(j+1)}}\phi^{(2)}\Big]
    \bigg\}
   \phi^{(3)\nu({s-2-2k})}_{2| 0,2k}\prod_{r=1}^{j-k}\eta_{\nu_{s-2(j+1)+2r-1}\nu_{s-2(j+1)+2r}}\bigg] \nonumber  .
  \end{eqnarray}
   jointly with the action (\ref{Scon00}) (also with ones for the scalars),  for free fields subject to the traceless constraints  (\ref{traceres}) may be served as interacting action in triplet formulation for the fields in question.

   When the triplet  $\big|\Phi^{(3)}_l\rangle$, $l=0,1,2$ is
 expressed in terms of only single field,  from the al\-geb\-raic equation of motion $ \big|\Phi^{(3)}_1\rangle=\check{l}_1 \big|\Phi^{(3)}\rangle-\check{l}_1^+\big|\Phi^{(3)}_2\rangle$  and due to ((\ref{traceres}):
              $\check{l}{}^{(3)}_{11}\big|\Phi^{(3)}\rangle=-  \big|\Phi^{(3)}_2\rangle$)
              \begin{eqnarray}  \label{phi2phires}
           && \big|\phi^{(3)}_{2|0,k}\rangle_{s-k-2}= \big|\phi^{(3)}_{0|0,k+2}\rangle_{s-k-2}    - 2 l^{(3)}_{11}\big|\phi^{(3)}_{0|0,k}\rangle_{s-k} \\
            && \Leftrightarrow \ \ \phi^{(3)\nu({s-k-2})}_{2| 0,k}=  \phi^{(3)\nu({s-k-2})}_{0| 0,k+2}-\phi_{0| 0,k}^{(3)\nu({s-k-2})\mu}{}_{\mu} \nonumber
              \end{eqnarray}
       (for $ k=0,1,...,s-2$)     it follows  the
 representation of free Lagrangian  (\ref{SclsrsingleF}), (\ref{gaugetrsing}) in terms of  single massive  $\phi^{(3)\nu({s})} $ and auxiliary  $\phi^{(3)\nu({s-k})}_{0,k}$ fields in $b^{(3)+}$-independent vector $\big|{\Phi}^{(3)}\rangle_s$
\begin{eqnarray}
 &&  \label{SclsrsingleFf}  \mathcal{S}^m_{C|s}\left({\Phi}^{(3)}\right)  =  {}_{s}\langle \Phi^{(3)}  \big|    \Big(  l_0-\check{l}_1^{+}\check{l}_1  -(\check{l}{}^{+}_1)^2\check{l}_{11}
 -\check{l}^{+}_{11}(\check{l}_1)^2  -\check{l}{}^{+}_{11}(l_0 +  \check{l}_1\check{l}_1^{+}) \check{l}_{11}    \Big)
       \big|{\Phi}^{(3)}\rangle_s, \\
 &&     \delta  \big|\Phi^{(3)}\rangle_{s}  =  \check{l}_1^{+}  |\Xi^{(3)}\rangle_{s-1}\ \   \mathrm{and}\ \
            \check{l}_{11}\big(\check{l}_{11}|{\Phi}^{(3)}\rangle,\,  |\Xi^{(3)}\rangle\big) = (0,0 ). \label{gaugetrsingf}
\end{eqnarray}
 In turn, the vertex (\ref{S[n]1ind1fred}) with use of   (\ref{phi2phires}) turn to one depending only on initial $\big|\phi^{(3)}\rangle_{s}$  field and Stukelberg
            $\big|\phi^{(3)}_{0|0,2k}\rangle_{s}\equiv \big|\phi^{(3)}_{0,2k}\rangle_{s}$, $k>0$  resulting in irreducible gauge theory with Lagrangian formulation $S^{(m)_3}_{[1](s)_3}$
            \begin{equation}\label{singledS1}
              S^{(m)_3}_{[1](s)_3} = \mathcal{S}^m_{C|s}\left[{\phi^{(3)}}\right] +  \int  d^d x \sum_{i=1}^2  \phi^{(i)}\Box \phi^{(i)}+ S^{(m)_3}_{1|(s)_3}\big[\phi^{(1)},\phi^{(2)},\phi^{(3)} \big]
            \end{equation}
             subject to the gauge symmetry with  constraints (\ref{gaugetrsingf}) and deformed  gauge transformations
             \begin{eqnarray} \label{cubgtrex3gicomp11redsin}
         &\hspace{-0.5em} & \hspace{-0.5em}  \delta_{[1]} \phi^{(1)}(x_1) =
          -g\hspace{-0.15em} \int \hspace{-0.15em}d^dx \bigg[\hspace{-0.15em}\sum_{j \geq0}^{[s-1/2]}\hspace{-0.15em}t_j\hspace{-0.15em}\sum_{k \geq0}^{j} \hspace{-0.15em}\sum_{u=0}^{s-1-2j}\hspace{-0.15em}
     C_j^{k,0} \frac{(s-1-2j)!}{u!(s-1-2j-u)!}\Xi^{(3)\nu({s-1-2k})}_{0,2k} (x)  \label{cubgtrex3gicomp1111} \\
   &\hspace{-0.5em}&  \hspace{-0.5em}\quad \times \frac{1}{2^{2(j-k)}} \prod_{r=1}^{j-k}\eta_{\nu_{s-1-2j+2r-1}\nu_{s-1-2j-+2r}}\Big[\partial_{\nu_{u+1}}...\partial_{\nu_{s-1-2j}}\phi^{(2)}\Big] \Big[\partial_{\nu_0}...\partial_{\nu_u}  \Big]
    \bigg]\delta^{(d)}\big(x -  x_{1}\big)
     \nonumber \\
    &\hspace{-0.5em}&  \hspace{-0.5em} \delta_{[1]} \phi^{(2)} (x_2) = -  \delta_{[1]} \phi^{(1)} (x_1)|_{[ \phi^{(1)} (x_1) \to  \phi^{(2)} (x_2)]}. \label{cubgtrex2gicomp21redsin}
\end{eqnarray}

Then we use the notation for $l$-th trace degree
         $\phi^{(3)\mu({s-2l})\rho_1...\rho_{l}}_{0,0}{}_{\rho_1...\rho_{l}}\equiv \phi^{(3)\mu({s-2l})}_{0,0}$ and
         the symmetrizer  (\ref{A1}), (\ref{A2})  $ S^{\nu(k)}_{\mu(k)}\equiv S^{\nu_1...\nu_k}_{\mu_1......\mu_k}$    with the property: $\prod_{j=1}^ka^{+\nu_j} = \frac{1}{k!}S^{\nu(k)}_{\mu(k)}\prod_{i=1}^ka^{+\mu_i}$ to present the vector (\ref{soldecomptrace})
    $\big|\phi_{2k}\rangle_{s-2k}$ in the tensor form with definite rationals $\alpha^0_{k|l}$ for $k\geq 1$
   \begin{equation*}
       \phi^{(3)\nu({s}-2k)}_{0,2k} = (-1)^k\sum_{l=1}^{[s-1/2]-k}\frac{\alpha^0_{k|l}}{(s-2(k+l))!}   S^{\nu(s-2k)}_{\mu(s-2k)} \prod_{t=1}^{l}\eta^{\mu_{s-2(k+l-t)-1}\mu_{s-2((k+l-t)}}\phi^{(3)\mu({s}-2k-2l)}_{0,0}.
\end{equation*}
Finally, in terms of  only ungauged  unconstrained  initial
            $\phi^{(3)\nu({s})}_{0,0}$ and auxiliary $\phi^{(3)\nu({s-3})}_{ 0,1} \equiv {\widetilde{\phi}}{}^{(3)\nu({s-3})}_{ 0,1}$  fields  surviving after resolution of the constraints (\ref{decompmas}) and rest interacting scalars $\phi^{(i)}$ the interacting  Lagrangian action
              \begin{eqnarray}
              &\hspace{-0.5em}&  \hspace{-0.5em}S^{(m)_3}_{[1](s)_3}\left[\phi^{(i)},{\phi^{(3)}}, \phi^{(3)}_1\right]  =  \mathcal{S}^m_{C|s}\left[{\phi}, {\phi}_1\right] +  \hspace{-0.15em}\int   \hspace{-0.15em}d^d x \sum_{i=1}^2  \phi^{(i)}\Box \phi^{(i)}+ S^{(m)_3}_{1|(s)_3}\big[\phi^{(1)},\phi^{(2)},\phi^{(3)} \big],   \label{SinghHun}\\
              &\hspace{-0.5em}& \hspace{-0.5em} S^{(m)_3}_{1|(s)_3}\big[\phi^{(i)},\phi^{(3)} \big] = - 2g \int d^dx \bigg[\sum_{j \geq0}^{[s/2]}t_j\sum_{k \geq0}^{j}
     \frac{(-1)^kj!(s-2j)!}{2^{2(j-k)}k!(j-k)!}
    \bigg\{\sum_{u=0}^{s-2j} \frac{(-1)^u}{u!(s-2j-u)!} \label{SinghHunint} \\
 &\hspace{-0.5em}& \hspace{-0.5em} \quad \times\Big[\partial_{\nu_0}...\partial_{\nu_u}\phi^{(1)}  \Big] \Big[\partial_{\nu_{u+1}}...\partial_{\nu_{s-2j}}\phi^{(2)}\Big]
    \bigg\}\prod_{r=1}^{j-k}\eta_{\nu_{s-2j+2r-1}\nu_{s-2j+2r}}\Big(\phi^{(3)\mu({s})}_{0,0}\delta_{k,0}\nonumber \\
  &\hspace{-0.5em}& \hspace{-0.5em}   +  \sum_{l=1}^{[s-1/2]-k}\frac{\alpha^0_{k|l}S^{\nu(s-2k)}_{\mu(s-2k)} }{(s-2(k+l))!}   \prod_{t=1}^{l}\eta^{\mu_{s-2(k+l-t)-1}\mu_{s-2(k+l-t)}}\phi^{(3)\mu({s}-2k-2l)}_{0,0}\Big)  \nonumber\\
   &\hspace{-0.5em}& \hspace{-0.5em} +  \sum_{j \geq0}^{[s/2]-1}t_j\sum_{k \geq0}^{j}
     \frac{(-1)^k(j+1)!(s-2(j+1))!}{2^{2(j-k)}k!(j-k)!}     \bigg\{\sum_{u=0}^{s-2(j+1)} \frac{(-1)^u}{u!(s-2(j+1)-u)!} \Big[\partial_{\nu_0}...\partial_{\nu_u}\phi^{(1)}  \Big]  \nonumber\\
   &\hspace{-0.5em}& \hspace{-0.5em} \quad\times \Big[\partial_{\nu_{u+1}}...\partial_{\nu_{s-2(j+1)}}\phi^{(2)}\Big]
    \bigg\} \prod_{r=1}^{j-k}\eta_{\nu_{s-2(j+1)+2r-1}\nu_{s-2(j+1)+2r}}\Big(-\phi^{(3)\mu({s-2})}_{0,0}\delta_{k,0}\nonumber \\
   &\hspace{-0.5em}& \hspace{-0.5em}   +
    \sum_{l=1}^{[s-3/2]-k}\frac{\alpha^0_{k+1|l}S^{\nu(s-2(k+1))}_{\mu(s-2(k+1))} }{(s-2(k+l+1))!}   \prod_{t=1}^{l}\eta^{\mu_{s-2(k+1+l-t)-1}\mu_{s-2(k+1+l-t)}}\phi^{(3)\mu({s}-2k-2-2l)}_{0,0}\nonumber\\
    &\hspace{-0.5em}& \hspace{-0.5em}
     -\eta_{\nu_{s-2k}\nu_{s-2k-1}}\sum_{l=1}^{[s-1/2]-k}\frac{\alpha^0_{k|l}S^{\nu(s-2k)}_{\mu(s-2k)} }{(s-2(k+l))!}   \prod_{t=1}^{l}\eta^{\mu_{s-2(k+l-t)-1}\mu_{s-2(k+l-t)}}\phi^{(3)\mu({s}-2k-2l)}_{0,0}
    \Big)\bigg] \nonumber
  \end{eqnarray}
  (for  $\mathcal{S}^m_{C|s}\left[{\phi}, {\phi}_1\right] $ defined in (\ref{resfv}), (\ref{SinghH} with definite rationals $\alpha^0_{k|l}$ from (\ref{soldecomptrace}) and  without involving of the field $\phi^{(3)\nu({s-3})}_{ 0,1}$ into interaction)   determines ungauge theory with accuracy up to the first order in $g$.

           Note, for constrained fields: double traceless $\phi_{0,0}^{(3)\nu({s})}$ and traceless $\phi_{0,1}^{(3)\nu({s-3})} $,
          there are only two coupling constants $t_0, t_1$ due to the absence of double traces in (\ref{SinghHun}), (\ref{SinghHunint}).

For  spin   $s=4$ the interacting action (\ref{SinghHunint}) for unconstrained  massive field $\phi_{0,0}^{(3)\nu({4})}$
 (for $\phi_{0,2}^{(3)\nu({2})} = -\frac{2}{d} \eta^{\nu(2)}\phi^{(3)\mu\nu}_{0,0}{}_{\mu\nu}$)
  takes the form
 \begin{eqnarray}
              && \hspace{-0.5em} S^{(0,0,m)}_{1|(0,0,4)}\big[\phi^{(i)},\phi^{(3)} \big] = - 2g \int d^dx \bigg\{
              t_0 \phi^{(3)\nu({4})}_{0,0}\left( \phi^{(1)} \partial_{\nu_{1}}...\partial_{\nu_{4}} - 4 \Big[\partial_{\nu_{1}}\phi^{(1)}\Big] \partial_{\nu_{2}}...\partial_{\nu_{4}}\right. \label{SinghHunint14}\\
               && \hspace{-0.5em}\qquad \left.+ 6\Big[\partial_{\nu_1}\partial_{\nu_2}\phi^{(1)}  \Big] \partial_{\nu_{3}}\partial_{\nu_{4}} -  4\Big[\partial_{\nu_1}\partial_{\nu_2}\partial_{\nu_3}\phi^{(1)}  \Big] \partial_{\nu_{4}}\right. \nonumber \\
             && \hspace{-0.5em}\qquad \left. + \Big[\partial_{\nu_1}\partial_{\nu_2}\partial_{\nu_3}\partial_{\nu_{4}}\phi^{(1)}  \Big]  \right)\phi^{(2)}+ \Big(\widetilde{t}_1 \phi^{(3)\nu(2)\nu}_{0,0}{}_{\nu}-\frac{2(t_0+t_1)}{d} \eta^{\nu_1\nu_2}\phi^{(3)\mu\nu}_{0,0}{}_{\mu\nu}\Big)\Big( \phi^{(1)} \partial_{\nu_{1}}\partial_{\nu_{2}}
 \nonumber \\
             && \hspace{-0.5em}\qquad -2\Big[\partial_{\nu_{1}}\phi^{(1)}\Big] \partial_{\nu_{2}}+\Big[\partial_{\nu_1}\partial_{\nu_2}\phi^{(1)}  \Big] \Big)\phi^{(2)}+ \widetilde{t}_2\phi^{(3)\nu_1\nu_2}_{0,0}{}_{\nu_1\nu_2}\phi^{(1)}\phi^{(2)}  \bigg\}\nonumber , \\
             . && \hspace{-0.5em}\quad
\mathrm{where}\ \ \widetilde{t}_1=  \frac{t_1 }{4}-t_0, \ \   \widetilde{t}_2\equiv -\frac{1}{4}\left( 6t_1 + \frac{7 t_2}{2}\right) \label{newconst}
            \end{eqnarray}
       are redetermined coupling constants with the dimensions: $([t_0], [\widetilde{t}_1], [\widetilde{t}_2])$ = $(\frac{d+2}{2},\frac{d-2}{2},  \frac{d-6}{2})$.
                    {We stress, that there are no any terms in the interacting action with divergences both for the model with $s=4$ and for arbitrary spin $s$.
                    In case of using the double-traceless massive field the last term with $\widetilde{t}_2$ in (\ref{SinghHunint14})  vanishes.}

           The interacting action for cubic  vertex of massless field of helicity $s$ and two massless scalars coincides with the action (\ref{SinghHun}) , (\ref{SinghHunint}) for putting $k=0$ with  $j\geq 1$  for the first term and $j=0$  for the second one for double traceless $\phi_{0,0}^{(3)\nu({s})}$. The  gauge transformations for the field  present  usual gradient one with traceless parameter  $\xi^{(3)\nu({s-1})}\equiv \Xi^{(3)\nu({s-1})}_{0,0}$, whereas in the deformed transformations for the scalars $\phi^{(i)}$ (\ref{cubgtrex3gicomp1111}), (\ref{cubgtrex2gicomp21redsin})  one should put $k=j=0$.

\subsubsection{Case  $(m,s)$, $(0,\lambda_i)$ for $ \lambda_i \leq 1$} \label{BRSTsolgen2mmm31}

Second, for the case of interaction of fields with  $(0,1)$, $(0,0)$, $(m,s)$, i.e massless vector $\phi^{(1)}_\mu$ and scalar $\phi^{(2)}$  fields and massive irreducible tensor field  $\phi^{(3)}_{\mu(s)}$, the cubic vertex will be uniquely determined
in the form (with coupling  constants $t_j$)
\begin{eqnarray}\label{genvertex2}
   {V}{}^{(3)}_{m|(1,0,s)} &=&\sum_{j \geq0}^{[s/2]}t_j U^{(s)}_{j}  \mathcal{L}^{(3)}_{s-1-2j} \mathcal{L}^{(31)+}_{11|1}= \sum_{j \geq0}^{[s/2]}t_j U^{(s)}_{j}\sum_{i=0}^{[s-1/2]-j} (-1)^{i}({L}^{(3)})^{s-1-2j-2i}\\
  & \times& (\hat{p}^{(3)})^{2i} \frac{(s-1-2j)!}{i! 2^i(s-1-2j-2i)!} \frac{(b^{(3)+})^i}{C(i,h^{(3)})}\Big(a^{(1)\mu+}a^{(3)+}_{\mu} \nonumber \\
   & - & \frac{1}{2m_3^2} L^{(1)}L^{(3)}-\frac{1}{2m_3}d^{(3)+} L^{(1)}
- \frac{1}{2}\mathcal{P}^{(1)+}_1\eta_1^{(3)+} -
\frac{1}{2}\mathcal{P}^{(3)+}_1\eta_1^{(1)+}\Big) ,\nonumber
\end{eqnarray}
The values of  $h^{(i)}(s_i)$ are equal to $h^{(1)}(0)=h^{(2)}(0)-1= -(d-4)/2$ and  $h^{(3)}(s)=-s-(d-5)/2$.
Having used the decomposition of \emph{mixed-trace} operator ${L}^{(31)+}_{11}$  (\ref{Lrr+3})  in powers of ghosts
 \begin{eqnarray}
  {L}^{(31)+}_{11} &=& {L}^{(31)+}_{11||0} + {L}^{(31)+}_{11|\mathrm{gh}} \equiv  {L}^{(31)+}_{11}|_{(\mathcal{P}=\eta^+=0)}+  \big({L}^{(31)+}_{11} - {L}^{(31)+}_{11||0} \big).\label{L31dec}
\end{eqnarray}
a component ghost-independent  form  of the interacting action  reads with account of (\ref{Sungh0m})--(\ref{1gtrind}) and (\ref{S[n]1ind103a})
  \begin{eqnarray}\label{singledS10s}
              S^{(m)_3}_{[1](1,0,s)}& =& \mathcal{S}^m_{s}\left[{\chi^{(3)}}\right] +  \mathcal{S}^0_{1}\left[{\chi^{(1)}}\right]+  \int  d^d x\big\{ \phi^{(2)}\Box \phi^{(2)}\big\}+ S^{(m)_3}_{1|(1,0,s)}\big[\chi^{(1)},\phi^{(2)},\chi^{(3)} \big], \\
              \mathcal{S}^0_{1}\left[{\chi^{(1)}}\right]&=& \left({}_{1}\langle \phi^{(1)}  \big|    {}_{0}\langle \phi^{(1)}_1\big|   \right) \left(\begin{array}{cc}
  l^{(1)}_0 &   -{l}{}^{{(1)}+}_1  \\
        -{l}^{{(1)}}_1 &   1  \end{array}\right)
       \left(  \begin{array}{l}\big|\phi^{(1)}\rangle_1 \\ \big|{\phi^{(1)}_1}\rangle_{0}   \end{array} \right) ,
                \nonumber\\ \label{S[n]1ind103}
              S^{(m)_3}_{1|(1,0,s)}&=& \sum_{J=(0, 2, 32, 22)} S^{(J)}_1\left[\phi^{(1)}, \phi^{(2)}, \Phi^{(3)}_J\right]
            \end{eqnarray}
(for $\Phi^{(3)}_0\equiv \Phi^{(3)}$)       For  deformed gauge transformations of massless fields we have with account for (\ref{+cubgtrex3gicomp10sa})--(\ref{cubgtrex2gicompa})
     \begin{eqnarray}
         && \delta_{[1]} \big| \phi^{(1)} \rangle_{r}  = \delta_{1| \Xi^{(3)}}  \big| \phi^{(1)} \rangle_{r}  + \delta_{1| \Xi^{(3)}_{12}}  \big| \phi^{(1)} \rangle_{r},\ \ r=0,\,1,\label{+cubgtrex3gicomp10s} \\
     && \delta_{[1]} \big| \phi^{(2)} \rangle_{0}  = \delta_{1| \Xi^{(3)}} \big| \phi^{(2)} \rangle_{0}+\delta_{1| \Xi^{(3)}_{12}} \big| \phi^{(2)} \rangle_{0}
+\delta_{1| \Xi^{(1)}} \big| \phi^{(2)} \rangle_{0} ,\label{cubgtrex2gicomp}
       \end{eqnarray}
and also for the components from the  field $\big| \chi^{(3)}\rangle_s$ described in (\ref{cubgtrex3gicomp10sa})--(\ref{6cubgtrex3gicomp10sa})
\begin{eqnarray}
         &\hspace{-0.7em}& \hspace{-0.7em}\delta_{1} \big| \Phi^{(3)}_P \rangle_{s-...}\hspace{-0.1em}   =\hspace{-0.1em}   \delta_{1|\Xi^{(1)}} \big| \Phi^{(3)}_P \rangle_{s-...},\ \ P\in (0, 1, 2, 22, 32,13 ),\label{cubgtrex3gicomp10s}  \\
&\hspace{-0.7em}&\hspace{-0.7em}
 \delta_{1} \big| \Phi^{(3)}_{J}\rangle_{s-...}   = \ 0 ,  \ \ J=\{12, 11, 31, 21\}\nonumber
\end{eqnarray}
At the same time,  deformed first-level  gauge transformations are trivial (for $ \delta_{1} \hspace{-0.1em}\big( \big|\Lambda^{(3)} \rangle_{s}, \hspace{-0.12em} \big|\Lambda^{(1)} \rangle_{1}\big)$ $=0$). Note, that for trace-deformed  vertex representation  (see subsection \ref{BRSTsolgen1mm})  with the operators $ \widetilde{\mathcal{L}}{}^{(3)}_{k}$  (\ref{LrZ2m})  for the zero-level gauge  parameter $\big|\Lambda^{(1)} \rangle_1$ the first-level transformations look (with respective  modifications the previous gauge transformations and the interacting  part of action)
     \begin{eqnarray}\label{11cubgtrex3gicomp10s}
         && \delta_{1} \big|\Xi^{(1)} \rangle_{0}   =  -
\frac{g }{2}\prod_{i=2}^3 \delta^{(d)}\big(x_{1} -  x_{i}\big){}_{0}\langle \phi^{(2)}\big| \Big\{{}_{s-3}\langle \Xi^{(3)1}
   K^{(3)}
    \big|\sum_{j \geq0}^{[s-3/2]}t_j (\check{L}{}^{(3)+}_{11})^j \\
    && \ \times (s-2-2j)\textstyle\left[\frac{(s-3-2j)}{2}\right] \widetilde{\mathcal{L}}^{(3)0}_{s-3-2j}|0\rangle\nonumber .
\end{eqnarray}
In deriving representations above we have used the relations (\ref{genvertexaux0}), (\ref{genvertexaux1}), (\ref{Usj}), (\ref{L31dec}).

Third, for the case [$(0,1)$, $(0,1)$, $(m,s)$] of interacting massive field $\phi^{(3)}_{\mu(s)}$ with massless vectors $\phi^{(i)}_\mu$, the cubic vertex will be determined according to (\ref{genvertex}), (\ref{Vmets})
\begin{eqnarray}\label{genvertex3}
  &\hspace{-0.5em}&\hspace{-0.5em} {V}{}^{(3)}_{m|(1,1,s)} \ = \sum_{j \geq0}^{[s/2]}t_j U^{(s)}_{j}  \Big\{ \mathcal{L}^{(3)}_{s-2j} \mathcal{L}^{(12)+}_{11|1}\hspace{-0.15em}+\mathcal{L}^{(3)}_{s-2-2j}\mathcal{L}^{(23)+}_{11|1}\mathcal{L}^{(31)+}_{11|1}\Big\}\hspace{-0.15em} = {V}{}^{(3)1}_{m|(1,1,s)}\hspace{-0.15em}+{V}{}^{(3)2}_{m|(1,1,s)},\\
   \label{genvertex31}
  &\hspace{-0.5em}&\hspace{-0.5em} {V}{}^{(3)1}_{m|(1,1,s)} \  = \  \sum_{j \geq0}^{[s/2]}t_j U^{(s)}_{j}\sum_{i=0}^{[s/2]-j} (-1)^{i}({L}^{(3)})^{s-2j-2i} (\hat{p}^{(3)})^{2i} \frac{(s-2j)!}{i! 2^i(s-2j-2i)!} \\
   &\hspace{-0.5em}&\hspace{-0.5em}  \quad \times \frac{(b^{(3)+})^i}{C(i,h^{(3)})}\Big(a^{(1)\mu+}a^{(2)+}_{\mu} +\frac{1}{2m_3^2} L^{(1)}L^{(2)}-
\frac{1}{2}\mathcal{P}^{(1)+}_1\eta_1^{(2)+} -
\frac{1}{2}\mathcal{P}^{(2)+}_1\eta_1^{(1)+}\Big) ,\nonumber \\
\label{genvertex32}
 &\hspace{-0.5em}&\hspace{-0.5em}  {V}{}^{(3)2}_{m|(1,1,s)} \  = \  \sum_{j \geq0}^{[s/2]}t_j U^{(s)}_{j}\sum_{i=0}^{[s/2]-j-1} (-1)^{i}({L}^{(3)})^{s-2-2j-2i} (\hat{p}^{(3)})^{2i} \frac{(s-2-2j)!}{i! 2^i(s-2-2j-2i)!} \\
  &\hspace{-0.5em}&\hspace{-0.5em}  \quad  \times \frac{(b^{(3)+})^i}{C(i,h^{(3)})}
   \Big({L}^{(23)+}_{11}{L}^{(31)+}_{11|1} -\big[{W}^{(3)}_{(23)|0},{L}^{(31)+}_{11|1}\big\}\frac{b^{(3)+}}{h^{(3)}}\Big),\nonumber
   \end{eqnarray}
   with
   \begin{eqnarray}
&\hspace{-0.5em}&\hspace{-0.5em}  \big[{W}^{(3)}_{(23)|0},{L}^{(31)+}_{11|1}\big\}\  = \  \big[\big[{\widehat{L}}{}^{(3)}_{11}, {L}^{(23)+}_{11}\big\},{L}^{(31)+}_{11}\big\}\label{genvertex321}
 \\ &\hspace{-0.5em}&\hspace{-0.5em} \phantom{ \big[{W}^{(3)}_{(23)|0},{L}^{(31)+}_{11|1}\big\}} =  \frac{1}{2}\Big[\big(2a^{(2)\mu+}- \frac{1}{m_3^2} L^{(2)}\widehat{p}{}^{(3)\mu}\big)\big(a^{(1)+}_{\mu}- \frac{1}{2m_3^2} L^{(1)}\widehat{p}{}^{(3)}_{\mu}\big)\nonumber
 \\
&\hspace{-0.5em}&\hspace{-0.5em} \phantom{ \big[{W}^{(3)}_{(23)|0},{L}^{(31)+}_{11|1}\big\}} +\frac{1}{2}\big(-i\frac{1}{2m_3^2} L^{(2)}\widehat{\mathcal{P}}{}^{(3)}_0+ \mathcal{P}^{(2)+}_1\big)\eta_1^{(1)+}+ \frac{1}{2}\mathcal{P}^{(1)+}_1\eta_1^{(2)+} +\frac{1}{2m^2_3} L^{(2)} L^{(1)}\Big],
    \nonumber
\end{eqnarray}
where we have  used  definitions (\ref{Lrr+1})--(\ref{Lrr+3})  for $L^{(i{}i+1)}_{11}$  and (\ref{defW}) for ${W}^{(3)}_{(23)|0}$.

As the result, the interacting action and deformed  gauge transformations may be found according to the above developed  procedure for the  triples of fields
with $(0,\lambda_1)$,  $(0, 0)$, $(m,s)$ for $\lambda_1=0,1$. We stress that the gauge   transformations for the massless fields become to be non-Abelian and reducible, whereas the gauge symmetry  for the massive field is deformed by remains with untouched reducibility relation with accuracy up  the first order in $g$

\subsubsection{Case $(m,s)$, $(0,\lambda_i)$ for $ (\lambda_2 +s)\leq 1$} \label{BRSTsolgen2mmm312}

Firstly, we stress that there is no non-trivial interaction for massless field of helicity $\lambda_1\equiv \lambda \in \mathbb{Z}$ with massless and massive scalars except for "trace" vertex (without derivatives) for even $\lambda=2r$
\begin{eqnarray}\label{genvertex4}
   {V}{}^{(3)}_{m|(\lambda,0,0)} &=& \delta_{\lambda, 2[\lambda/2]}U^{(\lambda)}_{[\lambda/2]},
. \end{eqnarray}
which however vanishes after passing to Fronsdal (single-field) formulation for massless tensor. $\phi^{\mu(\lambda)}$

For the case of $(\lambda_2, s)=(1, 0)$ or $(\lambda_2, s)=(0, 1)$  non-trivial  solutions for the vertex exist.
For the latter case the vertex has the representation
\begin{eqnarray}\label{genvertex5}
   {V}{}^{(3)}_{m|(\lambda,0,1)} &=&t  \delta_{\lambda, 2[\lambda/2]}U^{(\lambda)}_{[\lambda/2]} {L}^{(3)}+  t_1  \delta_{\lambda, 2[\lambda/2]+1}U^{(\lambda)}_{[\lambda/2]} \mathcal{L}^{(31)+}_{11}= {V}{}^{(3)1}_{m|(\lambda,0,1)}+{V}{}^{(3)2}_{m|(\lambda,0,1)},
   \end{eqnarray}

\subsection{Vertices for  Fields with  $(0,\lambda_1)$, $(m, 0)$, $(m,s_3)$} \label{BRSTsolgen2mmmm}

In this subsection we derive ghost-independent form for the cubic vertices $V_{}$ for  two massive HS fields with $(m,s_i)$, $i=2,3$ for coinciding masses  and one  massless  HS field
$(0,\lambda_1)$ restricting by the values  $\lambda_1=s_1=0$.

The vertex is  $1$-parameter family and determined according to general prescription (\ref{genvertexm2mn})--(\ref{Vmets0mm})
\begin{eqnarray}\label{genvertexm2mnm}
   {V}{}^{(3)}_{(\bar{m})_2|(0,0,s)} &=&  \sum_{(j) =0}^{[s/2]} U^{(s)}_{j}\mathcal{L}^{(3)}_{s-2j} = \sum_{(j) =0}^{[s/2]} U^{(s)}_{j}\sum_{i=0}^{[s/2]-j} (-1)^{i}({\check{L}}^{(3)})^{s-2j-2i} \\
   &\times & (\hat{p}^{(3)} + m^2 )^{2i} \frac{(s-2j)!}{i! 2^i(s-2(j+i))!} \frac{(b^{(3)+})^i}{C(i,h^{(3)})}  \nonumber ,
   \end{eqnarray}
where the operator ${\check{L}}^{(3)}\ $ (\ref{LrL2ie}) is additive $d^{(3)+}$ extension  of the standard operator ${{L}}^{(3)}$ (\ref{LrZ}):
\begin{eqnarray}\label{cL3d}
   && \qquad {\check{L}}^{(3)} -{{L}}^{(3)}= - m d^{(3)+}  .
\end{eqnarray}
The quantity  (\ref{genvertexm2mnm}) appears by the extension of the vertex $ {V}{}^{(3)}_{m|(0,0,s)}$  (\ref{genvertex1})  from the subsection \ref{BRSTsolgen2mmm3}
 for the interaction two massless scalars with massive HS field and contains it for $q=0$ in
 \begin{eqnarray}\label{genvertexm2mnext}
   {V}{}^{(3)}_{(\bar{m})_2|(0,0,s)} &=& \sum_{(j) =0}^{[s/2]} U^{(s)}_{j}\sum_{i=0}^{[s/2]-j} \sum_{q=0}^{s-2j-2i}(-1)^{i+q}
   ({L}^{(3)})^{s-2j-2i-q}(m d^{(3)+})^q \\
   &\times & (\hat{p}^{(3)} + m^2 )^{2i} \frac{(s-2j)!}{i! q!2^i(s-2j-2i-q))!} \frac{(b^{(3)+})^i}{C(i,h^{(3)})}  \nonumber ,
   \end{eqnarray}
(for $h^{(1)}(0)=h^{(2)}(0)+1/2= -(d-6)/2$ and  $h^{(3)}(s)=-s-(d-5)/2$)  The representation above
permits one to determine
the interacting action in the ghost-independent form
 \begin{eqnarray} \label{S[n]1indmmmg}
 &\hspace{-0.5em}&\hspace{-0.5em} S^{(\bar{m})_2}_{[1]|(0,0,s)}[\phi^{(i)}, \chi^{(3)}]  =   \int \hspace{-0.15em} d^d x\hspace{-0.2em} \sum_{i=1}^2 \phi^{(i)}(\Box+\delta_{2,i}m^2) \phi^{(i)}+
 \mathcal{S}^m_{s}[\chi^{(3)}] +S^{(\bar{m})_2}_{1|(0,0,s)}[\phi^{(i)}, \chi^{(3)}], \\
 \label{S[n]1indmmm}
  &\hspace{-0.5em}&\hspace{-0.5em} S^{(\bar{m})_2}_{1|(0,0,s)}[\phi^{(i)}, \chi^{(3)}] \ = \    - g \prod_{i=2}^3 \delta^{(d)}\big(x_{1} -  x_{i}\big) \Big( \Big\{{}_{0}\langle \phi^{(2)}
  \big| {}_{0}\langle \phi^{(1)}\big|\Big[{}_{s}\langle \Phi^{(3)} K^{(3)}
    \big|
  \end{eqnarray}
     \begin{eqnarray}
    &\hspace{-0.5em}& \hspace{-0.5em} \qquad  \times\sum_{j \geq0}^{[s/2]}t_j (\check{L}{}^{(3)+}_{11})^j  \mathcal{\check{L}}^{(3)0}_{s-2j}
    + {}_{s-2}\langle \Phi^{(3)}_2 K^{(3)}
    \big|\sum_{j \geq0}^{[s-2/2]}t_j(j+1) (\check{L}{}^{(3)+}_{11})^j  \mathcal{\check{L}}^{(3)0}_{s-2(j+1)} \nonumber \\
    &\hspace{-0.5em}& \hspace{-0.5em} \qquad + {}_{s-4}\langle \Phi^{(3)}_{32} K^{(3)}
    \big|\sum_{j \geq0}^{[s-4/2]}t_j(j+1)(j+2) (\check{L}{}^{(3)+}_{11})^j  \mathcal{\check{L}}^{(3)0}_{s-2(j+2)}\nonumber \\
    &\hspace{-0.5em}& \hspace{-0.5em} \qquad - {}_{s-6}\langle \Phi^{(3)}_{22} K^{(3)}
    \big|\sum_{j \geq0}^{[s-6/2]}t_j(j+1)(j+2)(j+3) (\check{L}{}^{(3)+}_{11})^j  \mathcal{\check{L}}^{(3)0}_{s-2(j+3)}\Big]\Big\}|0\rangle +  h.c. \Big) \nonumber  ,
  \end{eqnarray}
  (for $\mathcal{L}^{(3)0}_{s-2(j+...)}\equiv \mathcal{L}^{(3)}_{s-2(j+...)}|_{\eta^+=0}$) together with deformed gauge transformations
 \begin{eqnarray}
    \hspace{-0.7em}     &\hspace{-0.7em}& \hspace{-0.7em} \delta_{[1]} \big| \phi^{(1)} \rangle_{0}  =  -
g \prod_{i=2}^3 \delta^{(d)}\big(x_{1} -  x_{i}\big){}_{0}\langle \phi^{(2)}\big| \Big\{{}_{s-1}\langle \Xi^{(3)}
   K^{(3)}
    \big|\hspace{-0.25em}\sum_{j \geq0}^{[s-1/2]}\hspace{-0.25em}t_j(s-2j) (\check{L}{}^{(3)+}_{11})^j \mathcal{\check{L}}^{(3)0}_{s-1-2j}\label{cubgtrex3gim} \\
\hspace{-0.7em}    &\hspace{-0.7em}& \hspace{-0.7em}\quad  - {}_{s-5}\langle \Xi^{(3)}_{12}
   K^{(3)}
    \big|  \sum_{j \geq0}^{[s-5/2]}t_j(j+1)(j+2)(s-2(j+2)) (\check{L}{}^{(3)+}_{11})^j \mathcal{\check{L}}^{(3)0}_{s-5-2j}|\Big\}|0\rangle ; \nonumber\\
    \hspace{-0.7em}&\hspace{-0.7em}&\hspace{-0.7em} \delta_{[1]} \big| \phi^{(2)} \rangle_{0}  =
g \prod_{i=1,3}\delta^{(d)}\big(x_{2} -  x_{i}\big)\hspace{-0.15em} {}_{0}\langle \phi^{(1)}\big| \Big\{{}_{s-1}\langle \Xi^{(3)}
   K^{(3)}
    \big|\hspace{-0.25em} \sum_{j \geq0}^{[s-1/2]}\hspace{-0.25em} t_j(s-2j) (\check{L}{}^{(3)+}_{11})^j  \mathcal{\check{L}}^{(3)0}_{s-1-2j}\label{cubgtrex2gim} \\
 \hspace{-0.7em}   &\hspace{-0.7em}&\hspace{-0.7em} \quad  - {}_{s-5}\langle \Xi^{(3)}_{12}
   K^{(3)}
    \big|  \sum_{j \geq0}^{[s-5/2]}t_j(j+1)(j+2)(s-2(j+2)) (\check{L}{}^{(3)+}_{11})^j  \mathcal{\check{L}}^{(3)0}_{s-5-2j}\Big\}|0\rangle . \nonumber
\end{eqnarray}
The action functional  for free massive HS field $\mathcal{S}^m_{s}[\chi^{(3)}]$ with respective reducible transformations  are presented by the formulas (\ref{Sungh0m}), (\ref{1gtrind}). Note, the interacting part of action and deformed  gauge transformations contain operators with "check" $ \mathcal{\check{L}}^{(3)0}_{s-1-2j}$ depending on $d^{(3)+}$ by the rule (\ref{cL3d}) opposite to the operators $\mathcal{{L}}^{(3)0}_{s-1-2j}$ for interaction of  two massless scalars with massive HS field. It means on the component level after gauge fixing procedure that  the auxiliary field $\phi^{(3)\nu({s-3})}_{0| 0,3}$ will be included therefore into the vertex on the equal footing with basic field$\phi^{(3)\nu({s})}_{0| 0,0}$.

Emphasize, that the inclusion  the constraints $L^{(i)}_{11}$ responsible
for the traces into BRST operator means that the standard condition
of vanishing double traces of the fields is fulfilled only on-shell
as the consequence of free equations of motion. Off-shell the
(double) traces of the fields do not vanish. At the same time the vanishing of double (single) traces of the fields (gauge parameters) for interacting higher spin fields
  is modified as compared to the case of free dynamics however with preservation of irreducibility for  any interacting (basic) fields,   As the result, the
trace constraints come into the cubic vertices whereas the respective ghost oscillators enter into the vertex generating operators (see e.g. (\ref{LrZ2m})) beyond these trace conditions..

\section{Conclusion}

\label{Discus} 

To sum up, we have constructed the generic cubic vertices for a
first-stage reducible gauge-invariant Lagrangian formulations of
totally-symmetric massless and massive  higher spin  fields with
arbitrary integer helicities and spins in $d$-dimensional  Minkowski
space-time in three different cases: for two massless fields of
helicities $\lambda_1, \lambda_2$ and massive field of spin $s_3$;
for one massless field of helicity $\lambda_1$ and two massive
fields  of spins $s_2, s_3$, first, with coinciding masses, second,
with different masses $m_2\ne m_3$. The procedure is realized in the
framework of the BRST approach, developing our {earlier} results for cubic
vertices for irreducible massless fields \cite{BRcub},
\cite{Rcubmasless} to higher spin field theories with complete BRST
operator, which includes all the constraints that determine an
irreducible massless or massive  higher spin representation on equal
footing. This approach allows to preserve  the irreducibility of
Poincare group representation for each interacting  higher spin
field and as consequence to provide preserving the  number of
physical degrees of freedom on the cubic level up to the first power
in the deformation parameter $g$.

To determine cubic vertices being consistent with a deformed gauge
invariance, we have realized an additive deformation of classical
actions for three copies of the respecting massless and massive higher spin fields and the gauge
transformations for the fields and gauge parameters, while requiring
for the deformed action  to be invariant in a linear approximation
with respect to  $g$, and for the gauge algebra
to be closed on a deformed mass shell up to the second order in $g$.
These requirements, as for {as for} massless case \cite{BRcub}, result in a system of generating equations for the
cubic vertices, containing the total BRST invariance operator
condition $\mathcal{Q}(V^3,\widetilde{V}^3) =0$,
$\mathcal{Q}(\widetilde{V}^3,\widehat{V}{}^3) =0$ (\ref{g1operV3}),
 the spin condition, and the
condition (\ref{closuregtr}) for the gauge algebra closure. The
cubic vertex, in the particular case of coinciding operators,
${V}^3=\widetilde{V}^3 =\widehat{V}{}^3$, satisfies the  equations
(\ref{g1Lmod}), and their solutions are found  using a respective
set of spin- and BRST-closed forms within a classification of
vertices  with respect to values of   polynomials of the fourth
order $D(m_1,m_2,m_3)$ and the first order $P(m_1,m_2,m_3)$ in power
of mass, considered, firstly  for $d=4$ in \cite{Metsaev-mass}, and
extracting the cases of real ($D>0$), virtual ($D<0$)  processes,
and real process ($D=0$) with vanishing transfer of momentum.  For
two massless and one massive higher-spin fields the modified
BRST-closed differential forms (\ref{LrZ1}) , (\ref{LrL2e}),
(\ref{Lr12312}), (\ref{Lr1231}),  (\ref{LrL23k1})  constructed from
ones in \cite{BRST-BV3}, and the new forms (\ref{trform}) related to
the trace operator constraints (having dependence on additional
oscillators, $b^{(i)+}$, $d^{(3)+}$ $\eta_{11}^{(i)+}$,
$\mathcal{P}_{11}^{(i)+}$)  compose the parity invariant  cubic
vertex $ {V}{}^{(3)|m}_{(s)_3}$ (\ref{genvertex}). The vertex has a
non-polynomial structure and  presents  $(3+1)$-parameters family to
be  enumerated by the natural parameters $(j_1,j_2,j_3)$  respecting
the orders of traces incoming into the vertex, and $k$ enumerating
the order of derivatives in it\footnote{{For another elaboration of inclusion the trace
constraints in Maxwell-like Lagrangians for interacting massless higher
spin fields an constant curvature spaces with multiple traces see \cite{FranciaMM}}.}.
As a result parity invariant  cubic
vertices for irreducible fields may involve terms with less
space-time derivatives as compared  with \cite{BRST-BV3}. This
vertex maybe equivalently presented in the polynomial form with
non-commuting BRST-closed generating elements:
$\mathcal{L}^{(3)}_{1}$  (\ref{LrZ10})  and
$\mathcal{L}^{(i{}i+1)+}_{11|1}$ (\ref{Lr12312}),  (\ref{Lr1231})
for $i-1,2,3$, which depend in addition on annihilation oscillators
as compared to the standard receipt \cite{reviews3},
\cite{BRST-BV3}. The vertex admit trace-deformed generalization
leading to the  change of the  standard trace restrictions on fields
and gauge parameters  imposing  off-shell in constrained BRST
approach that is revealed in BRST-closed modification of  two form
${ \mathcal{L}}{}^{(3)}_{2}$ (\ref{LrZ2m}) by "trace
$\eta^{(3)+}_{11}$   ghost. It means that after performing the
gauge-fixing procedure and partial resolution of the interacting
equations of motion  the final trace restrictions for the initial
higher-spin fields should not coincide with standard ones derived
from the Lagrangian  formulations for free fields.

For the case of one massless and two massive fields with coinciding
masses (when $D=P=0$) the solution for the vertex contains more
BRST-closed generating differential forms given by three sets
(\ref{LrL2ie}): Yang--Mills type form $\mathcal{Z}$
(\ref{LrZLLtot})  mixed-trace forms  $ \mathcal{L}^{(2{}3)+}_{11|k}
$  (\ref{LrL23k1k1})  with new trace forms  the parity invariant
cubic vertex ${V}{}^{(3)|{(\bar{m})_2}}_{(s)_3}$
(\ref{genvertexm2mn}), (\ref{Vmets0mm}) is constructed. It presents
$(3+2)$-parameter family to be enumerated by the natural parameters
(corresponding for traces) $(j_1,j_2,j_3)$  and $k_{\min}$,
$k_{\max}$   corresponding for order of derivatives. Again, the
vertex admits a polynomial representation in terms of non-commuting
BRST-closed  generating operators.

For the variant of one massless and two massive fields with
different masses (when $D>0$) derived spin- and BRST-closed vertex
${V}{}^{(3)|(m)_2}_{(s)_3}$  (\ref{genvertexm2m3n}), (\ref{Vmetsm2})
was constructed as the product of differential $(2+3)$ sets of
differential $\mathcal{L}^{(i)}_{k_i} $  (\ref{LrL2idiff})  and
mixed-trace  $  \mathcal{L}^{(i{}i+1)+}_{11|k}$, for $i=1,2,3$
(\ref{Lr1ii+11mmT}), (\ref{LrL23k1k2}) and respective new  trace forms
$U^{(s_i)}_{j_i}$ of the rank $j=1,2,..., [s_i/2]$. The vertex
represents the  $(3+2)$-parameter family to be  enumerated by the
natural parameters $(j_1,j_2,j_3)$  and  $\tau_2, \tau_3$.  A
polynomial representation for the vertex also exists

From the obtained solutions it follows, first, the possibilities  to
construct cubic vertices and in\-ter\-acting first-stage reducible
Lagrangian formulations for mentioned three cases including the
fields with all helicities and spins, e.g. for triples with  two
massless and one massive fields
\begin{equation}\label{V3allspins}
{V}{}^{(3)|m}_{\sum(s)_3}= \sum_{(\lambda_1, \lambda_2,s_3)\geq 0} {V}{}^{(3)|m}_{(s)_3}
\end{equation}
 with the same  mass for massive fields $\phi^{(3)\mu(s_3)}(x)$ with different values of spin  $s_3$.

{Second, a condition that the cubic approximation for interacting model will be the final term (without higher order vertices) in both Lagrangian and gauge transformations is based on the non-trivial solution of the operator equation on the vertex $\big|{V}{}^{(3)}\rangle^{(m)_3}_{(s)_{3}}$ in the second order in deformation constant $g$:
}\begin{eqnarray}\label{V3V3}
 && \left(\left\{\mathcal{V}(i_1,j_1;i_2,j_2)+ \mathcal{V}(j_1, i_2,;i_1,j_2)\right\} -(i_1,j_1)\right)+ \left(\mathcal{V}(i_2, i_1;j_1j_2)-(i_1,i_2)\right) =0,  \\
  && \texttt{for} \   \int d\eta^{(3)}_0 {}^{(m)_3}_{(s_1,s_2,s_3)}\langle{V}{}^{(3)}\big|K^{(3)}\big|{V}{}^{(3)}\rangle^{(m)_3}_{(s^{\prime}_{1},s^{\prime}_{2},s_{3})} \equiv \mathcal{V}(s_1,s_2;s^{\prime}_{1},s^{\prime}_{2}), \ \ i_1,j_1,i_2,j_2=1,2,3
  \nonumber
\end{eqnarray}
{which should be considered additionally to the system(\ref{g1Lmod}).}

The inclusion of trace constraints into the complete  BRST operator
has led to a larger content of configuration spaces in Lagrangian
formulations for interacting massless and massive  fields of integer
spins in question (in comparison with the constrained BRST approach
\cite{BRST-BV3}), which has permitted the appearance of new trace
operator components $U^{(s_i)}_{j_i}$ \ in the cubic vertex. In this
regard, the correspondence between the obtained vertices
$|{V}{}^{(3)}\rangle$ and the respective vertices
$|{V}{}^{M(3)}\rangle$ of \cite{BRST-BV3} is not unique due to the
fact that the tracelessness conditions for the latter vertex are not
satisfied: ${L}{}^{(i)}_{11}$ $|{V}{}^{M(3)}\rangle \ne 0$ as it was
in details discussed in the Appendix~\ref{appconBRST}. Both vertices
for the same set of higher-spin fields  will correspond to each
other, first, after extracting the irreducible components
$\big|  \overline{V}{}^{(3)}_c\rangle^{(m)_3}_{(s)_{3}} \equiv |{V}{}^{M(3)}_{irrep}\rangle$ from $|{V}{}^{M(3) }\rangle$,
satisfying ${L}{}^{(i)}_{11}|{V}{}^{M(3)}_{irrep}\rangle = 0$ according to (\ref{L11V+}).
{ We pay attention, that the form of the cubic vertices for irreducible (massless and massive)
higher-spin fields within approach with incomplete BRST operator is  firstly obtained by the equation (\ref{L11V+}).}    Then,
after eliminating the auxiliary fields and gauge parameters by
partially fixing the gauge and using the equations of motion, the
vertex $|{V}{}^{(3)}\rangle $ will transform to
$|{\breve{V}}{}^{(3)}\rangle$ in a triplet formulation of
\cite{BRST-BV3}, so that, up to total derivatives, the vertices
$|{V}{}^{M(3)}_{irrep}\rangle$ and $|{\breve{V}}{}^{(3)}\rangle$
must coincide. At the same time different
representation for the vertices with the same set of fields, among
them with trace-deformed generalization (\ref{LrZ2m}) leads to
different local representations of the interacting Lagrangian
formulations  as it was shown with generation of non-trivial
deformed first-level gauge transformation for vector gauge parameter
(\ref{11cubgtrex3gicomp10s}) for the interacting massless vector and
scalar fields with massive  $(m,s)$ field. We stress, following to
the Appendix~\ref{appconBRST} results, that imposing of only
traceless constraints on fields and gauge parameters (\ref{L11})
represents the necessary but not sufficient condition for the
consistency of deformed (on cubic level) Lagrangian dynamics for
interacting higher spin fields with spins $s_1, s_2, s_3$ within
constrained BRST approach. In addition to BRST closeness one should
valid the traceless conditions for the cubic vertices
(\ref{gencubBRSTc}) that guarantees the preservation of Poincare
group irreducibility  for the interacting higher spin fields in
question. Without it, the
number of physical degrees of freedom, which is determined by one of independent
initial data for the equations of motion (partial differential equations) due to  (\ref{cubgtr1c})
for the interacting model is differed than one  from that for the undeformed model
with vanishing traceless constraints evaluated on respecting equations of motion.

To illustrate the generic cubic vertices solutions we have also
elaborated a number of examples for the interacting Lagrangian for
the unconstrained fields with special value of spins. The  basic
result have achieved with  cubic interaction for triple of fields
with $(0,\lambda_1)$,  $(0,\lambda_2)$, $(m,s)$  in
Section~\ref{BRSTsolgen2mmm3} on a basis of
appendix~\ref{Singhcvomp} for different  Lagrangian formulations of
free  massive higher-spin field from BRST representation  and
Appendix~\ref{Singhcvomp1} for respective interacting  component and
tensor representations.  The resulting interacting model is given in
ghost-independent (\ref{S[n]1ind9})-- (\ref{cubgtrex2gi}) and tensor
(\ref{S[n]1ind1f}) representations  with  deformed  gauge
transformations for the massless scalars (\ref{cubgtrex3gicomp11}), (\ref{cubgtrex3gicomp11a})
(\ref{cubgtrex2gicomp21a})  and untouched for massive higher-spin
field. An application of  the gauge-fixing procedure admissible from
the free formulations permits to present the  interacting Lagrangian
both in triplet tensor  form (\ref{S[n]1ind1fred}), (\ref{Scon00})
with off-shell traceless constraints (\ref{traceres}) with
interacting action depending on 2 set of fields  with irreducible
deformed gauge transformations.(\ref{cubgtrex3gicomp1111}),
(\ref{cubgtrex2gicomp21redsin}),  then in the tensor form
(\ref{singledS1}) with only massless scalars $\phi^{(i)}$, $i=1,2$,
basic massive field $\phi^{(3)\nu({s})}$ and set of auxiliary fields
$\phi_{ 0,2k}^{(3)\nu({s-2k})}$  and in ungauge form for only
quartet of unconstrained  fields $\phi^{(i)}(x)$,
$\phi^{(3)\nu({s})}$, $ \phi_{0,1}^{(3)\nu({s-3})}$,
(\ref{SinghHun}), (\ref{SinghHunint}) (\ref{SinghH}),  or in terms
of double-traceless initial and  traceless auxiliary  tensor fields.
{This result appears by new one and explicitly   demonstrated by
the interacting action (\ref{SinghHunint14}) for massive spin $s=4$
field.} The example on the stage of triple and singlet fields
formulations  admits  massless limit, so that for the  triple of
fields $(0,0)$, $(0,0)$, $(0,s)$ we get non-trivial cubic  vertex
with deformed gauge transformations for the scalars according to
\cite{zinoviev}. The ghost-independent forms for the interacting
Lagrangian formulations have developed also for the set of 4
triples: [$(0,1)$, $(0,\lambda_2)$  $(m,s)$] for $\lambda_2=0,1$;
[$(0,\lambda_1)$, $(0,0)$, $(m,1)$] and for massless scalar with
massive scalar and massive field of spin $s_3$ with coinciding
masses [$(0,0)$, $(m, 0)$, $(m,s_3)$]. The interacting first-stage
reducible Lagrangian for the fields with   [$(0,1)$, $(0,0)$,
$(m,s)$] are given by (\ref{singledS10s}), (\ref{S[n]1ind103})
whereas the deformed part of the gauge transformations in
(\ref{+cubgtrex3gicomp10s})--(\ref{cubgtrex2gicomp}) for massless
vector and scalar and for massive tensor component
(\ref{cubgtrex3gicomp10s}), (\ref{cubgtrex3gicomp10sa})--(\ref{6cubgtrex3gicomp10sa}).  The cubic
vertex for two massless vectors and massive $(m,s)$ tensor was
presented by the relations (\ref{genvertex3})--(\ref{genvertex321}).
For the case fields with [$(0,0)$, $(m, 0)$, $(m,s_3)$] the cubic
vertex,  interacting Lagrangian and deformed reducible gauge
transformations for only the scalars are given by
(\ref{genvertexm2mnext}),  (\ref{S[n]1indmmmg}), (\ref{S[n]1indmmm})
and (\ref{cubgtrex3gim}), (\ref{cubgtrex2gim}) respectively. Note,
the interaction of massless field of helicity $\lambda_1$ with
massless and massive scalars is trivial (\ref{genvertex4}) which
however vanishes after passing to the Fronsdal (single-field)
formulation. {We stress, that there are no any terms in any obtained
interacting vertices and therefore in the interacting part of the
action with divergences by construction. It means that on mass-shell
after gauge-fixing  determined by the Lorentz-like or (in general,
$R_\xi$-type, see e.g.   \cite{2010.15741})  gauge  to derive the
non-degenerate quantum action for the interacting  model in question
the vertices do not vanish.}

There are many possibilities to apply and  to develop
the suggested method. Among them we can highlight a finding cubic vertices, first,  for
irreducible massless and for massive half-integer higher spin fields on flat
backgrounds, second,  for mixed-symmetric higher
spin fields, third, for higher spin
supersymmetric fields, where in all cases the vertices should include any powers
of traces. The construction in question may be generalized
to determine cubic vertices for irreducible higher spin fields on
anti-de-Sitter spaces, having in mind the  bypassing of a flat limit absence
for many of  the cubic vertices  in the formulation \cite{FradkinVasiliev},
\cite{FradkinVasiliev1}, because of  one-to-one correspondence of cubic vertices in flat and anti-de-Sitter spaces in the
Fronsdal formulation demonstrated for specific cases in \cite{BoulangerLeclercqSundell} and more
generally in \cite{FranciaMM}, \cite{JoungTaronna}.
On this way we may to use explicitly the ambient formalism of embedding $d$-dimensional anti-de-Sitter space in $(d+1)$-dimensional Minkowski space \cite{0607248_AdS_amb} (see, as well \cite{BBGG} and references therein) to uplift obtained covariant cubic vertices in anti-de-Sitter space.

In this connection, it is appropriate to point out some
features of the BRST construction for higher spins in the space
(A)dS in comparison with the Minkowski space. Here we should stress,
that the description of irreducible representations for the (A)dS group with both integer and half-integer
spins in (A)dS space is completely different as compared with ones
for the Poincare group in flat space-time even for free theories. In
all known cases the Lagrangian constructions  for the same higher
spin field obtained within the constrained (incomplete) BRST
approach with additional non-differential constraints and within
approach with complete BRST operator do not coincide. The Lagrangian
formulations for both integer and half-integer spins in AdS  spaces
in the BRST approach with a complete BRST operator, have been
successfully formulated for massless and massive particles of
integer spins in \cite{BPT}, \cite{BKL} {(recently for mixed-symmetric case  \cite{RYT2})} and for massive particles of
half-integer spins in \cite{BKR}. Problems related with approach using incomplete BRST operator
 have not been {discussed in details} even for a free field of a given
higher spin.

One should also note the problems of
constructing the fourth and higher vertices and related various
problems of locality (see the discussion initiated in  \cite{ProkushkinVasiliev}, then in \cite{T1}, \cite{DT},
\cite{T2} with recent analysis  \cite{RoibanTseytlin} and also \cite{DGKV}, \cite{Vasil}, \cite{Didenko},
\cite{Didenko1}), where the BRST approach can possibly be useful.
The construction and  {quantum} loop calculations with the
BRST quantum action for the models with derived cubic vertices can
be realized within BRST approach following to \cite{2010.15741}.
 We plan to address all of the mentioned problems in {the}
forthcoming works.

\vspace{-1ex}

\paragraph{Acknowledgements}
The authors are grateful to K.B. Alkalaev, S.A. Fedoruk, V.A.Krykh\-tin, R.R.
Metsaev, D.S. Ponomarev,  K.V. Stepanyantz, M.~Tsulaia, M.A. Vasiliev, Yu.M. Zinoviev for useful discussions and
comments.  The work of I.L.B was partially supported
by Russian Science Foundation, project No 21-12-0012. Work of A.A.R was partially supported by the Ministry of Education of Russian Federation, project No QZOY-2023-0003.

\appendix
\section*{Appendix}

\section{Reduction to Singh-Hagen Lagrangian}\label{Singhcvomp}
\renewcommand{\theequation}{\Alph{section}.\arabic{equation}}
\setcounter{equation}{0}

In this appendix, we deduce the Lagrangian formulation  for free massive HS field $(m,s)$ in terms of
only initial field $\phi_{\mu(s)}$.

From the action $\mathcal{S}^m_{0|s}[|\chi\rangle_s]$ (\ref{PhysStatetot})  and reducible gauge transformations (\ref{gauge trasnform})
we have, first, in the  ghost-independent form
\begin{eqnarray}
&&  \mathcal{S}^{m}_{0|s}  \ = \
 \mathcal{S}^m_{C|s}[\chi_c]  - \bigg\{\Big({}_{s}\langle \Phi \big| K \check{L}{}^{+}_{11} +{}_{s-2}\langle \Phi_2  \big|K \label{Sungh0m}\\ && \phantom{\mathcal{S}_{s}  =} -{}_{s-3}\langle \Phi_{31}  \big| K\check{l}{}_1-{}_{s-4}\langle \Phi_{32}  \big|  K\check{L}{}_{11} \Big) \big|\Phi_{11}\rangle_{s-2}+ \Big(-{}_{s-2}\langle \Phi _2 \big| K \check{L}{}^{(3)+}_{11} \nonumber \\
       &&  \phantom{\mathcal{S}_{s}  =}   -{}_{s-6}\langle \Phi_{22}  \big|K\check{L}{}_{11} +{}_{s-3}\langle \Phi_{31}  \big|  K    \check{l}{}^{+}_1-{}_{s-4}\langle \Phi_{32}  \big|  K \Big) \big|\Phi_{12}\rangle_{s-4} \nonumber\\
       &&  \phantom{\mathcal{S}_{s}   =}+ \Big({}_{s-4}\langle \Phi_{32} \big| K \check{l}{}^{+}_{1}   + {}_{s-6}\langle \Phi_{22}  \big|K \check{l}{}_{1} - {}_{s-3}\langle \Phi_{21}  \big| K \check{L}{}^{+}_{11}\nonumber\\
             &&  \phantom{\mathcal{S}_{s}  =} +\frac{1}{2}{}_{s-5}\langle \Phi_{13}  \big|  K \Big) \big|\Phi_{13}\rangle_{s-5} + {}_{s-3}\langle \Phi_{21}  \big| K\Big(l_0 \big|\Phi_{31}\rangle_{s-3}  -  \check{L}{}_{11} \big|\Phi_{1}\rangle_{s-1} \Big)  \nonumber
         \end{eqnarray}
            \begin{eqnarray}
             &&  \phantom{\mathcal{S}_{s}  =}  -  \frac{1}{2} \Big({}_{s-6}\langle \Phi_{22}  \big| K l_0 \big|\Phi_{22}\rangle_{s-6} - {}_{s-4}\langle \Phi_{32}  \big|K l_0 \big|\Phi_{32}\rangle_{s-4}  \Big) +h.c.\bigg\} ,\nonumber
       \end{eqnarray}
      where  the functional $\mathcal{S}^m_{C|s}$   is  the action  for triplet formulation for massive  fields of spins  $s,s-2,...., 1(0)$ (following to \cite{franciasag}) or for the HS field of spin  $s$ with additional off-shell traceless constraint \cite{Klishevich}, but with  $ b^{+}, d^{+}$-dependence  in the triplet $ |\chi_c\rangle_s$ =  $|\chi\rangle_s\big|_{(\eta_{11}^{+}, \mathcal{P}_{11}^{+})=0}$:     
    \begin{eqnarray}
  && \mathcal{S}^m_{C|s}[\chi_c]  = \int d\eta_0 {}_s\langle\chi_c| KQ_c|\chi_c\rangle_s\label{Scon0}\\
   && \phantom{\mathcal{S}_{C|s}} = \left({}_{s}\langle \Phi  \big|    {}_{s-2}\langle \Phi_2\big| {}_{s-1}\langle \Phi_1\big|   \right) K\left(\begin{array}{ccc}
  l_0 &   0 & -\check{l}{}^{+}_1  \\
 0 & -l^{}_0 &  \check{l}{}^{}_1    \\
        -\check{l}{}^{}_1 & \check{l}{}^{+}_1&  1  \end{array}\right)
       \left(  \begin{array}{l}\big|\Phi\rangle_s\\ \big|{\Phi_2}\rangle_{s-2} \\ \big|{\Phi_1}\rangle_{s-1}   \end{array} \right) ,
                \nonumber
\end{eqnarray}
for $Q_c = Q\big|_{(\eta^{+}_{11}, \mathcal{P}^{+}_{11})=0}$.
The initial gauge transformations for the field $\big|\chi \rangle_{s}$,  and gauge parameter $\big| \Lambda \rangle_{s}$
\begin{eqnarray}
              &&\hspace{-0.7em} \delta_0 \big|\Phi\rangle_{s}\ =\  \check{l}{}^{+}_1|\Xi^{}\rangle_{s-1} + \check{L}{}^{+}_{11}  |\Xi_1\rangle_{s-2}, \label{0gtrind}\\
               &&\hspace{-0.7em}  \delta_0 \big|\Phi_1\rangle_{s-1}\ = \  l_0|\Xi\rangle_{s-1} + \check{L}{}^{+}_{11} |\Xi_{01}\rangle_{s-3},   \label{0gtrind111}\\
              &&  \hspace{-0.7em} \delta_0 \big|\Phi_2\rangle_{s-2}\  =\  \check{l}{}_1 |\Xi\rangle_{s-1}+ \check{L}{}^{+}_{11}  |\Xi_{11}\rangle_{s-4}  - |\Xi_{1}\rangle_{s-2}, \label{0gtrind112}\\ && \delta_0\big|\Phi_{21}\rangle_{s-3} \  =  \  \check{l}{}_1|\Xi_1\rangle_{s-2}-\check{l}{}_1^{+}|\Xi_{11}\rangle_{s-4}-|\Xi_{01}\rangle_{s-3},  \label{01gtrind}\\
              &&\hspace{-0.7em} \delta_0 \big|\Phi_{22}\rangle_{s-6} \ = \ - \check{L}{}_{11}|\Xi_{11}\rangle_{s-4} + \check{l}{}_1|\Xi_{12}\rangle_{s-5}   ,  \label{0gtrind122}\\
              && \hspace{-0.7em}\delta_0 \big|\Phi_{31}\rangle_{s-3} \ = \  \check{L}{}_{11}|\Xi\rangle_{s-1} + \check{L}{}^{+}_{11}|\Xi_{12}\rangle_{s-5} ,   \label{02gtrind}\\
              &&    \hspace{-0.7em}  \delta_0 \big|\Phi_{32}\rangle_{s-4}\  =\  \check{L}{}_{11} |\Xi_1\rangle_{s-2}  - \check{l}{}^{+}_1  |\Xi_{12}\rangle_{s-5} +  |\Xi_{11}\rangle_{s-4}  ,   \label{0gtrind222}\\ &&   \delta_0 \big|\Phi_{11}\rangle_{s-2} \ = \  l_0|\Xi_1\rangle_{s-2} -  \check{l}{}^{+}_{1}|\Xi_{01}\rangle_{s-3} , \label{0gtrind1},\\
              &&    \hspace{-0.7em} \delta_0 \big|\Phi_{12}\rangle_{s-4}\  =\  l_0|\Xi_{11}\rangle_{s-4}  -  \check{l}{}_1  |\Xi_{01}\rangle_{s-3}  , \label{0gtrind1123}\\
               && \hspace{-0.7em} \delta_0 \big|\Phi_{13}\rangle_{s-5} \ = \  l_0|\Xi_{12}\rangle_{s-5} -  \check{L}{}_{11}|\Xi_{01}\rangle_{s-3} , \label{0gtrind11},\\
              &\hspace{-0.7em}&\hspace{-0.9em} \delta_0 \left(|\Xi\rangle, |\Xi_{1}\rangle,|\Xi_{11}\rangle, |\Xi_{12}\rangle, |\Xi_{01}\rangle\right)  =  \left( - l^{+}_{11}+1/2(d^{+})^2 -b^+  ,\,\check{l}{}^{+}_1 ,\,  \check{l}{}_1,\,  \check{L}{}_{11} ,  l_0\right) \big|\Xi^{1}\rangle_{s-3} \label{1gtrind}
                      .  \end{eqnarray}
Second, we gauge away $b^+$-dependence from zero-level gauge parameter $|\Xi\rangle$ by means of all degrees of freedom of the first-level gauge parameter $|\Xi^{1}\rangle$ due to structure of $\check{L}{}^{+}_{11}$ trace operator, so that the theory becomes by the irreducible gauge theory. Third, analogously  we gauge away   $b^+$-dependence from the fields $\big|\Phi\rangle$, $\big|\Phi_1\rangle$, $\big|\Phi_2\rangle$, $\big|\Phi_{31}\rangle$ with use of all degrees of freedom
of  the gauge parameters $|\Xi_1\rangle$, $|\Xi_{01}\rangle$, $|\Xi_{11}\rangle$, $|\Xi_{12}\rangle$.   The residual non-vanishing  gauge transformations take the form
\begin{eqnarray}
              && \delta_0\big( \big|\Phi\rangle_{s},  \big|\Phi_1\rangle_{s-1},  \big|\Phi_2\rangle_{s-2},  \big|\Phi_{31}\rangle_{s-3}\big) |_{b^+=0}\ =\
              \big(\check{l}{}^{+}_1,  l_0,  \check{l}{}_1,   \check{L}{}_{11}\big) |\Xi^{}\rangle_{s-1} |_{b^+=0}. \label{0gtrindred}
               .  \end{eqnarray}
               Fourth, from the equations of motion, $Q|\chi\rangle_s=0$,  for the rest 6 fields $\big|\Phi_{21}\rangle$, $\big|\Phi_{22}\rangle$, $\big|\Phi_{32}\rangle$, $\big|\Phi_{11}\rangle$, $\big|\Phi_{12}\rangle$, $\big|\Phi_{13}\rangle$ encountered in $|\chi\rangle_s$ (\ref{spinctotsym}) with $\mathcal{P}_{11}^+$  ghost operator, we get that they income in the equations  with  $\check{L}{}^{+}_{11}$ operator and, therefore vanish.
            Fifth, we gauge away the field  $\big|\Phi_{31}\rangle$ by means of residual gauge transformations, so that the condition of its non-appearance leads to the constraint $\check{L}_{11}|_{b^{(+)}=0} |\Xi^{}\rangle|_{b^+=0}=0$.
             Sixth, from the residual equations of motions for the triplet $ \big|\Phi\rangle_{s},  \big|\Phi_1\rangle_{s-1},  \big|\Phi_2\rangle_{s-2}$ at $\eta_{11}^+$ we get the traceless constraints
            \begin{equation}\label{traceres}
              \check{l}_{11}\big|\Phi\rangle_s+  \big|\Phi_2\rangle_{s-2}=0,\quad  \check{l}_{11}|_{b^{(+)}=0}\big|\Phi_k\rangle_{s-k}=0,\ k=1,2., \check{l}_{11}\equiv \check{L}_{11}|_{b^{(+)}=0}
            \end{equation}
            As the result, we get  the triplet Lagrangian  formulation for the massive of spin $s$ field with auxiliary $3s-4$ fields subject to  the constraints (\ref{traceres})
            \begin{eqnarray}
  && \mathcal{S}^m_{C|s}[\chi_c]  = \int d\eta_0 {}_s\langle\chi_c| Q_c|\chi_c\rangle_s\label{Scon00}\\
   && \phantom{\mathcal{S}_{C|s}} = \left({}_{s}\langle \Phi  \big|    {}_{s-2}\langle \Phi_2\big| {}_{s-1}\langle \Phi_1\big|   \right) \left(\begin{array}{ccc}
  l_0 &   0 & -{\check{l}}{}^{+}_1  \\
 0 & -l^{}_0 &  {\check{l}}{}^{}_1    \\
        -{\check{l}}{}^{}_1 & {\check{l}}{}^{+}_1&  1  \end{array}\right)
       \left(  \begin{array}{l}\big|\Phi\rangle_s\\ \big|{\Phi_2}\rangle_{s-2} \\ \big|{\Phi_1}\rangle_{s-1}   \end{array} \right) ,
                \nonumber\\
                  &&
    \phantom{\mathcal{S}_{C|s}} \delta \left( \big|\Phi\rangle_{s} ,      \big|\Phi_1\rangle_{s-1},    \big|\Phi_2\rangle_{s-2}\right) = \left( \check{l}_1^+ ,l_0,\check{l}_1\right) |\Xi\rangle_{s-1} , \qquad \check{l}_{11} |\Xi^{}\rangle=0. \label{gtrd}
\end{eqnarray}
After expressing the field $ \big|\Phi_2\rangle$ for the  first constraint in (\ref{traceres}): $ \big|\Phi_2\rangle=-\check{L}_{11}\big|\Phi\rangle$  and expressing
  the field $ \big|\Phi_1\rangle$ from the algebraic equation of motion: $ \big|\Phi_1\rangle=\check{l}_1 \big|\Phi\rangle-\check{l}_1^+\big|\Phi_2\rangle$
 it follows the Lagrangian in the single vector form with $s-1$ auxiliary fields
\begin{eqnarray}
 &&  \label{SclsrsingleF}  \mathcal{S}^m_{C|s}\left({\phi},... \right)  =  {}_{s}\langle \Phi  \big|    \left(  l_0-\check{l}_1^+\check{l}_1  -(\check{l}_1^+)^2\check{l}_{11}
 -\check{l}_{11}^+\check{l}_1^2  -\check{l}_{11}^+(l_0 +  \check{l}_1\check{l}_1^+) \check{l}_{11}    \right)
       \big|{\Phi}\rangle_s, \\
 &&     \delta  \big|\Phi\rangle_{s}  =  \check{l}_1^+  |\Xi\rangle_{s-1}\ \   \mathrm{and}\ \
            \check{l}_{11}\big(\check{l}_{11}|{\Phi}\rangle,\,  |\Xi\rangle\big) = (0,0 ) ,\label{gaugetrsing}
\end{eqnarray}
The Lagrangian formulation (\ref{SclsrsingleF}), (\ref{gaugetrsing}) has smooth massless limit for $m=d^{(+)}=0$ resulting to  Fronsdal formulation \cite{Fronsdal} in the form of single field $\big|\phi\rangle_{s}=\big|\Phi\rangle_{s}|_{d^+=0}$   with $(0,s)$.

Now, as it was shown in \cite{BKrykhtin150703723} the Lagrangian formulation after resolution of the traceless constraints in terms of real constraints with decomposing of   $|\Phi\rangle_{s}|$ in powers of  $d^+$-independent vectors as well as the gauge parameter $ |\Xi\rangle_{s-1}$ we get that only four fields $\phi_{k|(\mu){s-k}}$, $k=0,1,2,3$   (with physical one at $k=0$) are independent from each other and two gauge parameters  $\xi_{l|(\mu){s-l-1}}$, $l=0,1$:
\begin{multline}\label{decompmas}
   \big|\xi\rangle_{s-2l-1}=(2l_{11})^l\big|\xi_0\rangle_{s-1}, \qquad  \big|\xi\rangle_{s-2l-2}=(2l_{11})^l\big|\xi_1\rangle_{s-2}, \ l= 1,...,[s/2]-1, \\ 
   \big|\phi\rangle_{s-2k}=k(2l_{11})^{k-1}\big|\phi_2\rangle_{s-2}-(k-1)(2l_{11})^{k}\big|\phi_0\rangle_{s},  \ k= 1,...,[s/2], \\ 
   \big|\phi\rangle_{s-2k-1}=k(2l_{11})^{k-1}\big|\phi_3\rangle_{s-3}-(k-1)(2l_{11})^{k}\big|\phi_1\rangle_{s-1},
                                                                                                                     \end{multline}
Substituting in (\ref{SclsrsingleF}) the fields $\big|\phi\rangle_{s-r}$, $r\geq 4$ in terms of unrestricted quartet of  fields $\big|\phi_k\rangle_{s-k}$, $k=0,1,2,3$ from (\ref{decompmas}) we get
 the action $\mathcal{S}^m_{s}\left({\phi_0},\phi_1,\phi_2,\phi_3 \right) = \mathcal{S}^m_{C|s}\left({\phi},... \right)|_{|\phi\rangle_{s-r}=|\phi(\phi_k)\rangle_{s-r}}$, to be invariant with respect to the gauge transformations
 with two unrestricted gauge parameters
 \begin{eqnarray}
   \delta \big|\phi_0\rangle_{s} &=& l_1^+ \big|\xi_0\rangle_{s-1},\qquad \qquad\qquad \hspace{-1em}\delta \big|\phi_2\rangle_{s-2} \ = \  l_1^+(2l_{11}) \big|\xi_0\rangle_{s-1}+m\big|\xi_1\rangle_{s-2},  \label{gtr0}\\
  \delta \big|\phi_1\rangle_{s-1} & =& l_1^+ \big|\xi_1\rangle_{s-2}+m \big|\xi_0\rangle_{s-1},\quad \hspace{-1em}\delta \big|\phi_3\rangle_{s-3} \ = \ l_1^+(2l_{11}) \big|\xi_1\rangle_{s-2}+m(2l_{11}) \big|\xi_0\rangle_{s-1}. \label{gtr1}
 \end{eqnarray}
 Instead,  the gauge-fixing procedure for the tower of the gauge transformations for all fields and gauge parameters (starting from the first column in (\ref{gtr0}), (\ref{gtr1}))
 \begin{eqnarray}
       \delta \big|\phi_2\rangle_{s-2} &=& l_1^+ \big|\xi_2\rangle_{s-3}+m \big|\xi_1\rangle_{s-2},\label{gtr0m} \\
        \delta \big|\phi_3\rangle_{s-3} & = & l_1^+ \big|\xi_3\rangle_{s-4}+m \big|\xi_2\rangle_{s-3}, \qquad  \big|\xi_{2}\rangle_{s-3} = 2l_{11}\big|\xi_{0}\rangle_{s-1},\label{gtr1m}\\
   \delta \big|\phi_4\rangle_{s-4} & = & l_1^+ \big|\xi_4\rangle_{s-5}+m \big|\xi_3\rangle_{s-4}, \qquad  \big|\xi_{3}\rangle_{s-4} = 2l_{11}\big|\xi_{1}\rangle_{s-2} \label{gtr2m}\\
   \ldots & & \ldots\ldots\ldots\ldots\ldots\ldots \nonumber \\
  \delta \big|\phi_{s-1}\rangle_{1} & = & l_1^+ \big|\xi_{s-1}\rangle_{0}+m \big|\xi_{s-2}\rangle_{1},  \qquad  \big|\xi_{s-2}\rangle_{1} = 2l_{11}\big|\xi_{s-4}\rangle_{3}, \label{gtrs-1m}\\
   \delta \big|\phi_{s}\rangle_{0} & = & m \big|\xi_{s-1}\rangle_{0}, \qquad \qquad  \big|\xi_{s-1}\rangle_{0} = 2l_{11}\big|\xi_{s-3}\rangle_{2} \label{gtrsm}
 \end{eqnarray}
 implies    from (\ref{gtrsm}), (\ref{gtrs-1m}) the removing of the fields $\big|\phi_{s}\rangle_{0}$, $\big|\phi_{s-1}\rangle_{1}$  by means of use of $\big|\xi_{s-1}\rangle_{0}$, $\big|\xi_{s-2}\rangle_{1}$ so that the next gauge parameters $\big|\xi_{s-m}\rangle_{m-1}$, $m=3,4$ become by traceless. Then, moving up to (\ref{gtr1}) we successively   removing of all traceless parts from the  auxiliary  fields  $\big|\phi_k\rangle_{s-k}$, $k\geq 1$ by means of respective use of traceless $\big|\xi_{k-1}\rangle_{s-k}$ for $k=s-2,..., 1$. As the result,  the system of the linear equations for the rest ungauge fields  in (\ref{decompmas}) after changing the fields $\big|{\phi}\rangle_{s-2k}$, $\big|{\phi}\rangle_{s-2k-1}$ on its trace parts except for $\big|\phi_0\rangle$ takes the form
 \begin{eqnarray}
 \hspace{-1em} &\hspace{-1em}&\hspace{-1em} (l_{11})^{[s/2]-i\theta_s-1}\big|{\phi}_{i+2}\rangle_{s-i-2}= 2\frac{([s/2]-i\theta_s -1)}{[s/2]-i\theta_s}(l_{11})^{[s/2]-i\theta_s}\big|{\phi}_i\rangle_{s-i}, \quad i=0, 1,  \label{phi2phi0}   \\
  \hspace{-1em} &\hspace{-1em}&\hspace{-1em}  l_{11}^+ \big|\widetilde{\phi}\rangle_{s-2k-2-i}=k(2l_{11})^{k-1}\big|\phi_{2+i}\rangle_{s-2-i}-\hspace{-0.15em}(k-1)(2l_{11})^{k}\big|\phi_i\rangle_{s-i},  \ k=\hspace{-0.15em} 1,...,\textstyle[\frac{s}{2}]-\hspace{-0.15em}1-\hspace{-0.15em}i\theta_s  \label{phi2phi01}
 \end{eqnarray}
  (for $\theta_s\equiv \delta_{s-2[s/2],0}$). In (\ref{phi2phi0}), (\ref{phi2phi01}) and below we using the decomposition  of the fields  $\big|\phi_j\rangle$   into sum of traceless fields $\big|{\phi}^{l}_j\rangle$
  \begin{equation}\label{decomptrace}
    \big|\phi_j\rangle_{s-j} = \big|{\phi}^0_j\rangle_{s-j} + \sum_{l=1}^{[s-j/2]}(2l_{11}^+)^l \big|{\phi}^{l}_j\rangle_{s-2l-j}\equiv \big|{\phi}^0_j\rangle_{s-j} + l_{11}^+\big|\widetilde{\phi}_{j}\rangle_{s-j-2}
  \end{equation}
 (for $\big|{\phi}^0_j\rangle_{s-j} = 0$, $j\geq 1$).  One can show that all fields $\big|{\phi}^{l}_j\rangle_{s-2l-j}$ for $j\geq 2$, therefore total fields $\big|\widetilde{\phi}_{j}\rangle_{s-j-2}$, are expressed in terms of $\big|\phi_0\rangle_{s}$, $\big|\widetilde{\phi}_1\rangle_{s-3}$ and its traces from the  system (\ref{phi2phi0}), (\ref{phi2phi01} for $k< [\frac{s-1-i}{2}]\equiv k_{\max}$
  \begin{equation}\label{soldecomptrace}
    \big|\phi_{2k+i}\rangle_{s-2k-i} =  \sum_{l=1}^{k_{\max}-k}\alpha^i_{k|l}(2l_{11}^+)^l(2l_{11})^{k+l}  \big|{\phi}_i\rangle_{s-i}, \ i=0,1, \ k\geq 1,
  \end{equation}
 with definite rationals $\alpha^0_{k|l}, \alpha^1_{k|l}$.
  Thus, the only $s$ traceless   fields: $[s/2]+1$ ones in  $\big|{\phi}_0\rangle_{s}$  and $[(s-1)/2]$  in $\big|\widetilde{\phi}_1\rangle_{s-3}$;
 compose   the residual field  vector,
\begin{equation}\label{resfv}
   \big|\widetilde{\Phi}\rangle_{s} =  \big|\phi_0\rangle_{s} +  d^+l^+_{11} \big|{\widetilde{\phi}}_1\rangle_{s-3} +
   \sum_{i=0}^1\sum_{k=1}^{k_{\max}}\sum_{l=1}^{k_{\max}-k}\frac{(d^+)^{2k+i}}{(2k+i)!}
   \alpha^i_{k|l}(2l_{11}^+)^l(2l_{11})^{k+l}  \big|{\phi}_i\rangle_{s-i}.
\end{equation}
 The respective action will be equivalent to Singh-Hagen action  \cite{SinghHagen}
 \begin{eqnarray}
   &&  \mathcal{S}^m_{C|s}\left({\phi}, {\phi}_1\right)  =  {}_{s}\langle \widetilde{\Phi}  \big|    \left(  l_0-\check{l}_1^+\check{l}_1  -(\check{l}_1^+)^2\check{l}_{11}
 -\check{l}_{11}^+\check{l}_1^2  -\check{l}_{11}^+(l_0 +  \check{l}_1\check{l}_1^+) \check{l}_{11}    \right)
       \big|{\widetilde{\Phi}}\rangle_s.  \label{SinghH}\end{eqnarray}.
We want to emphasize, there is a lot of equivalent Lagrangians  with the same or different number of tensor fields in the configuration spaces. In this connection, the Singh-Hagen action can be considered as only the one of possible representative from the set of equivalent Lagrangians. The main requirement for any such a Lagrangian formulation for the given field with mass and spin consists in the fact that its dynamics must be in one-to-one correspondence with one determined by the initial set of relations (\ref{irrepint}) on the physical field selecting the element of irreducible Poincare group representation. Thus, we present in the appendix many equivalent Lagrangian formulations in various configuration spaces  for free massive particle of spin $s$. No one of them is not more fundamental than all other.
At present, there exists a general BRST formulation allowing to derive any of them (see e.g. the quartet formulation with greater number of fields then in Singh-Hagen formulation).

 \section{Component interacting Lagrangian and gauge transformations for $(m,s), (0,0), (0,0)$}\label{Singhcvomp1}
\renewcommand{\theequation}{\Alph{section}.\arabic{equation}}
\setcounter{equation}{0}

    In oscillator  $a^{(i)(+)}_{\mu}, b^{(i)(+)}, d^{(i)(+)}$-dependent  form with use of the representations (\ref{Phiphi}), (\ref{Xiphi})  and its duals  for the fields and gauge parameters  the interacting part of the action (\ref{S[n]1ind9}) looks
  \begin{eqnarray}\label{S[n]1ind1}
  &&\hspace{-0.5em}  S^{(m)_3}_{1|(s)_3}[\phi^{(1)},\phi^{(2)}, \chi^{(3)}] \ = \    - g \int \prod_{i=1}^3 d^dx_i\prod_{i=2}^3 \delta^{(d)}\big(x_{1} -  x_{i}\big) \Big\{ \phi^{(2)}(x_2)
   \phi^{(1)}(x_1) \langle0|
   \end{eqnarray}
      \vspace{-1.0ex}
    \begin{eqnarray} 
   &&\hspace{-0.5em} \quad \times
   \Big(\sum_{k'=0}^{s}\frac{(d^{(3)})^{k'}}{k'!}\sum_{l'=0}^{[(s-k')/2]}\frac{(b^{(3)})^{l'}}{l'!}  C(l',h(s))
   \frac{(-\imath)^{s-k'-2l'}}{({s-k'-2l'})!}{\phi}{}^{(3)\mu({s-k'-2l'})}_{l',k'}(x_3)\prod_{i'=1}^{{s-k'-2l'}} a^{(3)}_{\mu_:{i'}}
  \nonumber  \\
   && \hspace{-0.5em} \quad  \times  \sum_{j \geq0}^{[s/2]}t_j \sum_{k \geq0}^{j}C^{k,l}_j(d{}^{(3)+})^{2k}\left(-\frac{1}{2}\right)^{k} \sum_{l \geq0}^{j-k}(b^{(3)+})^l (l{}^{(3)+}_{11})^{j-k-l} \nonumber \\
   && \hspace{-0.5em} \quad  \times \sum_{i=0}^{[s/2-j]} (-1)^{i}\{\widehat{p}{}^{(3)\nu}a^{(3)+}_{\nu}\}^{s-2j-2i} (\hat{p}^{(3)})^{2i} \frac{(s-2j)!}{i! 2^i(s-2j-2i)!} \frac{(b^{(3)+})^i}{C(i,h(s))}|0\rangle  \nonumber
\\
    && \hspace{-0.5em} \quad  +
   \sum_{k'=0}^{s-2}\frac{(d^{(3)})^{k'}}{k'!}\sum_{l'=0}^{[(s-k')/2-1]}\frac{(b^{(3)})^{l'}}{l'!} C(l',h(s))
   \frac{(-\imath)^{s-2-k'-2l'}}{({s-2-k'-2l'})!}{\phi}{}^{(3)\mu({s-k'-2l'})}_{2|l',k'}(x_3) \hspace{-0.3em}\prod_{i'=1}^{{s-2-k'-2l'}} \hspace{-0.5em} a^{(3)}_{\mu_:{i'}}
  \nonumber  \\
   && \hspace{-0.5em} \quad  \times
        \sum_{j \geq0}^{[s/2-1]}t_j (j+1)    \sum_{k \geq0}^{j} \sum_{l \geq0}^{j-k}C^{k,l}_j\times  \nonumber \\
   && \hspace{-0.5em} \quad  \times (l{}^{(3)+}_{11})^k (b^{(3)+})^l(d{}^{(3)+})^{2(j-k-l)}\left(-\frac{1}{2}\right)^{j-k-l}\sum_{i=0}^{[s/2-j-1]} (-1)^{i}({L}^{(3)})^{s-2(j+1)-2i}\times \nonumber \\
   && \hspace{-0.5em} \quad  \times (\hat{p}^{(3)})^{2i} \frac{(s-2(j+1))!}{i! 2^i(s-2(j+1)-2i)!} \frac{(b^{(3)+})^i}{C(i,h(s))}   \nonumber
   \\
   && \hspace{-0.5em} \quad  + {}_{s-4}\langle \Phi^{(3)}_{32} K^{(3)}
    \big|\sum_{j \geq0}^{[s-4/2]}t_j(j+1)(j+2) (\check{L}{}^{(3)+}_{11})^j  \mathcal{L}^{(3)}_{s-2(j+2)}\nonumber \\
    && \hspace{-0.5em} \quad  - {}_{s-6}\langle \Phi^{(3)}_{22} K^{(3)}
    \big| \hspace{-0.3em}\sum_{j \geq0}^{[s/2-3]} \hspace{-0.3em}t_j(j+1)(j+2)(j+3) (\check{L}{}^{(3)+}_{11})^j  \mathcal{L}^{(3)}_{s-2(j+3)}\Big)|0\rangle +  h.c. \bigg\}\nonumber  ,
   \end{eqnarray}
(for $C^{k,l}_j\equiv \frac{j!}{k!l!(j-k-l)!}$),
  and also for the gauge transformations (\ref{cubgtrex3gi}), (\ref{cubgtrex2gi})
 \begin{eqnarray}\label{cubgtrex3gicomp1}
         &\hspace{-0.5em} &\hspace{-0.5em}  \delta_{[1]} \big| \phi^{(1)} \rangle_{0}  =  -
 g  \int d^dx_2 \phi^{(2)}(x_2)\bigg\{\int d^dx_3
      \langle0|
  \sum_{l'=0}^{[s-1-k'/2]}\sum_{l \geq0}^{j-k}\sum_{i=0}^{[(s-1)/2]-j}C(l',h(s)) \frac{(b^{(3)})^{l'}}{l'!} \\
   &\hspace{-0.5em} & \hspace{-0.5em} \quad  \times \frac{(-1)^{i} j!}{k!l!(j-k-l)!}  \frac{(b^{(3)+})^{l+i}}{C(i,h(s))}
   \sum_{k'=0}^{s-1}\sum_{j \geq0}^{[(s-1)/2]}t_j \frac{(d^{(3)})^{k'}}{k'!}(d{}^{(3)+})^{2k}\left(-\frac{1}{2}\right)^{k} \nonumber\\
   &\hspace{-0.5em}& \hspace{-0.5em} \quad  \times \sum_{k \geq0}^{j}
   \frac{(-\imath)^{s-1-k'-2l'}}{({s-1-k'-2l'})!}{\Xi}{}^{(3)\mu({s-1-k'-2l'})}_{l',k'}(x_3)\prod_{i'=1}^{{s-1-k'-2l'}} a^{(3)}_{\mu_{i'}}
   (l{}^{(3)+}_{11})^{j-k-l} \nonumber\\
       &\hspace{-0.5em}& \hspace{-0.5em} \quad  \times \{\widehat{p}{}^{(3)\nu}a^{(3)+}_{\nu}\}^{s-1-2j-2i}(\hat{p}^{(3)})^{2i} \frac{(s-1-2j)!}{i! 2^i(s-1-2j-2i)!}|0\rangle \nonumber
     \\
    &\hspace{-0.5em}& \hspace{-0.5em} \quad  - {}_{s-5}\langle \Xi^{(3)}_{12}
   K^{(3)}
    \big|  \sum_{j \geq0}^{[s-5/2]}t_j (j+1)(j+2)(\check{L}{}^{(3)+}_{11})^j \big(\mathcal{L}^{(3)}_{s-5-2j}\big)^\prime\Big\}|0\rangle \prod_{i=2}^3 \delta^{(d)}\big(x_{1} -  x_{i}\big) ; \nonumber
   \\
 \label{cubgtrex2gicomp2}   && \hspace{-0.5em}\delta_{[1]} \big| \phi^{(2)} \rangle_{0}  = - \delta_{[1]} \big| \phi^{(1)} \rangle_{0}\vert_{\textstyle \big(\phi^{(1)}(x_1)\to \phi^{(2)}(x_2)\big)}.
\end{eqnarray}
In (\ref{S[n]1ind1}) two last summands with vectors ${}_{s-6}\langle \Phi^{(3)}_{32}|$,  ${}_{s-6}\langle \Phi^{(3)}_{22}| $ have the same forms as for two first ones,  with  ${}_{s}\langle \Phi^{(3)}_{}|$,  ${}_{s-2}\langle \Phi^{(3)}_{2}| $,  as well as the  last term  with ${}_{s-5}\langle \Xi^{(3)}_{12}|$ in (\ref{cubgtrex3gicomp1})  is written similar as for the first term with ${}_{s-1}\langle \Xi^{(3)}_{}|$.

To calculate (\ref{S[n]1ind1})--(\ref{cubgtrex2gicomp2}) we use the list of formulas with  oscillator's pairing and trivial reality conditions for complex-conjugated   tensor fields $\overline{\phi}{}^{(i)\mu({s-k-2l})}_{...|l,k}(x)$ and gauge parameters $\overline{\Xi}{}^{(i)\mu({s-1-k-2l})}_{...|l,k}(x)$  for bra-vectors to (\ref{Phiphi}), (\ref{Xiphi})
\begin{eqnarray}\label{A1}
& \hspace{-0.7em}& \hspace{-0.7em}\langle 0| \prod_{i'=1}^{{s'}} a^{(3)}_{\mu_{i'}}\prod_{i=1}^{{s}}\{\widehat{p}{}^{(3)\nu_i}a^{(3)+}_{\nu_i}\}(l{}^{(3)+}_{11})^k|0\rangle = \delta_{s',s+2k} \frac{(-1)^{s'}}{2^k}S^{\nu_{1}...\nu_{s'}}_{\mu_{1}...\mu_{s'}}\prod_{i=1}^{s}\widehat{p}^{(3)}_{\nu_i}\prod_{i=1}^{k}\eta_{\nu_{s+2i-1}\nu_{s+2i}} ,
 \\
& \hspace{-0.5em}& \hspace{-0.5em}  S^{\nu_{1}...\nu_{s}}_{\mu_{1}...\mu_{s}}= \sum_{P(\nu_{1},...,\nu_{s})}\prod_{i=1}^{s}\delta_{\mu_{i}}^{\nu_{i}},\label{A2}
 \\
   & \hspace{-0.5em}& \hspace{-0.5em} \langle 0| \sum_{k'=0}^{s'} (A^{(3)})^{k'}\sum_{k=0}^{s} (A{}^{(3)+})^{k} |0\rangle =s! \delta_{s',s}, \quad A\in \{b,d\}.\label{A3}
\end{eqnarray}
(with symmetrizer $S^{\nu_{1}...\nu_{s}}_{\mu_{1}...\mu_{s}}$). Thus, e.g. for the first integrand $A_1$ in $S^{(m)_3}_{1|(s)_3}$  from the representation
\begin{equation}\label{aux1}
-g \int \prod_{i=1}^3 d^dx_i A_1 \prod_{i=2}^3 \delta^{(d)}\big(x_{1} -  x_{i}\big)
\end{equation}
 we have
\begin{eqnarray}\label{aux2}
  &&  \hspace{-0.5em} A_1 \ =\ \phi^{(2)}(x_2)
   \phi^{(1)}(x_1)
   \langle0|
  \sum_{l'=0}^{[(s-k')/2]}\sum_{l \geq0}^{j-k}\sum_{i=0}^{[s/2-j]}C(l',h(s)) \frac{(b^{(3)})^{l'}}{l'!}  \frac{j!}{k!l!(j-k-l)!} \\
   && \hspace{-0.5em} \quad  \times  \frac{(-1)^{i}(b^{(3)+})^{l+i}}{C(i,h(s))}|0\rangle \langle0|
   \sum_{k'=0}^{s}\sum_{j \geq0}^{[s/2]}t_j \frac{(d^{(3)})^{k'}}{k'!}(d{}^{(3)+})^{2k}\left(-\frac{1}{2}\right)^{k}|0\rangle \nonumber\\
   &&  \hspace{-0.5em} \quad  \times \langle0|\sum_{k \geq0}^{j}
   \frac{(-\imath)^{s-k'-2l'}}{({s-k'-2l'})!}{\phi}{}^{(3)\mu({s-k'-2l'})}_{l',k'} \hspace{-0.2em}(x_3) \hspace{-0.3em}\prod_{i'=1}^{{s-k'-2l'}} \hspace{-0.5em} a^{(3)}_{\mu_{i'}}
   (l{}^{(3)+}_{11})^{j-k-l} \{\widehat{p}{}^{(3)\nu}a^{(3)+}_{\nu}\}^{s-2j-2i}|0\rangle \nonumber  \\
   &&  \hspace{-0.5em} \quad  \times (\hat{p}^{(3)})^{2i} \frac{(s-2j)!}{i! 2^i(s-2j-2i)!} \nonumber  \\
   && \hspace{-0.5em}  \ =\ \phi^{(2)}(x_2)
   \phi^{(1)}(x_1)
     \sum_{l'=0}^{[(s-k')/2]}\sum_{l \geq0}^{j-k}\sum_{i=0}^{[s/2-j]}C(l',h(s)){\delta_{l',l+i}}\frac{(-1)^{i}}{C(i,h(s))}\frac{j!}{k!l!(j-k-l)!}   \nonumber\\
   &&  \hspace{-0.5em}\quad  \times
   \sum_{k'=0}^{s}\sum_{j \geq0}^{[s/2]}t_j\delta_{k',2k} \left(-\frac{1}{2}\right)^{k}  \sum_{k \geq0}^{j}
   \frac{{(\imath)^{s-k'-2l'}}}{2^k}{\phi}{}^{(3)\mu({s-k'-2l'})}_{l',k'}(x_3)  \delta_{s-k'-2l',s-2l-2i-2k}   \nonumber  \\
   && \hspace{-0.5em} \quad  \times S^{\nu_{1}...\nu_{s-2l-2i-2k}}_{\mu_{1}...\mu_{s-k'-2l'}}\prod_{p=1}^{s-2j-2i}\widehat{p}^{(3)}_{\nu_p}\prod_{r=1}^{j-k-l}\eta_{\nu_{s-2j-2i+2r-1}\nu_{s-2j-2i+2r}} (\hat{p}^{(3)})^{2i} \frac{(s-2j)!}{i! 2^i(s-2j-2i)!} \nonumber  ,
     \end{eqnarray}
     and simplifying
       \begin{eqnarray} \label{aux3}
   &&  \hspace{-0.5em}A_1 \ = \ \phi^{(2)}(x_2)
   \phi^{(1)}(x_1)\sum_{j \geq0}^{[s/2]}t_j\sum_{k \geq0}^{j}
     \sum_{l \geq0}^{j-k}\sum_{i=0}^{[s/2-j]}\frac{C(l+i,h(s))}{C(i,h(s))}  \frac{j!(s-2j)!}{k!l!i! (s-2j-2i)!}  (\hat{p}^{(3)})^{2i}   \\
   && \hspace{-0.5em} \quad \times
    \frac{\left(-{1}\right)^{j-k-l+i}}{(j-k-l)!}
   \frac{{(\imath)^{s-2k-2l-2i}}}{2^{2(j-k-l)+i}}\phi^{(3)\nu({s-2l-2i-2k})}_{l+i,2k}(x_3)  \prod_{i=1}^{s-2j-2i}\widehat{p}^{(3)}_{\nu_i}   \prod_{r=1}^{j-k-l}\eta_{\nu_{s-2j-2i+2r-1}\nu_{s-2j-2i+2r}} \nonumber  .
  \end{eqnarray}
  Then, using respective  binomial and polynomial  decompositions
  \begin{eqnarray}\label{p3nut2}
  &&\hspace{-0.5em}  \int \prod_{j=1}^{3} d^dx_j\phi^{(1)}(x_1)\phi^{(2)}(x_2)\phi^{(3)\nu({s})}_{p,t}(x_3)
\prod_{i=1}^{t}\widehat{p}^{(3)}_{\nu_i} \prod_{j=2}^{3}\delta^{(d)}(x_1-x_i)
\\&&\hspace{-0.5em}  \quad
=\int d^dx (-i)^t\sum_{q=0}^t \frac{(-1)^qt!}{q!(t-q)!} \big(\partial_{\nu_0}...p_{\nu_q}  \phi^{(1)}(x)\big)\big(\partial_{\nu_q+1}...p_{\nu_t}\phi^{(2)}(x)\big)\phi^{(3)\nu({s})}_{p,t}(x),
  \nonumber\\
  \nonumber
    && \hspace{-0.5em}   (\hat{p}^{(3)})^{2i}= ({p}^{(1)2}-2{p}^{(1)}_{\mu}{p}^{(2)\mu}+{p}^{(2)2})^{i} = (-1)^i \sum_{q=0}^i \sum_{t=0}^{i-q}C_i^{q,t}
    \big(\partial_1^2\big)^q\big(-2\partial_{1\nu}\partial_{2}^{\nu}\big)^t\big(\partial_2^2\big)^{i-q-t}
  \end{eqnarray}
  (for $\partial_{\nu_0}\equiv 1$, $C_i^{q,t}=\frac{i!}{q!t!(i-t-q)!}$).
 we have for the first term in the action (\ref{S[n]1ind1})
\begin{eqnarray}
  && \hspace{-0.5em} -g \int \prod_{i=1}^3 d^dx_i A_1 \prod_{i=2}^3 \delta^{(d)}\big(x_{1} -  x_{i}\big) = -g \int d^dx \sum_{j \geq0}^{[s/2]}t_j\sum_{i=0}^{[s/2-j]}\sum_{k \geq0}^{j}
     \sum_{l \geq0}^{j-k} \frac{C(l+i,h(s))}{ C(i,h(s))}C_j^{k,l} \label{S[n]1ind1in}\\
   &&  \hspace{-0.5em}\times  \frac{(s-2j)!}{i!} \bigg\{\sum_{u=0}^{s-2j-2i} \frac{(-1)^u}{u!(s-2j-2i-u)!}\sum_{q=0}^i \sum_{t=0}^{i-q}C_i^{q,t} (-2)^t\ \Big[\partial_{\nu_0}...\partial_{\nu_u}\Big(\Box^{q}\partial_{\nu_{u+1}}...\partial_{\nu_{u+t}}\phi^{(1)}\Big)  \Big]  \nonumber\\
   &&  \hspace{-0.5em} \times \Big[\partial_{\nu_{u+t+1}}...\partial_{\nu_{s-2j-2i+t}}\Big(\partial^{\nu_{u+1}}...\partial^{\nu_{u+t}}\Box^{i-q-t}\phi^{(2)}\Big)\Big]
    \bigg\}
     \frac{1}{2^{2(j-k-l)+i}}\phi^{(3)\nu({s-2l-2i-2k})}_{l+i,2k}  \nonumber\\
   && \hspace{-0.5em} \quad  \times\prod_{r=1}^{j-k-l}\eta_{\nu_{s-2j-2i+2r-1}\nu_{s-2j-2i+2r}}    \nonumber.
   \end{eqnarray}
 Calculating by the receipt above, first, the similar rest three terms in the action (\ref{S[n]1ind1}), second the similar  terms in the gauge transformations (\ref{cubgtrex3gicomp1})   we get with accuracy up to overall factor $(-1)^ss!$
 the representations in the tensor form for the action
  (\ref{S[n]1ind1f})
  \begin{eqnarray}\label{S[n]1ind1fa}
  &&  \hspace{-0.5em} S^{(m)_3}_{1|(s)_3}= \hspace{-0.5em}\sum_{J=(0;2;32;22)} \hspace{-0.5em} S^{(J)}_1\left[\phi^{(1)}, \phi^{(2)}, \Phi^{(3)}_J\right]\hspace{-0.15em} =  -2g \int d^dx \bigg[\sum_{j \geq0}^{[s/2]}t_j\sum_{i=0}^{[s/2-j]}\sum_{k \geq0}^{j}
     \sum_{l \geq0}^{j-k} \frac{C(l+i,h(s))}{ C(i,h(s))} \\
   && \hspace{-0.5em} \times   C_j^{k,l} \frac{(s-2j)!}{i!} \bigg\{\sum_{u=0}^{s-2j-2i} \frac{(-1)^u}{u!(s-2j-2i-u)!}\sum_{q=0}^i \sum_{t=0}^{i-q} \frac{(-1)^t C_i^{q,t}}{2^{2(j-k-l)+i-t}} \Big[\partial_{\nu_0}...\partial_{\nu_u}\Big(\Box^{q}  \nonumber\\
   && \hspace{-0.5em} \times \partial_{\nu_{u+1}}...\partial_{\nu_{u+t}}\phi^{(1)}\Big)  \Big] \Big[\partial_{\nu_{u+t+1}}...\partial_{\nu_{s-2j-2i+t}}\Big(\partial^{\nu_{u+1}}...\partial^{\nu_{u+t}}\Box^{i-q-t}\phi^{(2)}\Big)\Big]
    \bigg\}
\phi^{(3)\nu({s-2l-2i-2k})}_{l+i,2k}  \nonumber\\
   && \hspace{-0.5em} \quad \times\prod_{r=1}^{j-k-l}\eta_{\nu_{s-2j-2i+2r-1}\nu_{s-2j-2i+2r}}.\nonumber
 \end{eqnarray}
 \begin{eqnarray}
    && \hspace{-0.5em}\quad +  \sum_{j \geq0}^{[s/2]-1}t_j\sum_{i=0}^{[s/2-j-1]}\sum_{k \geq0}^{j}
     \sum_{l \geq0}^{j-k} \frac{C(l+i,h(s))}{ C(i,h(s))}C_j^{k,l}   \frac{(j+1)!(s-2(j+1))!}{j!i! } \nonumber \\
   && \hspace{-0.5em}\times    \bigg\{\hspace{-0.2em}\sum_{u=0}^{s-2(j+1)-2i}\hspace{-0.7em} \frac{(-1)^u}{u!(s-2(j+1)-2i-u)!}\sum_{q=0}^i \sum_{t=0}^{i-q}  \frac{(-1)^t C_i^{q,t}}{2^{2(j-k-l)+i-t}} \Big[\partial_{\nu_0}...\partial_{\nu_u}\Big(\Box^{q}\partial_{\nu_{u+1}}...\partial_{\nu_{u+t}}\phi^{(1)}\Big)  \Big]  \nonumber\\
   && \hspace{-0.5em} \times \Big[\partial_{\nu_{u+t+1}}...\partial_{\nu_{s-2(j+1)-2i+t}}\Big(\partial^{\nu_{u+1}}...\partial^{\nu_{u+t}}\Box^{i-q-t}\phi^{(2)}\Big)\Big]
    \bigg\}
\phi^{(3)\nu({s-2-2l-2i-2k})}_{2| l+i,2k}  \nonumber
 \\
   &&\hspace{-0.5em} \quad \times\prod_{r=1}^{j-k-l}\eta_{\nu_{s-2(j+1)-2i+2r-1}\nu_{s-2(j+1)-2i+2r}} \nonumber \\
    &&\hspace{-0.5em}
     \quad +  \sum_{j \geq0}^{[s/2]-2}t_j\sum_{i=0}^{[s/2-j-2]}\sum_{k \geq0}^{j}
     \sum_{l \geq0}^{j-k} \frac{C(l+i,h(s))}{ C(i,h(s))}C_j^{k,l}   \frac{(j+2)!(s-2(j+2))!}{j!i! } \nonumber \\
   && \hspace{-0.5em}\times  \bigg\{\hspace{-0.3em}\sum_{u=0}^{s-2(j+2)-2i}\hspace{-0.7em} \frac{(-1)^u}{u!(s-2(j+2)-2i-u)!}\sum_{q=0}^i \sum_{t=0}^{i-q} \frac{ (-1)^t C_i^{q,t} }{2^{2(j-k-l)+i-t}} \Big[\partial_{\nu_0}...\partial_{\nu_u}\Big(\Box^{q}\partial_{\nu_{u+1}}...\partial_{\nu_{u+t}}\phi^{(1)}\Big)  \Big]  \nonumber\\
   && \hspace{-0.5em}\times \Big[\partial_{\nu_{u+t+1}}...\partial_{\nu_{s-2(j+2)-2i+t}}\Big(\partial^{\nu_{u+1}}...\partial^{\nu_{u+t}}\Box^{i-q-t}\phi^{(2)}\Big)\Big]
    \bigg\}
\phi^{(3)\nu({s-2(l+2)-2i-2k})}_{32| l+i,2k}  \nonumber
  \end{eqnarray}
            \begin{eqnarray}
   &&\hspace{-0.5em} \quad \times\prod_{r=1}^{j-k-l}\eta_{\nu_{s-2(j+2)-2i+2r-1}\nu_{s-2(j+2)-2i+2r}}\nonumber\\
     && \hspace{-0.5em}\quad - \sum_{j \geq0}^{[s/2]-3}t_j\sum_{i=0}^{[s/2-j-3]}\sum_{k \geq0}^{j}
     \sum_{l \geq0}^{j-k} \frac{C(l+i,h(s))}{ C(i,h(s))}C_j^{k,l}   \frac{(j+3)!(s-2(j+3))!}{j!i! }\nonumber \\
   && \hspace{-0.5em}\times  \bigg\{\hspace{-0.3em}\sum_{u=0}^{s-2(j+3)-2i}\hspace{-0.7em} \frac{(-1)^u}{u!(s-2(j+3)-2i-u)!}\sum_{q=0}^i \sum_{t=0}^{i-q} \frac{(-1)^t C_i^{q,t}}{2^{2(j-k-l)+i-t}} \Big[\partial_{\nu_0}...\partial_{\nu_u}\Big(\Box^{q}\partial_{\nu_{u+1}}...\partial_{\nu_{u+t}}\phi^{(1)}\Big)  \Big]  \nonumber\\
   && \hspace{-0.5em}\times \Big[\partial_{\nu_{u+t+1}}...\partial_{\nu_{s-2(j+3)-2i+t}}\Big(\partial^{\nu_{u+1}}...\partial^{\nu_{u+t}}\Box^{i-q-t}\phi^{(2)}\Big)\Big]
    \bigg\}
 \phi^{(3)\nu({s-2(l+3)-2i-2k})}_{22| l+i,2k}  \nonumber\\
   &&\hspace{-0.5em}  \quad \times\prod_{r=1}^{j-k-l}\eta_{\nu_{s-2(j+3)-2i+2r-1}\nu_{s-2(j+3)-2i+2r}}\bigg] \nonumber  ,
  \end{eqnarray}
(for the naturals $C^{k,l}_j\equiv \frac{j!}{k!l!(j-k-l)!}$ and usual convention $\partial_{\nu_0}\equiv \prod_{i=1}^0 \partial_{\nu_i}  \equiv 1$)
    and   for the gauge transformations  (\ref{cubgtrex3gicomp11})
     \begin{eqnarray}
         &\hspace{-0.5em}&\hspace{-0.5em} \delta_{[1]} \phi^{(1)}(x_1) =
          -g \int d^dx \bigg[\sum_{j \geq0}^{[(s-1)/2]}\hspace{-0.5em}t_j\hspace{-0.5em}\sum_{i=0}^{[(s-1)/2-j]}\hspace{-0.3em}\sum_{k \geq0}^{j}
     \sum_{l \geq0}^{j-k} \frac{C(l+i,h(s))}{ C(i,h(s))}C_j^{k,l} \frac{(s-1-2j)!}{i!}  \label{cubgtrex3gicomp11a} \\
   &\hspace{-0.5em}&\hspace{-0.5em} \times    \bigg\{\sum_{u=0}^{s-1-2j-2i} \frac{1}{u!(s-1-2j-2i-u)!}\sum_{q=0}^i \sum_{t=0}^{i-q}  \frac{(-1)^t C_i^{q,t}}{2^{2(j-k-l)+i-t}}\Xi^{(3)\nu({s-1-2l-2i-2k})}_{l+i,2k} (x) \nonumber\\
   &\hspace{-0.5em}&\hspace{-0.5em} \times \Big[\partial_{\nu_{u+t+1}}...\partial_{\nu_{s-1-2j-2i+t}}\Big(\partial^{\nu_{u+1}}...\partial^{\nu_{u+t}}\Box^{i-q-t}\phi^{(2)}\Big)\Big] \Big[\partial_{\nu_0}...\partial_{\nu_u}\Big(\Box^{q}\partial_{\nu_{u+1}}...\partial_{\nu_{u+t}}\Big)  \Big] \nonumber\\
   &\hspace{-0.5em}&\hspace{-0.5em} \quad \times\prod_{r=1}^{j-k-l}\eta_{\nu_{s-1-2j-2i+2r-1}\nu_{s-1-2j-2i+2r}}
    \bigg\}\delta^{(d)}\big(x -  x_{1}\big)
     \nonumber
      \end{eqnarray}
 \begin{eqnarray}
  &\hspace{-0.5em}& \hspace{-0.5em}\quad  -
  \sum_{j \geq0}^{[(s-5)/2]}t_j\sum_{i=0}^{[(s-5)/2-j]}\sum_{k \geq0}^{j}
     \sum_{l \geq0}^{j-k} \frac{C(l+i,h(s))}{ C(i,h(s))}C_j^{k,l} \frac{(s-5-2j)!}{i!}  \nonumber \\
   &\hspace{-0.5em}& \hspace{-0.5em}\times    \bigg\{\sum_{u=0}^{s-5-2j-2i} \frac{1}{u!(s-5-2j-2i-u)!}\sum_{q=0}^i \sum_{t=0}^{i-q} \frac{(-1)^t C_i^{q,t} }{2^{2(j-k-l)+i-t}}\Xi^{(3)\nu({s-5-2l-2i-2k})}_{12|{}l+i,2k} (x) \nonumber\\
   &\hspace{-0.5em}& \hspace{-0.5em}\times \Big[\partial_{\nu_{u+t+1}}...\partial_{\nu_{s-5-2j-2i+t}}\Big(\partial^{\nu_{u+1}}...\partial^{\nu_{u+t}}\Box^{i-q-t}\phi^{(2)}\Big)\Big] \Big[\partial_{\nu_0}...\partial_{\nu_u}\Big(\Box^{q}\partial_{\nu_{u+1}}...\partial_{\nu_{u+t}}\Big)  \Big]\nonumber\\
        &\hspace{-0.5em}&\hspace{-0.5em} \quad \times\prod_{r=1}^{j-k-l}\eta_{\nu_{s-5-2j-2i+2r-1}\nu_{s-5-2j-2i+2r}}
    \bigg\}\delta^{(d)}\big(x -  x_{1}\big)
  \bigg] \equiv  \delta_{1| \Xi^{(3)}} \phi^{(1)}(x_1) +\delta_{1| \Xi^{(3)}_{12}} \phi^{(1)}(x_1); \nonumber \\
    &\hspace{-0.5em}& \hspace{-0.5em}\delta_{[1]} \phi^{(2)} (x_2) = -  \delta_{[1]} \phi^{(1)} (x_1)|_{[ \phi^{(1)} (x_1) \to  \phi^{(2)} (x_2)]}. \label{cubgtrex2gicomp21a}
\end{eqnarray}

 \section{Component interacting Lagrangian formulation for $(m,s)$, $(0,\lambda_i)$ for $ \lambda_i \leq 1$}\label{Singhcvomp12}
\renewcommand{\theequation}{\Alph{section}.\arabic{equation}}
\setcounter{equation}{0}

For the  case of interaction of massless vector and scalar fields with massive fields:  $(0,1)$, $(0,0)$, $(m,s)$ from the vertex (\ref{genvertex2}) the interacting part of action (\ref{S[n]1ind103})  in the ghost-independent component form
 looks
\begin{eqnarray} \label{S[n]1ind103a}
              S^{(m)_3}_{1|(1,0,s)}&=& \hspace{-0.7em}\sum_{J=(0, 2, 32, 22)}\hspace{-0.7em} S^{(J)}_1\left[\hspace{-0.15em}\phi^{(1)}, \phi^{(2)}, \Phi^{(3)}_J\hspace{-0.15em}\right] =   - g \prod_{i=2}^3 \delta^{(d)}\big(x_{1} -  x_{i}\big) \bigg[\hspace{-0.15em}{}_{0}\langle \phi^{(2)} \big|\bigg(\hspace{-0.2em}
 {}_{1}\langle \phi^{(1)}\big|\Big\{{}_{s}\langle \Phi^{(3)}
    \big|  \\      
    &\times&K^{(3)}\sum_{j \geq0}^{[s-1/2]}t_j (\check{L}{}^{(3)+}_{11})^j  \mathcal{L}^{(3)0}_{s-1-2j}
    + {}_{s-2}\langle \Phi^{(3)}_2 K^{(3)}
    \big|\sum_{j \geq0}^{[s-3/2]}t_j(j+1) (\check{L}{}^{(3)+}_{11})^j  \mathcal{L}^{(3)0}_{s-1-2(j+1)}  \nonumber \\
    &+ & {}_{s-4}\langle \Phi^{(3)}_{32} K^{(3)}
    \big|\sum_{j \geq0}^{[s-5/2]}t_j(j+1)(j+2) (\check{L}{}^{(3)+}_{11})^j  \mathcal{L}^{(3)0}_{s-1-2(j+2)} \nonumber \\
    &-& {}_{s-6}\langle \Phi^{(3)}_{22} K^{(3)}
    \big|\sum_{j \geq0}^{[s-7/2]}t_j(j+1)(j+2)(j+3) (\check{L}{}^{(3)+}_{11})^j  \mathcal{L}^{(3)0}_{s-1-2(j+3)}\Big\}{L}^{(31)+}_{11 |0}
      \nonumber
       \\
    &-&
     \frac{1}{2m_3}{}_{0}\langle \phi^{(1)}_1\big|\bigg\{{}_{s-1}\langle \Phi^{(3)}_1 K^{(3)}
    \big|\sum_{j \geq0}^{[s-1/2]}t_j (\check{L}{}^{(3)+}_{11})^j \Big(( \mathcal{L}^{(3)}_{s-2-2j})^{\prime}\Big[\frac{\widehat{p}{}^{(3)\mu}{a}^{(3)+}_{\mu}}{m_3}\nonumber\\
    &+&{d}^{(3)+}\Big]+\frac{1}{m_3} \mathcal{L}^{(3)}_{s-1-2j}\Big) - {}_{s-5}\langle \Phi^{(3)}_{13} K^{(3)}
    \big|\sum_{j \geq0}^{[s-5/2]}t_j(j+1)(j+2) (\check{L}{}^{(3)+}_{11})^j \nonumber\\
    &\times & \Big(( \mathcal{L}^{(3)}_{s-2(j+3)})^{\prime}\Big[\frac{\widehat{p}{}^{(3)\mu}{a}^{(3)+}_{\mu}}{m_3}+{d}^{(3)+}\Big]+\frac{1}{m_3} \mathcal{L}^{(3)0}_{s-1-2(j+2)}\Big)\bigg\}\bigg)|0\rangle +h.c.\bigg]. \nonumber
            \end{eqnarray}
            The deformed gauge transformations (\ref{+cubgtrex3gicomp10s}), (\ref{cubgtrex2gicomp})  for the massless fields
                 \begin{eqnarray}
         && \delta_{[1]} \big| \phi^{(1)} \rangle_{1}  =  -
g \prod_{i=2}^3 \delta^{(d)}\big(x_{1} -  x_{i}\big){}_{0}\langle \phi^{(2)}\big| \Big\{{}_{s-1}\langle \Xi^{(3)}
   K^{(3)}
    \big|\sum_{j \geq0}^{[s-1/2]}t_j  (\check{L}{}^{(3)+}_{11})^j\Big[(s-1\label{+cubgtrex3gicomp10sa}
    \end{eqnarray}
                             \begin{eqnarray}
    && \quad  -2j)\mathcal{L}^{(3)0}_{s-2-2j}{L}^{(31)+}_{11 |0} -\frac{1}{2m_3^2}(\widehat{p}{}^{(1)\mu}a^{(1)+}_{\mu})\mathcal{L}^{(3)0}_{s-1-2j} \Big]- {}_{s-5}\langle \Xi^{(3)}_{12}
   K^{(3)}
    \big|  \sum_{j \geq0}^{[s-5/2]}t_j \frac{(j+2)!}{j!} \nonumber \\
    &&\quad
    \times(\check{L}{}^{(3)+}_{11})^j  \Big[(s-5-2j)\mathcal{L}^{(3)0}_{s-6-2j} {L}^{(31)+}_{11 |0} -\frac{1}{2m_3^2}(\widehat{p}{}^{(1)\mu}a^{(1)+}_{\mu})\mathcal{L}^{(3)0}_{s-5-2j} \Big]
    \Big\}|0\rangle ; \nonumber
     \\
    && \delta_{[1]} \big| \phi^{(1)}_1 \rangle_{0}  =  -
\frac{g}{2} \prod_{i=2}^3 \delta^{(d)}\big(x_{1} -  x_{i}\big){}_{0}\langle \phi^{(2)}\big| \Big\{{}_{s-1}\langle \Xi^{(3)}
   K^{(3)}
    \big|\sum_{j \geq0}^{[s-1/2]}t_j (s-2j) (\check{L}{}^{(3)+}_{11})^j\label{-cubgtrex3gicomp10s1a} \\
    && \quad  \times\mathcal{L}^{(3)0}_{s-1-2j} {}_{s-5}-\langle \Xi^{(3)}_{12}
   K^{(3)}
    \big|  \sum_{j \geq0}^{[s-5/2]}t_j (j+1)(j+2)(s-4-2j) (\check{L}{}^{(3)+}_{11})^j  \mathcal{L}^{(3)0}_{s-5-2j}|\Big\}|0\rangle ; \nonumber
   \\
    && \delta_{[1]} \big| \phi^{(2)} \rangle_{0}  =
g \prod_{i=1,3}\delta^{(d)}\big(x_{2} -  x_{i}\big)\bigg[ {}_{1}\langle \phi^{(1)}\big| \Big\{{}_{s-1}\langle \Xi^{(3)}
   K^{(3)}
    \big|\sum_{j \geq0}^{[s-1/2]}t_j (\check{L}{}^{(3)+}_{11})^j\Big[ (s-1\label{cubgtrex2gicompa} \\
    && \quad  -2j)\mathcal{L}^{(3)0}_{s-2-2j}{L}^{(31)+}_{11 |0} -\frac{1}{2m_3^2}(\widehat{p}{}^{(1)\mu}a^{(1)+}_{\mu})\mathcal{L}^{(3)0}_{s-1-2j} \Big]- {}_{s-5}\langle \Xi^{(3)}_{12}
   K^{(3)}
    \big|  \sum_{j \geq0}^{[s-5/2]}t_j \frac{(j+2)!}{j!} \nonumber  \\
    &&\quad
    \times(\check{L}{}^{(3)+}_{11})^j\Big[(s-5-2j)\mathcal{L}^{(3)0}_{s-6-2j} {L}^{(31)+}_{11 |0} -\frac{1}{2m_3^2}(\widehat{p}{}^{(1)\mu}a^{(1)+}_{\mu})\mathcal{L}^{(3)0}_{s-5-2j} \Big]
    \Big\}|0\rangle  \nonumber
      \end{eqnarray}
     \begin{eqnarray}
&& \quad +\frac{1}{2}{}_{0}\langle \phi^{(1)}_1\big| \Big\{{}_{s-1}\langle \Xi^{(3)}
   K^{(3)}
    \big|\sum_{j \geq0}^{[s-1/2]}t_j (s-2j) (\check{L}{}^{(3)+}_{11})^j \mathcal{L}^{(3)0}_{s-1-2j}\nonumber \\
    && \qquad  - {}_{s-5}\langle \Xi^{(3)}_{12}
   K^{(3)}
    \big|  \sum_{j \geq0}^{[s-5/2]}t_j (j+1)(j+2)(s-4-2j) (\check{L}{}^{(3)+}_{11})^j  \mathcal{L}^{(3)0}_{s-5-2j}|\Big\}|0\rangle  \nonumber\\
&& \qquad + \frac{1}{2m_3} {}_{0}\langle \Xi^{(1)}
 \Big\{\big|{}_{s-1}\langle \Phi^{(3)}_1  K^{(3)}\big| \sum_{j \geq0}^{[s-1/2]}t_j (s-2j) (\check{L}{}^{(3)+}_{11})^j    \Big(( \mathcal{L}^{(3)}_{s-2-2j})^{\prime}\Big[\frac{\widehat{p}{}^{(3)\mu}{a}^{(3)+}_{\mu}}{m_3}\nonumber\\
    &&\qquad +{d}^{(3)+}\Big]+\frac{1}{m_3} \mathcal{L}^{(3)0}_{s-1-2j}\Big)\nonumber\\
    &&  \hspace{-0.5em} - {}_{s-5}\langle \Phi^{(3)}_{13}
   K^{(3)}
    \big|  \sum_{j \geq0}^{[s-5/2]}t_j (j+1)(j+2)(s-2(j+2)) (\check{L}{}^{(3)+}_{11})^j  \nonumber\\
    && \hspace{-0.5em}  \times  \Big(( \mathcal{L}^{(3)}_{s-2(j+3)})^{\prime}\Big[\frac{\widehat{p}{}^{(3)\mu}{a}^{(3)+}_{\mu}}{m_3}+{d}^{(3)+}\Big]+\frac{1}{m_3} \mathcal{L}^{(3)0}_{s-1-2(j+2)}\Big)\Big\}\bigg]|0\rangle,   \nonumber
    \end{eqnarray}
and for the components from the massive field $\big| \chi^{(3)}\rangle_s$ (\ref{cubgtrex3gicomp10s})
\begin{eqnarray}
         &\hspace{-0.7em}& \hspace{-0.7em}\delta_{1} \big| \Phi^{(3)} \rangle_{s}\hspace{-0.1em}   =\hspace{-0.1em}    \frac{g}{2m_3}{}_{0}\langle A^{(12)}_{(\phi^{(2)}, \Xi^{(1)})}\big|
\sum_{j \geq0}^{[s-1/2]}t_j  (\check{L}{}^{(3)+}_{11})^j \mathcal{L}^{(3)0}_{s-1-2j}\Big[\frac{\widehat{p}{}^{(3)\mu}{a}^{(3)+}_{\mu}}{m_3}+{d}^{(3)+}\Big]|0\rangle,\label{cubgtrex3gicomp10sa} \\
    &\hspace{-0.7em}& \hspace{-0.7em} \delta_{1} \big| \Phi^{(3)}_2 \rangle_{s-2} \hspace{-0.1em}   = \hspace{-0.1em}  - \frac{g}{2m_3}{}_{0}\langle A^{(12)}_{(\phi^{(2)}, \Xi^{(1)})}\hspace{-0.1em} \big|\hspace{-0.4em}
\sum_{j \geq0}^{[s-3/2]}\hspace{-0.3em}t_j (j\hspace{-0.1em} +\hspace{-0.1em} 1) (\check{L}{}^{(3)+}_{11})^j \mathcal{L}^{(3)0}_{s-3-2j}\hspace{-0.15em}\Big[\frac{\widehat{p}{}^{(3)\mu}{a}^{(3)+}_{\mu}}{m_3}\hspace{-0.1em} +\hspace{-0.1em} {d}^{(3)+}\hspace{-0.15em}\Big]\hspace{-0.1em} |0\rangle,\label{2cubgtrex3gicomp10sa}\\
  &\hspace{-0.7em}& \hspace{-0.7em} \delta_{1} \big| \Phi^{(3)}_{22}\rangle_{s-6} \hspace{-0.1em}   =\hspace{-0.1em} -  \frac{g}{2m_3}{}_{0}\langle A^{(12)}_{(\phi^{(2)}, \Xi^{(1)})}\hspace{-0.1em} \big|\hspace{-0.4em}
\sum_{j \geq0}^{[s-7/2]}\hspace{-0.3em}t_j \hspace{-0.15em} \frac{(j\hspace{-0.1em} +\hspace{-0.1em} 3)!}{j!} (\check{L}{}^{(3)+}_{11})^j\mathcal{L}^{(3)0}_{s-7-2j}\hspace{-0.15em}\Big[\frac{\hspace{-0.1em} \widehat{p}{}^{(3)\mu}{a}^{(3)+}_{\mu}}{m_3}\hspace{-0.1em} +\hspace{-0.1em} {d}^{(3)+}
\hspace{-0.15em}\Big]\hspace{-0.1em}  |0\rangle,\label{3cubgtrex3gicomp10sa}
 \end{eqnarray}
 \begin{eqnarray}
  &\hspace{-0.7em}& \hspace{-0.7em} \delta_{1} \big| \Phi^{(3)}_{32}\rangle_{s-4}   =  -\frac{g}{2m_3}{}_{0}\langle A^{(12)}_{(\phi^{(2)}, \Xi^{(1)})}\hspace{-0.1em} \big|\hspace{-0.4em}
\sum_{j \geq0}^{[s-5/2]}\hspace{-0.3em}t_j \hspace{-0.15em}\frac{(j\hspace{-0.1em} +\hspace{-0.1em} 2)!}{j!} (\check{L}{}^{(3)+}_{11})^j \mathcal{L}^{(3)0}_{s-5-2j}\hspace{-0.15em}\Big[\hspace{-0.1em} \frac{\widehat{p}{}^{(3)\mu}{a}^{(3)+}_{\mu}}{m_3}\hspace{-0.1em} +\hspace{-0.1em} {d}^{(3)+}\hspace{-0.15em}\Big]\hspace{-0.1em}  |0\rangle \label{4cubgtrex3gicomp10sa}
\\
  &\hspace{-0.7em}& \hspace{-0.7em}  \delta_{1} \big| \Phi^{(3)}_{1}\rangle_{s-1}   =  \frac{g}{2}{}_{0}\langle A^{(12)}_{(\phi^{(2)}, \Xi^{(1)})}\big|
\sum_{j \geq0}^{[s-1/2]}t_j(\check{L}{}^{(3)+}_{11})^j \mathcal{L}^{(3)0}_{s-1-2j}|0\rangle,\label{5cubgtrex3gicomp10sa}
\\
  &\hspace{-0.7em}& \hspace{-0.7em}  \delta_{1} \big| \Phi^{(3)}_{13}\rangle_{s-5}   =   \frac{g}{2}{}_{0}\langle A^{(12)}_{(\phi^{(2)}, \Xi^{(1)})}\big|
\sum_{j \geq0}^{[s-5/2]}t_j \frac{(j + 2)!}{j!}(\check{L}{}^{(3)+}_{11})^j \mathcal{L}^{(3)0}_{s-5-2j}|0\rangle ,\label{6cubgtrex3gicomp10sa}
\end{eqnarray}
(for ${}_{0}\langle A^{(12)}_{(\phi^{(2)}, \Xi^{(1)})}\big| =\prod_{i=1}^2 \delta^{(d)}\big(x_{3} -  x_{i}\big){}_{0}\langle \phi^{(2)}\big| {}_{0}\langle \Xi^{(1)}
    \big|$).

 \section{On consistency of Lagrangian dynamics for interacting  higher-spin fields in constrained BRST approach}\label{appconBRST}
\renewcommand{\theequation}{\Alph{section}.\arabic{equation}}
\setcounter{equation}{0}

Let the field and gauge parameter vectors with given spins $s_1, s_2,s_3$ be traceless, and the incomplete
total BRST operator $Q_c^{tot}$ = $\sum_jQ_c^{(j)}$ forms with  traceless  constraints and constrained spin operators closed superalgebra:
\begin{eqnarray}\label{L11}
    && L^{(i)}_{11} \Big(|\chi^{(j)}_c\rangle_{s_j}, |\Lambda^{(j)}_c\rangle_{s_j}\Big) =0, \\
         &&   \sigma^{(i)}_{c} \Big(|\chi^{(i)}_c\rangle_{s_i}, |\Lambda^{(i)}_c\rangle_{s_i}\Big) = \Big(s_i-1+\frac{d+\theta_{m_i,0}}{2}\Big)\Big(|\chi^{(i)}_c\rangle_{s_i}, |\Lambda^{(i)}_c\rangle_{s_i}\Big), \label{s11} \\
      &&   (Q_c^{tot})^2 \ = \  [ Q_c^{tot}, \,  L^{(i)}_{11}\} \ =\ [ Q_c^{tot}, \,  \sigma^{(i)}_{c}\}= 0, \ \ [\sigma^{(i)}_{c},\,  L^{(j)}_{11}\}= -2\delta^{ij} L^{(j)}_{11} , \label{QsL11}
\end{eqnarray}
(for $i,j=1,2,3$) where with account for (\ref{extconstsp2}), (\ref{extconstsp21}), (\ref{extconstsp3})
\begin{eqnarray}\label{L11exp}
 &\hspace{-0.5em} &\hspace{-0.5em}   L^{(i)}_{11} = \widehat{L}^{(i)}_{11}\vert_{b^{(i)+}=b^{(i)}=0} = {l}^{(i)}_{11}- (1/2)(d^{(i)})^2+\eta^{(i)}_{1} \mathcal{P}^{(i)}_{1}\equiv
 \hat{l}^{(i)}_{11}+\eta^{(i)}_{1} \mathcal{P}^{(i)}_{1}  ,\\
 &\hspace{-0.5em} &\hspace{-0.5em} {\sigma}^{(i)}_c  = {\sigma}^{(i)}\vert_{b^{(i)+}=\eta^{(i)+}_{11} =\mathcal{P}^{(i)+}_{11}=0} =   g^{(i)}_0 + \theta_{m_i,0} d^{(i)+}d^{(i)} + \frac{1}{2}+ \eta_1^{(i)+}\mathcal{P}^{(i)}_{1}
-\eta^{(i)}_1\mathcal{P}_{1}^{(i)+}  ,
\label{extconstsp5}\\
 &\hspace{-0.5em} &\hspace{-0.5em}  |\chi^{(i)}_c\rangle_{s_i} = |\chi^{(i)}\rangle_{s_i}\vert_{b^{(i)+}=\eta^{(i)+}_{11} =\mathcal{P}^{(i)+}_{11}=0} = |\Phi^{(i)}\rangle_{s_i}-\mathcal{P}_1^{(i)+}\big(\eta_0^{(i)}|\Phi^{(i)}_1\rangle_{s_i-1}+\eta_1^{(i)+}|\Phi^{(i)}_2\rangle_{s_i-2}\big), \label{L11exp1} \\
  &\hspace{-0.5em} &\hspace{-0.5em}  |\Lambda^{(i)}_c\rangle_{s_i} = |\Lambda^{(i)}\rangle_{s_i}\vert_{b^{(i)+}=\eta^{(i)+}_{11} =\mathcal{P}^{(i)+}_{11}=0} = \mathcal{P}_1^{(i)+}|\Xi^{(i)}\rangle_{s_i-1}, \label{L11exp2}\\
  &\hspace{-0.5em} &\hspace{-0.5em} Q_c^{(i)} = {Q}^{(i)}\vert_{b^{(i)+}=\eta^{(i)+}_{11} =\mathcal{P}^{(i)+}_{11}=0} =
\eta^{(i)}_0l^{(i)}_0+\eta_1^{(i)+}\check{l}^{(i)}_1+\check{l}_1^{(i)+}\eta^{(i)}_1+
{\imath}\eta_1^{(i)+}\eta^{(i)}_1{\cal{}P}^{(i)}_0.
\label{Qctotsymcon}
\end{eqnarray}
Then, the Lagrangian formulation for free irreducible higher spin fields of spins $(s_1,s_2,s_3)$ (massless or massive) is determined by the gauge-invariant Lagrangian  with holonomic (traceless) constraints
\begin{eqnarray}
\label{PhysStatetotc}
\sum_i \mathcal{S}^{m_i}_{0C|s_i}[|\chi^{(i)}_c\rangle_{s_i}] = \sum_i\int d\eta^{(i)}_0 {}_{s_i}\langle\chi^{(i)}_c|
Q^{(i)}|\chi^{(i)}_c\rangle_{s_i}, \ \delta_0\big(|\chi^{(i)}_c\rangle_{s_i}, \, |\Lambda^{(i)}_c\rangle_{s_i}\big) = \big( Q^{(i)}|\Lambda^{(i)}_c\rangle_{s_i} , \, 0 \big).
\end{eqnarray}
Note, that any field representative ($\big|\widetilde{\chi}^{(i)}_c\rangle_{s_i}$) from the gauge orbit
\begin{equation}\label{goc}
\mathcal{O}_{0|\chi^{(i)}_c} = \big\{\big|\widetilde{\chi}^{(i)}_c\rangle_{s_i} \big|\, \  \big|\widetilde{\chi}^{(i)}_c\rangle_{s_i} =\big|{\chi}^{(i)}_c\rangle_{s_i} +  Q^{(i)}_c \big|\Lambda^{(i)}_c\rangle_{s_i}, \, \forall \big|\Lambda^{(i)}_c\rangle_{s_i}\big\}
\end{equation}
 remains by traceless if the field $\big|{\chi}^{(i)}_c\rangle$ and any gauge parameter $ \big|\Lambda^{(i)}_c\rangle$  are traceless because of the commutation of
 $L^{(i)}_{11}$ with $Q_c^{tot}$,\  (\ref{QsL11}).

In case of cubic interaction adopted for constrained BRST approach the interacting action and deformed gauge transformations are defined according to (\ref{S[n]}), (\ref{cubgtr}) without ghost and auxiliary oscillators associated with trace constraints
\begin{eqnarray}\label{S[n]c}
  &\hspace{-0.5em} &\hspace{-0.5em} S^{(m)_3}_{[1]C|(s)_3}[\chi^{(1)}_c,\chi^{(2)}_c, \chi^{(3)}_c] \ = \  \sum_{i=1}^{3} \mathcal{S}^{m_i}_{0C|s_i}   +
  g  \int \prod_{e=1}^{3} d\eta^{(e)}_0  \Big( {}_{s_{e}}\langle \chi^{(e)}_c  \big|  V^{(3)}_c\rangle^{(m)_3}_{(s)_{3}}+h.c. \Big)  , \\
   &\hspace{-0.5em} &\hspace{-0.5em} \delta_{[1]} \big| \chi^{(i)}_c \rangle_{s_i}  =  Q^{(i)}_c \big| \Lambda^{(i)}_c \rangle_{s_i} -
g \int \prod_{e=1}^{2} d\eta^{(i+e)}_0  \Big( {}_{s_{i+1}}\langle
\Lambda^{({i+1})}_c\big|{}_{s_{i+2}}
   \langle \chi^{({i+2})}_c\big|
\nonumber \\
   &\hspace{-0.5em} &\hspace{-0.5em}
\ \   \phantom{\delta_{[1]} \big| \chi^{(i)} \rangle_{s_i}}   +(i+1 \leftrightarrow i+2)\Big)
\big|\widetilde{V}{}^{(3)}_c\rangle^{(m)_3}_{(s)_{3}}  \label{cubgtrc}
\end{eqnarray}
with unknown vertex operators $| V^{(3)}_c\rangle^{(m)_3}_{(s)_{3}}$, $ |\widetilde{V}{}^{(3)}_c\rangle^{(m)_3}_{(s)_{3}}$.
In particular case, when these operators coincide: $V^{(3)}_c\rangle^{(m)_3}_{(s)_{3}} =  \widetilde{V}{}^{(3)}_c\rangle^{(m)_3}_{(s)_{3}}$,  the requirement of
consistent deformation for the classical action and initial gauge transformations means the validity of the equations (for $i=1,2,3$)
\begin{eqnarray}\label{gencubBRSTc}
 &&  Q_c^{tot}
\big|{V}{}^{(3)}_c\rangle^{(m)_3}_{(s)_{3}} =0,   \qquad     L^{(i)}_{11} \big|{V}{}^{(3)}_c\rangle^{(m)_3}_{(s)_{3}} \ =\ 0, \\
 && \sigma^{(i)}_c\big|{V}{}^{(3)}_c\rangle^{(m)_3}_{(s)_{3}}\ =\  \Big(s_i+\frac{d-2+\theta_{m_i,0}}{2}\Big)\big|{V}{}^{(3)}_c\rangle^{(m)_3}_{(s)_{3}} .
\label{gencubBRSTc1}
\end{eqnarray}

Let us verify  that   any representative ($\big|\widetilde{\chi}^{(i)}_c\rangle_{s_i}$ from arbitrary  gauge orbit  $\mathcal{O}_{[1]|\chi^{(i)}_c}$
\begin{equation}\label{goc1}
\mathcal{O}_{[1]|\chi^{(i)}_c} = \big\{\big|\widetilde{\chi}^{(i)}_c\rangle_{s_i} \big|\ \  \big|\widetilde{\chi}^{(i)}_c\rangle_{s_i} =\big|{\chi}^{(i)}_c\rangle_{s_i} + \delta_{[1]}|\chi^{(i)}\rangle_{s_i}, \, \forall \big|\Lambda^{(j)}_c\rangle_{s_i}, \ \ j=1,2,3 \big\}
\end{equation}
 for interacting fields remains by traceless
after applying the deformed gauge transformations (as it was done for the undeformed fields)
and also check the same for the deformed field equations. Then, it is sufficient to find that
\begin{eqnarray}
&\hspace{-0.5em} &\hspace{-0.5em}   L^{(i)}_{11}\delta_{[1]}|\chi^{(i)}_c\rangle_{s_i}= L^{(i)}_{11}Q_c^{(i)}|\Lambda^{(i)}_c\rangle_{s_i} - g \int d\eta^{{i+1}}_0d\eta^{{i+2}}_0\Big( {}_{s_{i+1}}\langle
\Lambda^{({i+1})}_c\big|{}_{s_{i+2}}
   \langle \chi^{({i+2})}_c\big|  \nonumber\\
   &\hspace{-0.5em} &\hspace{-0.5em}
\ \   \phantom{\delta_{[1]} \big| \chi^{(i)} \rangle_{s_i}} +(i+1 \leftrightarrow i+2)\Big)
L^{(i)}_{11}\big|{V}{}^{(3)}_c\rangle^{(m)_3}_{(s)_{3}} = 0, \label{cubgtrcc} \\
  &\hspace{-0.5em} &\hspace{-0.5em}  L^{(j)}_{11}\frac{\overrightarrow{\delta} S^{(m)_3}_{[1]C|(s)_3}}{\delta {}_{s_{i}}
   \langle \chi^{({i})}_c\big|}  = L^{(j)}_{11}Q_c^{(i)}|\chi^{(i)}_c\rangle_{s_i} + g  \int  \prod_{e=1}^2 d\eta^{(i+e)}_0   {}_{s_{i+e}}\langle \chi^{(i+e)}_c \big|
 L^{(j)}_{11} \big|  V^{(3)}_c\rangle^{(m)_3}_{(s)_{3}}= 0. \label{cubgtr1c}
\end{eqnarray}

We proved, that imposing of only traceless constraints on fields and gauge parameters (\ref{L11}) represents the necessary but not sufficient condition for the consistency
of deformed (on cubic level) Lagrangian dynamics.
Indeed, in this case the latter term in (\ref{cubgtrcc}), (but not in (\ref{cubgtr1c})) does not vanish leading to $ L^{(i)}_{11}\delta_{[1]}|\chi^{(i)}_c\rangle_{s_i}\ne 0$ (but $L^{(j)}_{11}{\overrightarrow{\delta} S^{(m)_3}_{[1]C|(s)_3}}/{\delta {}_{s_{i}}
   \langle \chi^{({i})}_c\big|}  =0$). Indeed, due to solutions for traceless constraints (\ref{L11exp}) the fields and gauge parameters take the form (for $\hat{l}^{(i)}_{11}\big(\big|\overline{\Phi}{}^{(i)}_1\rangle, \big|\overline{\Xi}{}^{(i)}\rangle\big) \equiv 0$)
   \begin{equation}\label{solphi}
   \hspace{-1em}\left(\hspace{-0.15em}\big|\overline{\chi}{}^{(i)}_c\rangle_{s_i},  \big|\overline{\Lambda}{}^{(i)}_c\rangle_{s_i}\hspace{-0.17em}\right) = \left(\hspace{-0.15em}\big|\overline{\Phi}{}^{(i)}\rangle_{s_i}- \mathcal{P}{}^{(i)+}_1\big[\eta{}^{(i)}_0 \big|\overline{\Phi}{}^{(i)}_1\rangle_{s_i-1}-\eta{}^{(i)+}_1 \big|\hat{l}^{(i)}_{11}\overline{\Phi}{}^{(i)}\rangle_{s_i-2}\big] ,  \mathcal{P}{}^{(i)+}_1 \big|\overline{\Xi}{}^{(i)}\rangle_{s_i-1}\hspace{-0.17em}\right)\hspace{-0.15em}.
   \end{equation}
In view of completeness of the inner product in the total Hilbert space in question the solutions (\ref{solphi}) generate (hermitian) projectors $P^{(j)}_{m|11}$, $m=0,1$ on the subspaces of traceless field  and gauge vectors:
\begin{equation}\label{projectors}
  \left({}_{s_i}\langle\overline{\chi}{}^{(i)}_c\big|, {}_{s_i}\langle\overline{\Lambda}{}^{(i)}_c\big| \right)\equiv \left({}_{s_i}\langle{\chi}{}^{(i)}_c\big|P^{(j)}_{0|11},\   {}_{s_i}\langle{\Lambda}{}^{(i)}_c\big|P^{(j)}_{1|11}\right),
\end{equation}
such that the following cubic vertices:
\begin{equation}\label{projectors}
   \left(\big|  \overline{V}{}^{(3)}_c\rangle^{(m)_3}_{(s)_{3}}, \widehat{V}{}^{(3)}_c\rangle^{(m)_3}_{(s)_{3}}\right)=  \left(\prod_{j=1}^3P^{(j)}_{0|11}\big|  V^{(3)}_c\rangle^{(m)_3}_{(s)_{3}} ,\   P^{(i+1)}_{0|11}P^{(i+2)}_{1|11}\big|  V^{(3)}_c\rangle^{(m)_3}_{(s)_{3}}\right),
\end{equation}
survive respectively in the action (\ref{S[n]c}) and in the gauge transformations (\ref{cubgtrc}).
Namely, the vertex $\big|  \overline{V}{}^{(3)}_c\rangle^{(m)_3}_{(s)_{3}}$ respects the irreducibility of   the triple of interacting higher spin fields, whereas
$\widehat{V}{}^{(3)}_c\rangle^{(m)_3}_{(s)_{3}}$ does not respect that property, because of
\begin{equation}\label{violtraceless}
L^{(i)}_{11}\big|\widehat{V}{}^{(3)}_c\rangle^{(m)_3}_{(s)_{3}} \ne 0.
\end{equation}
Therefore, if the cubic vertex is not vanishing at acting of  $ L^{(i)}_{11}$ constraints, then, first, the
deformed Lagrangian dynamics is contradictory; second, the
interacting fields do not belong to the Poincar\'{e} group
irreducible space of a certain mass and spin (they do not satisfy
the traceless condition(\ref{goc1}) along the any entire gauge
orbit $\mathcal{O}_{[1]|\chi^{(i)}_c}$); third, the supermatrix of second derivatives of the
deformed action with respect to all the fields evaluated on the
deformed  mass shell has  proper eigen-vectors not respecting traceless properties (\ref{violtraceless}), albeit for the undeformed action. Therefore, the
number of physical degrees of freedom, which is determined by one of independent initial data for the equations of motion (partial differential equations) due to  (\ref{cubgtr1c})   for the interacting model is differed with that for the undeformed model with vanishing traceless constraints evaluated on respecting equations of motion.

We emphasize, the problem of deriving covariant cubic vertices for  interacting
higher integer spin fields realizing irreducible Poincare group representations has not been completely solved in the BRST approach with an incomplete BRST operator
\cite{BRST-BV3}.

It is easy to find  $L^{(i)}_{11}$-traceless solution  $\big|  \overline{V}{}^{(3)}_c\rangle^{(m)_3}_{(s)_{3}} \equiv |{V}{}^{M(3)}_{irrep}\rangle$   of the equations (\ref{gencubBRSTc})   with  $Q_c^{tot}$-closed  vertex  $\big|  V^{(3)}_c\rangle^{(m)_3}_{(s)_{3}}$   as follows
\begin{eqnarray}\label{L11V+}
 && \big| \overline{ V}{}^{(3)}_c\rangle^{(m)_3}_{(s)_{3}} = \ \bigg(1 -
\sum_i  \frac{1}{s_i-2+d/2}L^{(i)+}_{11} L^{(i)}_{11}   \\
&& \quad + \sum_{i_1}\bigg[  \prod_{k=1}^2\frac{1}{k(s_{i_1}-1-k+d/2)}(L^{(i_1)+}_{11})^2 (L^{(i_1)}_{11})^2  +
\sum_{i_2>i_1} \prod_{k=1}^2\frac{1}{(s_{i_k}-2+d/2)}L^{(i_k)+}_{11} L^{(i_k)}_{11}\bigg] \nonumber\\
&& \quad - \sum_{i_1}\bigg[  \prod_{k=1}^3\frac{1}{k(s_{i_1}-1-k+d/2)}(L^{(i_1)+}_{11})^3 (L^{(i_1)}_{11})^3  +
\sum_{i_2>i_1} \prod_{k=1}^2\frac{1}{k(s_{i_1}-1-k+d/2)}\times\nonumber\\
 && \quad \times\frac{1}{(s_{i_2}-2+d/2)}(L^{(i_1)+}_{11})^2L^{(i_2)+}_{11}(L^{(i_1)}_{11})^2 L^{(i_2)}_{11} +  \prod_{k=1}^3\frac{1}{(s_{k}-2+d/2)}L^{(k)+}_{11} L^{(k)}_{11} \bigg] \nonumber\\
 && \quad + \ldots\ldots\ldots\ldots\ldots\ldots\ldots\ldots\ldots\ldots\ldots\ldots\ldots\ldots\ldots\ldots\nonumber\\
 && \quad
  + (-1)^{\textstyle\sum_i\left[\frac{s_i}{2}\right]}  \prod_{i=1}^3\bigg\{\prod_{k=1}^{\left[\frac{s_{i}}{2}\right]}\frac{1}{k(s_{i}-1-k+d/2)}(L^{(i)+}_{11})^{\left[\frac{s_{i}}{2}\right]}(L^{(i)}_{11})^{\left[\frac{s_{i}}{2}\right]} \bigg\}\bigg)\big|  V^{(3)}_c\rangle^{(m)_3}_{(s)_{3}} \nonumber .
\end{eqnarray}
The  substituting of found cubic vertex $\big| \overline{ V}{}^{(3)}_c\rangle$ (\ref{L11V+}) into (\ref{S[n]c}) and (\ref{cubgtrc}),  leads for the same properties of the Lagrangian formulation for interacting  fields with given spins as ones for undeformed  model for triple of free fields\footnote{In case of the BRST-BV approach with incomplete BRST operator when the field and transformed (into ghost field)  gauge parameter are combined together with theirs antifield vectors within unique generalized field-antifield vector, $|\chi^{(i)}_{g|c}\rangle_{s_i}$, the deformed minimal BRST-BV action,  $S^{(m)_3}_{[1]C|(s)_3}[\chi^{(1)}_{g|c},\chi^{(2)}_{g|c}, \chi^{(3)}_{g|c}]$ = $S^{(m)_3}_{[1]C|(s)_3}[\chi^{(1)}_{c},\chi^{(2)}_{c}, \chi^{(3)}_{c}]\big|_{(\chi^{(i)}_{c}\to \chi^{(3)}_{g|c})}$ obtained from deformed classical action (\ref{S[n]c}) (see for details \cite{ReshBRSTBV} to be applicable for the method with off-shell constraints) will completely select the traceless parts from the vertex $\big|  V^{(3)}_c\rangle^{(m)_3}_{(s)_{3}}$ in the form $\big|  \overline{V}{}^{(3)}_c\rangle^{(m)_3}_{(s)_{3}}$ (\ref{projectors}) or (\ref{L11V+})  without the vertex $\widehat{V}{}^{(3)}_c\rangle^{(m)_3}_{(s)_{3}}$ being by the source for destroying an irreducibility for interacting higher spin fields in the constrained BRST approach \cite{BRST-BV3}}.

Finally, when expressing of the  ghost-independent fields $|\Phi^{(i)}_k\rangle_{s_i-k}$, $k=1,2$  in the triplet (\ref{L11exp1})  in terms of the single Fronsdal field  $|\Phi^{(i)}\rangle_{s_i}$ (for simplicity for triple of massless fields, i.e. for vanishing $d^{(i)+}$) from undeformed equations of motion and traceless constraints  as
\begin{equation}\label{fronsexp}
     |\Phi^{(i)}_1\rangle_{s_i-1} = l^{(i)}_1|\Phi^{(i)}\rangle_{s_i} - l^{(i)+}_1|\Phi^{(i)}_2\rangle_{s_i-2},\quad
     l_{11}^{(i)}|\Phi^{(i)}\rangle_{s_i} = -|\Phi^{(i)}_2\rangle_{s_i-2}
\end{equation}
 we get Fronsdal  Lagrangians with single double  traceless fields $|\Phi^{(i)}\rangle_{s_i}$ for the actions $ \mathcal{S}^{m_i}_{0C|s_i}$, so that
 the components of the  cubic vertex $\big|  V^{(3)}_c\rangle^{(m)_3}_{(s)_{3}}$  after calculating of ghost-pairings in (\ref{S[n]c}) and (\ref{cubgtrc})
 should satisfy to the consequences from the traceless equations to get the interacting action which encodes the noncontradictory dynamics.
 Unfortunately, we can not find a solution of this problem elsewhere, see  e.g. in \cite{Manvelyan}, \cite{Manvelyan1}, \cite{Joung}  where the  cubic vertices have been constructed  in the Lorentz covariant form.

\end{document}